\pdfoutput=1

\documentclass[aps,prb,notitlepage,showkeys,showpacs,superscriptaddress,nofootinbib,longbibliography,preprintnumbers,11pt]{revtex4-2}

\usepackage[utf8]{inputenc}

\usepackage{here}
\usepackage{ulem} 

\usepackage{url}

\usepackage{dsfont}

\usepackage[top=30truemm,bottom=30truemm,left=25truemm,right=25truemm]{geometry}

\usepackage{spectralsequences}
\usepackage{tikz}
\usetikzlibrary{intersections}
\usetikzlibrary{calc}

\usepackage{xcolor}

\newcommand{\tr}[1]{\mathrm{tr}\left(#1\right)}

\usepackage{tikz-cd}

\usepackage{slashed}

\usepackage{amsthm}
\usepackage{thmtools}
\usepackage{blindtext}
\usepackage{braket}

\declaretheoremstyle[
       shaded={bgcolor=\color{rgb}{0.9,0.9,0.9}}  
]{theorem}

\declaretheoremstyle[
       shaded={bgcolor=\color{rgb}{0.9,0.9,0.9}}  
]{remark}

\declaretheoremstyle[
       shaded={bgcolor=\color{rgb}{0.9,0.9,0.9}}  
]{proposition}

\declaretheoremstyle[
       shaded={bgcolor=\color{rgb}{0.9,0.9,0.9}}  
]{axiom}

\declaretheoremstyle[
       shaded={bgcolor=\color{rgb}{0.9,0.9,0.9}}  
]{lemma}

\theoremstyle{plain}
\newtheorem{definition}{Definition}

\theoremstyle{plain}

\theoremstyle{plain}
\newtheorem{question}{Question}

\theoremstyle{plain}
\newtheorem{assumption}{Assumption}

\theoremstyle{plain}
\newtheorem{corollary}{Corollary}

\newcommand{\Hom}{\text{Hom}}
\newcommand{\End}{\text{End}}

\newcommand{\Rep}{\text{Rep}}

\def\1{\mathds{1}}
\def\id{{\mathrm{id}}}

\newcommand{\ldim}{\mathrm{ldim}}
\newcommand{\rdim}{\mathrm{rdim}}
\newcommand{\qdim}{\mathrm{qdim}}
\newcommand{\ind}{\mathrm{ind}}

\newcommand{\lrank}{\mathrm{rank}_l}
\newcommand{\rrank}{\mathrm{rank}_r}
\newcommand{\rank}{\mathrm{rank}}

\usepackage{mathrsfs}
\usepackage{amsmath,amssymb}

\usepackage{mathtools}

\usepackage{esvect}

\usepackage{adjustbox}

\numberwithin{equation}{section}
\numberwithin{definition}{section}

\usepackage[status=draft,inline,nomargin]{fixme}
\fxusetheme{color}
\FXRegisterAuthor{ki}{aki}{KI}

\usetikzlibrary{decorations.pathmorphing,shapes}
\newcounter{sarrow}

\usepackage{xurl}

\usepackage{hyperref}
\hypersetup{colorlinks=true,linktoc=page,citecolor=red,linkcolor=blue}

\begin{document}

\title{Remarks on non-invertible symmetries on a tensor product Hilbert space in 1+1 dimensions}

\author{Kansei Inamura}
\email{kansei.inamura@physics.ox.ac.uk}
\affiliation{Mathematical Institute, University of Oxford, Andrew Wiles Building, Woodstock Road, Oxford, OX2 6GG, UK}
\affiliation{Rudolf Peierls Centre for Theoretical Physics, University of Oxford, Parks Road, Oxford, OX1 3PU, UK}

\begin{abstract}
We propose an index of non-invertible symmetry operators in 1+1 dimensions and discuss its relation to the realizability of non-invertible symmetries on the tensor product of finite dimensional on-site Hilbert spaces on the lattice.
Our index generalizes the Gross-Nesme-Vogts-Werner index of invertible symmetry operators represented by quantum cellular automata (QCAs).
Assuming that all fusion channels of symmetry operators have the same index, we show that the fusion rules of finitely many symmetry operators on a tensor product Hilbert space can agree, up to QCAs, only with those of weakly integral fusion categories.
We also discuss an attempt to establish an index theory for non-invertible symmetries within the framework of tensor networks.
To this end, we first propose a general class of matrix product operators (MPOs) that describe non-invertible symmetries on a tensor product Hilbert space.
These MPOs, which we refer to as topological injective MPOs, include all invertible symmetries, non-anomalous fusion category symmetries, and the Kramers-Wannier symmetries for finite abelian groups.
For topological injective MPOs, we construct the defect Hilbert spaces and the corresponding sequential quantum circuit representations.
We also show that all fusion channels of topological injective MPOs have the same index if there exist fusion and splitting tensors that satisfy appropriate conditions.
The existence of such fusion and splitting tensors has not been proven in general, although we construct them explicitly for all examples of topological injective MPOs listed above.
\end{abstract}

\maketitle

\setcounter{tocdepth}{3}
\tableofcontents

\section{Introduction}
\label{sec: Introduction}
Symmetry is one of the guiding principles in theoretical physics.
In the last decade, the notion of symmetry has been generalized in various ways.
Prominent examples of generalized symmetries are those generated by non-invertible conserved operators.
Such symmetries are called non-invertible symmetries; see \cite{McGreevy:2022oyu, Cordova:2022ruw, Schafer-Nameki:2023jdn, Brennan:2023mmt, Shao:2023gho, Carqueville:2023jhb, Iqbal:2024pee, Costa:2024wks} for reviews.
Given the generalizations of the notion of symmetry, it is natural to ask what kinds of mathematical structures symmetries of quantum many-body systems generally have.

In relativistic quantum field theories (QFTs), symmetry operators/defects are defined by topological operators/defects \cite{Gaiotto:2014kfa}.
Based on this definition, it is expected that finite internal non-invertible symmetries in 1+1 dimensions are generally described by fusion categories \cite{Bhardwaj:2017xup, Chang:2018iay, Thorngren:2019iar}.
There are many examples of non-invertible symmetries in 1+1d, such as the Kramers-Wannier duality symmetry of the Ising CFT and more general symmetries generated by the Verlinde lines of diagonal rational conformal field theories \cite{Chang:2018iay, Thorngren:2021yso, Verlinde:1988sn, Petkova:2000ip, Fuchs:2002cm, Frohlich:2004ef, Frohlich:2006ch, Fuchs:2007tx}.
Notably, it is known that any fusion category can be realized as the symmetry of a topological QFT \cite{Thorngren:2019iar, Komargodski:2020mxz, Huang:2021zvu, Inamura:2021szw}.

On the lattice, however, the problem is more subtle.
This is because the symmetry structure on the lattice is strongly constrained by the fact that the state space is given by the tensor product of finite dimensional local Hilbert spaces.\footnote{There are also lattice models, such as the anyon chain models \cite{Feiguin:2006ydp, Buican:2017rxc, Aasen:2020jwb}, whose state space does not admit a tensor product decomposition. We will not consider such models in this paper, except for Section~\ref{sec: without tensor product decomposition}. We will also not consider the case where the local Hilbert space has an infinite dimension.}
Indeed, it was recently shown in \cite{Evans:2025msy} that a fusion category $\mathcal{C}$ can be realized as the symmetry on a tensor product Hilbert space if and only if $\mathcal{C}$ is integral, i.e., all objects of $\mathcal{C}$ have integral quantum dimensions.
Nevertheless, even if $\mathcal{C}$ is not integral, it sometimes admits a slightly modified realization on a tensor product Hilbert space.
For example, as clarified in \cite{Seiberg:2023cdc, Seiberg:2024gek}, the fusion rules of the Ising fusion category, which is an example of a non-integral fusion category, can be realized on a tensor product Hilbert space of qubits if we allow mixing with the lattice translation.
More specifically, the Kramers-Wannier duality operator $\mathsf{D}_{\mathrm{KW}}$ on the lattice obeys the fusion rule \cite{Seiberg:2023cdc, Seiberg:2024gek}
\begin{equation}
\mathsf{D}_{\mathrm{KW}} \mathsf{D}_{\mathrm{KW}} = T (1+U),
\end{equation}
where $T$ is the lattice translation by one site and $U$ is the spin-flip operator generating the $\mathbb{Z}_2$ symmetry of the Ising model.
More general examples of this type of realization were studied recently in \cite{Lu:2026rhb}.

The above example suggests that the fusion rules of some non-integral fusion categories can be realized on a tensor product Hilbert space if we allow mixing with lattice translations or more general quantum cellular automata (QCAs).\footnote{A QCA is a unitary operator that preserves the locality; see, e.g., \cite{Farrelly:2019zds, Arrighi:2019uor} for reviews.}
Based on this observation, it is natural to ask which class of fusion categories admits such a realization.

A recent conjecture proposed in \cite{Tantivasadakarn2025KITP, Tantivasadakarn2025INI} states that a fusion category symmetry $\mathcal{C}$ can be realized on a tensor product Hilbert space if we allow mixing with QCAs only when $\mathcal{C}$ is weakly integral, i.e., the total dimension of $\mathcal{C}$ is an integer.\footnote{Equivalently, the quantum dimension of every simple object of $\mathcal{C}$ is a square root of an integer \cite[Proposition 9.6.9]{EGNO2015}.}
To prove or disprove this conjecture, it would be inevitable to have a precise definition of non-invertible symmetries on a tensor product Hilbert space.
However, such a definition is still lacking, despite recent remarkable observations on general properties of non-invertible symmetries \cite{Okada:2024qmk, Tantivasadakarn:2025txn, Lootens:2023wnl, Ortiz:2025psr, Bartsch:2026wqq}.

In this paper, we provide two complementary approaches to the conjecture mentioned above.
In the first approach, without specifying a precise definition of symmetry operators, we postulate several physical assumptions on the corresponding defect Hilbert spaces.
Based on these assumptions, we study the constraints on possible fusion rules of symmetry operators on a tensor product Hilbert space.
As a key quantity, we introduce an index of non-invertible symmetry operators, generalizing the Gross-Nesme-Vogts-Werner (GNVW) index of QCAs \cite{Gross:2011yvb}.
We show that as long as all fusion channels of two symmetry operators have the same index, the symmetry operators on a tensor product Hilbert space can only realize the fusion rules of a weakly integral fusion category, even if we allow mixing with QCAs.
This result provides a connection between the conjecture in \cite{Tantivasadakarn2025KITP, Tantivasadakarn2025INI} and the yet-to-be-established index theory of non-invertible symmetries.

As another approach, we also discuss the above problem within the framework of tensor networks.
Specifically, we first propose a general class of tensor network operators that we consider to be indecomposable symmetry operators on a tensor product Hilbert space.
These are matrix product operators (MPOs) with some special properties, and are referred to as topological injective MPOs.\footnote{We refer the reader to, e.g., \cite{Molnar:2022nmh, Lootens:2021tet, Lootens:2022avn, Vancraeynest-DeCuiper:2025msv} for various MPO representations of non-invertible symmetries.}
For these MPOs, we define an index, which agrees with the one mentioned in the previous paragraph, and write down sufficient conditions for all fusion channels to have the same index.
As a corollary, it follows that topological injective MPOs satisfying these sufficient conditions can only realize the fusion rules of a weakly integral fusion category, even if we allow mixing with QCAs.
We note that the sufficient conditions have not been proven yet in the general setting.
Nevertheless, we will show that all examples of topological injective MPOs discussed in this paper satisfy the sufficient conditions.

We emphasize that general properties of the index of non-invertible symmetry operators have not been fully understood.
In particular, it has not been proven that every fusion channel of two non-invertible symmetry operators has the same index.
As such, the conjecture in \cite{Tantivasadakarn2025KITP, Tantivasadakarn2025INI} is still open.

\vspace*{\baselineskip}
\noindent{\bf Structure of the paper.}
This paper is organized as follows.
In Section~\ref{sec: no mixing with QCAs}, we discuss non-invertible symmetries realized on a tensor product Hilbert space without mixing with QCAs.
Based on the physical assumptions spelled out in Section~\ref{sec: Physical assumptions}, we show that the fusion rules of a unitary fusion category $\mathcal{C}$ can be realized exactly on a tensor product Hilbert space only when $\mathcal{C}$ is integral, recovering the mathematical result in \cite{Evans:2025msy}.
In Section~\ref{sec: mixing with QCAs}, we generalize this result to the case when the fusion rules mix with QCAs.
Assuming that every fusion channel of two symmetry operators has the same index, we show that the fusion rules of $\mathcal{C}$ can be realized up to QCAs on a tensor product Hilbert space only when $\mathcal{C}$ is weakly integral.
As in Section~\ref{sec: no mixing with QCAs}, the discussions in Section~\ref{sec: mixing with QCAs} are based on the physical assumptions in Section~\ref{sec: Physical assumptions}.
In Section~\ref{sec: Tensor network formulation}, we provide a tensor network formulation of non-invertible symmetry operators on a tensor product Hilbert space.
More specifically, we introduce topological injective MPOs and construct the associated defect Hilbert spaces.
We also show that topological injective MPOs admit sequential quantum circuit representations.
In Section~\ref{sec: Homogeneity of index TN}, we give sufficient conditions for every fusion channel of two topological injective MPOs to have the same index.
In Section~\ref{sec: Examples}, we discuss various examples of topological injective MPOs, including all invertible symmetry operators and the Kramers-Wannier duality operator.
We will see that all of these examples satisfy the sufficient conditions given in Section~\ref{sec: Homogeneity of index TN}.
In Section~\ref{sec: Summary and outlook}, we conclude with a more detailed summary of the results and a brief outlook.
Several technical details are relegated to appendices.

\vspace*{\baselineskip}
\noindent{\bf Conventions.}
The vector spaces and fusion categories considered in this paper are defined over $\mathbb{C}$.
In particular, the fusion categories are always supposed to be unitary.
The on-site Hilbert space, denoted by $\mathcal{H}_{\mathrm{o}}$, is supposed to be finite-dimensional.
All QCAs and MPOs considered in this paper are translationally invariant.
The fixed point of the transfer matrix of an injective MPO always refers to the eigenvector with the largest eigenvalue, even if the eigenvalue is not normalized to $1$.

\subsection{Physical assumptions}
\label{sec: Physical assumptions}
Before moving on to the main text, we first specify the physical assumptions on which the discussions in Sections~\ref{sec: no mixing with QCAs} and \ref{sec: mixing with QCAs} rely.
The discussions from Section~\ref{sec: Tensor network formulation} onwards do not rely on these assumptions.

\begin{figure}[t]
\centering
\includegraphics[scale=1, trim={10, 10, 10, 10}]{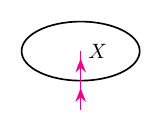}
$\qquad \qquad$
\includegraphics[scale=1, trim={10, 10, 10, 10}]{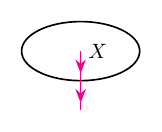}
\caption{A schematic picture of defect Hilbert spaces $\mathcal{H}_{\mathsf{D}_X}^l$ (left) and $\mathcal{H}_{\mathsf{D}_X}^r$ (right).}
\label{fig: defect Hilbert spaces}
\end{figure}

\begin{assumption}[Existence of defect Hilbert spaces] \label{assump: defect Hilbert space}
For every (possibly non-invertible) symmetry operator $\mathsf{D}_X$, there exist defect Hilbert spaces $\mathcal{H}^l_{\mathsf{D}_X}$ and $\mathcal{H}^r_{\mathsf{D}_X}$, which are obtained by modifying the original Hilbert space $\mathcal{H}$ locally around the defect locus.
Physically, $\mathcal{H}^l_{\mathsf{D}_X}$ is the Hilbert space in the presence of a defect $X$ oriented upwards, while $\mathcal{H}^r_{\mathsf{D}_X}$ is the Hilbert space in the presence of a defect $X$ oriented downwards; see Figure \ref{fig: defect Hilbert spaces}.
We assume that the defect $X$ can be moved by unitary operators.
In particular, the defect Hilbert spaces do not depend on the position of a defect up to isomorphism.
\end{assumption}

\begin{assumption}[Product of symmetry operators] \label{assump: Product of symmetry operators}
For every pair of (possibly non-invertible) symmetry operators $\mathsf{D}_X$ and $\mathsf{D}_Y$, the product $\mathsf{D}_X \mathsf{D}_Y$ is also a symmetry operator.
The defect Hilbert spaces associated with $\mathsf{D}_X \mathsf{D}_Y$ are isomorphic to the Hilbert spaces in the presence of two defects $X$ and $Y$ apart from each other.\footnote{Since the insertion of a defect is a local modification due to Assumption~\ref{assump: defect Hilbert space}, the Hilbert space in the presence of two defects is well-defined as long as the distance between the defects is sufficiently large.}
\end{assumption}

\begin{assumption}[Sum of symmetry operators] \label{assump: Sum of symmetry operators}
For every pair of (possibly non-invertible) symmetry operators $\mathsf{D}_X$ and $\mathsf{D}_Y$, the sum $\mathsf{D}_X + \mathsf{D}_Y$ is also a symmetry operator.
The defect Hilbert spaces associated with $\mathsf{D}_X + \mathsf{D}_Y$ are isomorphic to the direct sum of the defect Hilbert spaces associated with each summand, i.e., $\mathcal{H}^l_{\mathsf{D}_X+\mathsf{D}_Y} \cong \mathcal{H}^l_{\mathsf{D}_X} \oplus \mathcal{H}^l_{\mathsf{D}_Y}$ and $\mathcal{H}^r_{\mathsf{D}_X+\mathsf{D}_Y} \cong \mathcal{H}^r_{\mathsf{D}_X} \oplus \mathcal{H}^r_{\mathsf{D}_Y}$.
\end{assumption}

\begin{assumption}[Compatibility with the unitary structure] \label{assump: Compatibility with the unitary structure}
For every (possibly non-invertible) symmetry operator $\mathsf{D}_X$, its Hermitian conjugate $\mathsf{D}_X^{\dagger}$ is also a symmetry operator.
The defect Hilbert spaces associated with $\mathsf{D}_X^{\dagger}$ are isomorphic to those associated with $\mathsf{D}_X$ with the opposite orientation, i.e., $\mathcal{H}^l_{\mathsf{D}_X^{\dagger}} \cong \mathcal{H}^r_{\mathsf{D}_X}$ and $\mathcal{H}^r_{\mathsf{D}_X^{\dagger}} \cong \mathcal{H}^l_{\mathsf{D}_X}$.
\end{assumption}

Throughout the paper, except for Section~\ref{sec: without tensor product decomposition}, we assume that any defect Hilbert space is given by the tensor product of finite dimensional Hilbert spaces on a one-dimensional chain.
In particular, the original Hilbert space without a defect is given by
\begin{equation}
\mathcal{H} = \bigotimes_{i =1, 2, \cdots, N} \mathcal{H}_{\mathrm{o}}^{(i)},
\label{eq: untwisted Hilbert space}
\end{equation}
where $N$ is the number of sites and $\mathcal{H}_{\mathrm{o}}^{(i)}$ is a finite dimensional Hilbert space on site $i$.
For simplicity, in what follows, we assume that $\mathcal{H}_{\mathrm{o}}^{(i)}$ is independent of site $i$.
Accordingly, the on-site Hilbert space will be denoted simply by $\mathcal{H}_{\mathrm{o}}$.
We also assume that the symmetry operators are translationally invariant, which implies that the local modification around the defect does not depend on the position of the defect.

Due to Assumption~\ref{assump: defect Hilbert space}, the defect Hilbert spaces $\mathcal{H}_{\mathsf{D}_X}^l$ and $\mathcal{H}_{\mathsf{D}_X}^r$ are obtained by modifying the original Hilbert space $\mathcal{H}$ locally around the defect.
Since $\mathcal{H}$ is supposed to have the tensor product decomposition as in \eqref{eq: untwisted Hilbert space}, the local modification amounts to replacing the Hilbert space on a finite region around the defect with another Hilbert space.
Therefore, the defect Hilbert spaces can be generally written as
\begin{equation}
\mathcal{H}_{\mathsf{D}_X}^l \cong L_X \otimes \mathcal{H}_{\mathrm{o}}^{\otimes N-n_X^l}, \qquad
\mathcal{H}_{\mathsf{D}_X}^r \cong R_X \otimes \mathcal{H}_{\mathrm{o}}^{\otimes N-n_X^r},
\label{eq: general defect Hilbert space}
\end{equation}
where $n_X^l$ and $n_X^r$ are positive integers and $L_X$ and $R_X$ are finite dimensional Hilbert spaces.
We note that $n_X^l$, $n_X^r$, $L_X$, and $R_X$ are independent of the position of the defect because of the translational invariance.
We also note that these quantities are independent of the system size $N$ because the modification is assumed to be local.
Equation~\eqref{eq: general defect Hilbert space} means that $\mathcal{H}_{\mathsf{D}_X}^l$ is obtained by replacing the Hilbert space on $n_X^l$ sites with $L_X$, and $\mathcal{H}_{\mathsf{D}_X}^r$ is obtained by replacing the Hilbert space on $n_X^r$ sites with $R_X$.
In the case of invertible symmetries represented by QCAs, a general construction of defect Hilbert spaces is given in \cite{Seifnashri:2023dpa}.

\section{Non-invertible symmetries without mixing with QCAs}
\label{sec: no mixing with QCAs}
In this section, we consider non-invertible symmetries in 1+1d that do not mix with QCAs.
Based on the physical assumptions stated in Section~\ref{sec: Physical assumptions}, we will show that a unitary fusion category symmetry $\mathcal{C}$ can be realized without mixing with QCAs on a tensor product Hilbert space only when $\mathcal{C}$ is integral, i.e., only when every object of $\mathcal{C}$ has an integral quantum dimension.
This statement and its converse were recently proved rigorously in \cite{Evans:2025msy} by a different method.
Throughout this section, we use the notations introduced in Section~\ref{sec: Physical assumptions}.

\subsection{Left and right dimensions}
\label{sec: Left and right dimensions}
Following \cite{Seiberg:2024gek}, we first introduce key quantities that we will call the left and right dimensions.

\begin{definition}
For each symmetry operator $\mathsf{D}_X$ on a tensor product Hilbert space, we define the left and right dimensions of $\mathsf{D}_X$ by the ratios
\begin{equation}
\ldim(\mathsf{D}_X) \coloneq \frac{\dim(\mathcal{H}_{\mathsf{D}_X}^l)}{\dim(\mathcal{H})}, \qquad
\rdim(\mathsf{D}_X) \coloneq \frac{\dim(\mathcal{H}_{\mathsf{D}_X}^r)}{\dim(\mathcal{H})},
\label{eq: ldim rdim}
\end{equation}
where $\mathcal{H}_{D_X}^l$ and $\mathcal{H}_{D_X}^r$ are the defect Hilbert spaces and $\mathcal{H}$ is the Hilbert space without a defect.
\end{definition}
In \cite{Seiberg:2024gek}, the above quantities are called lattice quantum dimensions. 
However, in this paper, we reserve the term ``lattice quantum dimension" for another quantity that we will define in Section~\ref{sec: Lattice quantum dimension and index}.

The left and right dimensions are well-defined because the dimensions of the defect Hilbert spaces do not depend on the position of the defect due to Assumption~\ref{assump: defect Hilbert space}.
Based on the general form~\eqref{eq: general defect Hilbert space} of the defect Hilbert spaces, the left and right dimensions can also be written in terms of local quantities as
\begin{equation}
\ldim(\mathsf{D}_X) = \frac{\dim(L_X)}{\dim(\mathcal{H}_\mathrm{o})^{n_X^l}}, \qquad
\rdim(\mathsf{D}_X) = \frac{\dim(R_X)}{\dim(\mathcal{H}_\mathrm{o})^{n_X^r}}.
\end{equation}
We note that both $\ldim(\mathsf{D}_X)$ and $\rdim(\mathsf{D}_X)$ are positive rational numbers by definition.
Furthermore, they are independent of the system size $N$ because $n_X^l$, $n_X^r$, $L_X$, and $R_X$ are independent of $N$ as mentioned in Section~\ref{sec: Physical assumptions}.
We also note that due to Assumption~\ref{assump: Compatibility with the unitary structure}, the left and right dimensions of $\mathsf{D}_X$ and $\mathsf{D}_X^{\dagger}$ are related by $\ldim(\mathsf{D}_X^{\dagger}) = \rdim(\mathsf{D}_X)$ and $\rdim(\mathsf{D}_X^{\dagger}) = \ldim(\mathsf{D}_X)$.

\vspace*{\baselineskip}
\noindent{\bf Multiplicativity and additivity.}
Due to Assumption~\ref{assump: Product of symmetry operators}, we can show that the left and right dimensions are multiplicative under the composition of symmetry operators, that is,
\begin{equation}
\ldim(\mathsf{D}_X\mathsf{D}_Y) = \ldim(\mathsf{D}_X)\ldim(\mathsf{D}_Y), \qquad
\rdim(\mathsf{D}_X\mathsf{D}_Y) = \rdim(\mathsf{D}_X)\rdim(\mathsf{D}_Y).
\label{eq: ldim rdim multiplicative}
\end{equation}
To show the above equation, we note that Assumption~\ref{assump: Product of symmetry operators} implies that the defect Hilbert spaces associated with the composite operator $\mathsf{D}_X \mathsf{D}_Y$ are given by
\begin{equation}
\mathcal{H}_{\mathsf{D}_X \mathsf{D}_Y}^l \cong L_X \otimes L_Y \otimes \mathcal{H}_{\mathrm{o}}^{\otimes N - n_X^l - n_Y^l}, \qquad
\mathcal{H}_{\mathsf{D}_X \mathsf{D}_Y}^r \cong R_X \otimes R_Y \otimes \mathcal{H}_{\mathrm{o}}^{\otimes N - n_X^r - n_Y^r}.
\end{equation}
Therefore, the left and right dimensions of $\mathsf{D}_X\mathsf{D}_Y$ can be computed as
\begin{equation}
\begin{aligned}
\ldim(\mathsf{D}_X\mathsf{D}_Y) &= \frac{\dim(L_X \otimes L_Y)}{\dim(\mathcal{H}_\mathrm{o})^{n_X^l+n_Y^l}} = \ldim(\mathsf{D}_X)\ldim(\mathsf{D}_Y), \\
\rdim(\mathsf{D}_X\mathsf{D}_Y) &= \frac{\dim(R_X \otimes R_Y)}{\dim(\mathcal{H}_\mathrm{o})^{n_X^r+n_Y^r}} = \rdim(\mathsf{D}_X)\rdim(\mathsf{D}_Y),
\end{aligned}
\end{equation}
which shows \eqref{eq: ldim rdim multiplicative}.
Furthermore, due to Assumption~\ref{assump: Sum of symmetry operators}, the left and right dimensions are additive under the sum of symmetry operators, that is,
\begin{equation}
\ldim(\mathsf{D}_X+\mathsf{D}_Y) = \ldim(\mathsf{D}_X)+\ldim(\mathsf{D}_Y), \qquad
\rdim(\mathsf{D}_X+\mathsf{D}_Y) = \rdim(\mathsf{D}_X)+\rdim(\mathsf{D}_Y).
\label{eq: ldim rdim additive}
\end{equation}
This immediately follows from isomorphisms $\mathcal{H}_{\mathrm{D}_X + \mathrm{D}_Y}^l \cong \mathcal{H}_{\mathsf{D}_X}^l \oplus \mathcal{H}_{\mathsf{D}_Y}^l$ and $\mathcal{H}_{\mathrm{D}_X + \mathrm{D}_Y}^r \cong \mathcal{H}_{\mathsf{D}_X}^r \oplus \mathcal{H}_{\mathsf{D}_Y}^r$.

\vspace*{\baselineskip}
\noindent{\bf Fusion rules.}
As a consequence of the multiplicativity~\eqref{eq: ldim rdim multiplicative} and additivity~\eqref{eq: ldim rdim additive}, it follows that the left and right dimensions are one-dimensional representations of the fusion rules of the symmetry operators on the lattice.
To see this, we consider symmetry operators that obey the following fusion rules:
\begin{equation}
\mathsf{D}_X \mathsf{D}_Y = \sum_{Z \in S_{XY}} \mathsf{D}_Z.
\label{eq: fusion rule}
\end{equation}
Here, $S_{XY}$ is the set of fusion channels, which may contain a single label $Z$ multiple times, meaning that the fusion multiplicity can be greater than one.
Since the left and right dimensions are multiplicative~\eqref{eq: ldim rdim multiplicative} and additive~\eqref{eq: ldim rdim additive}, the fusion rules~\eqref{eq: fusion rule} imply
\begin{equation}
\ldim(\mathsf{D}_X)\ldim(\mathsf{D}_Y) = \sum_{Z \in S_{XY}} \ldim(\mathsf{D}_Z), \qquad
\rdim(\mathsf{D}_X)\rdim(\mathsf{D}_Y) = \sum_{Z \in S_{XY}} \rdim(\mathsf{D}_Z).
\label{eq: ldim rdim fusion rule}
\end{equation}
Thus, as pointed out in \cite{Seiberg:2024gek}, the left and right dimensions are one-dimensional representations of the fusion rules of the symmetry operators.
We emphasize that these one-dimensional representations are positive and rational by definition.
As we will see in the next subsection, this fact imposes a strong constraint on the possible symmetry structure that can be realized exactly on a tensor product Hilbert space.

\vspace*{\baselineskip}
\noindent{\bf The case of invertible symmetries.}
As an example, let us consider the case of invertible symmetries, whose symmetry operators are represented by QCAs.
For any QCA $U$, the multiplicativity~\eqref{eq: ldim rdim multiplicative} of the left and right dimensions imply
\begin{equation}
\ldim(U) \ldim(U^{\dagger}) = \rdim(U) \rdim(U^{\dagger}) = 1.
\end{equation}
On the other hand, due to Assumption~\ref{assump: Compatibility with the unitary structure}, the left and right dimensions of $U$ and $U^{\dagger}$ are related by $\ldim(U^{\dagger}) = \rdim(U)$ and $\rdim(U^{\dagger}) = \ldim(U)$.
Therefore, the above equation implies
\begin{equation}
\ldim(U)\rdim(U) = 1.
\label{eq: qdim QCA}
\end{equation}
Namely, the left dimension of a QCA is the inverse of its right dimension.
For instance, when $U$ is the lattice translation operator by one site, the left and right dimensions are given by $\dim(\mathcal{H}_{\mathrm{o}})$ and $\dim(\mathcal{H}_{\mathrm{o}})^{-1}$, respectively \cite{Seifnashri:2023dpa, Seiberg:2024gek}.
More generally, as we will discuss in Section~\ref{sec: The case of invertible symmetries}, the left and right dimensions of a general QCA are given by its GNVW index and its inverse.

\subsection{Exact fusion rules imply integrality}
\label{sec: Exact fusion rule implies integrality}
In this subsection, based on the physical assumptions in Section~\ref{sec: Physical assumptions}, we show that a unitary fusion category symmetry $\mathcal{C}$ in 1+1d can be realized exactly on a tensor product Hilbert space only when $\mathcal{C}$ is integral.
This result agrees with the mathematical result in \cite{Evans:2025msy}.
To make the statement more precise, let us suppose that there exists a finite set of symmetry operators $\{\mathsf{D}_X \mid X \in \mathrm{Irr}(\mathcal{C})\}$ whose fusion rules are exactly given by those of $\mathcal{C}$, i.e.,
\begin{equation}
\mathsf{D}_X \mathsf{D}_Y = \sum_{Z \in X \otimes Y} \mathsf{D}_Z.
\label{eq: fusion rule of C}
\end{equation}
Here, $\mathrm{Irr}(\mathcal{C})$ denotes the set of (isomorphism classes of) simple objects of $\mathcal{C}$, and $\otimes$ denotes the tensor product in $\mathcal{C}$.
The summation on the right-hand side of \eqref{eq: fusion rule of C} is taken over all fusion channels, including the fusion multiplicities.
In what follows, we will show that $\mathcal{C}$ must be an integral fusion category, i.e., a fusion category in which all objects have integral quantum dimensions.

To show the above statement, we first recall that the left and right dimensions defined by \eqref{eq: ldim rdim} are positive one-dimensional representations of the fusion rules of the symmetry operators, cf. \eqref{eq: ldim rdim fusion rule}.
Thus, equation~\eqref{eq: fusion rule of C} implies that the left and right dimensions are positive one-dimensional representations of the fusion rules of $\mathcal{C}$.
It is known that a positive one-dimensional representation of the fusion rules of any unitary fusion category $\mathcal{C}$ is unique and is given by the quantum dimension \cite[Proposition 3.3.6]{EGNO2015}.
Therefore, both the left and right dimensions of $\mathsf{D}_X$ agree with the quantum dimension of $X$, i.e.,
\begin{equation}
\ldim(\mathsf{D}_X) = \rdim(\mathsf{D}_X) = \qdim_{\mathcal{C}}(X),
\label{eq: ldim=rdim=qdim}
\end{equation}
where $\qdim_{\mathcal{C}}(X)$ denotes the quantum dimension of $X \in \mathrm{Irr}(C)$.
Since the left and right dimensions are rational numbers by definition, the above equation implies that the quantum dimension of any simple object of $\mathcal{C}$ is a rational number.
On the other hand, the quantum dimension of any object of a unitary fusion category is known to be an algebraic integer \cite[Proposition 3.3.4]{EGNO2015}.
Hence, it follows that $\qdim_{\mathcal{C}}(X)$ is a rational algebraic integer for all $X \in \mathrm{Irr}(\mathcal{C})$.
Since any rational algebraic integer must be an integer (see, e.g., \cite[Chapter 2]{Marcus2018}), we conclude that $\mathcal{C}$ is integral.

\vspace*{\baselineskip}
\noindent{\bf Remark.}
The above proof does not guarantee that the fusion rules of all integral unitary fusion categories can be realized exactly on a tensor product Hilbert space.
Nevertheless, it was recently shown in \cite{Evans:2025msy} that any integral fusion category can indeed be realized as a symmetry on a tensor product Hilbert space.\footnote{More precisely, in \cite{Evans:2025msy}, Evans and Jones constructed a symmetry action of an arbitrary integral fusion category on a tensor product quasi-local algebra on an infinite chain.}
Basic examples of integral unitary fusion categories are non-anomalous fusion categories \cite{Chang:2018iay, Thorngren:2019iar}, i.e., those that admit a fiber functor \cite{Thorngren:2019iar}.
Physically, non-anomalous fusion categories describe the symmetries that admit a symmetry protected topological phase \cite{Thorngren:2019iar}, i.e., a symmetric gapped phase with a unique ground state on a circle.
It is known that any non-anomalous unitary fusion category is equivalent to the category of representations of a finite dimesional semisimple Hopf algebra \cite{EGNO2015}.
Concrete realizations of this class of fusion category symmetries on a tensor product Hilbert space are given in \cite{Inamura:2021szw, Molnar:2022nmh, Jia:2024bng}.

\section{Non-invertible symmetries mixed with QCAs}
\label{sec: mixing with QCAs}
In this section, we discuss non-invertible symmetries in 1+1d that mix with non-trivial QCAs.
We first define the lattice quantum dimension and the index of non-invertible symmetry operators using the left and right dimensions defined in Section~\ref{sec: Left and right dimensions}.
We then show that the lattice quantum dimension is a positive one-dimensional representation of the fusion rules if and only if all fusion channels of two indecomposable symmetry operators have the same index.
When all fusion channels have the same index, we say that the index is homogeneous.
As a corollary, we show that if the index is homogeneous, the fusion rules of the symmetry operators on a tensor product Hilbert space can agree, up to QCAs, with those of a unitary fusion category $\mathcal{C}$ only when $\mathcal{C}$ is weakly integral.
We will also comment on the case where the Hilbert space does not admit a tensor product decomposition, and discuss why the above constraint does not apply to such a case.

\subsection{Lattice quantum dimension and index}
\label{sec: Lattice quantum dimension and index}
Let us begin with the definitions of the lattice quantum dimension and the index.

\begin{definition}
For each symmetry operator $\mathsf{D}_X$, we define the lattice quantum dimension and the index of $\mathsf{D}_X$ by
\begin{equation}
\qdim_{\mathrm{lat}}(\mathsf{D}_X) \coloneq \sqrt{\ldim(\mathsf{D}_X) \rdim(\mathsf{D}_X)}, \qquad
\ind(\mathsf{D}_X) \coloneq \sqrt{\frac{\ldim(\mathsf{D}_X)}{\rdim(\mathsf{D}_X)}},
\label{eq: qdim ind}
\end{equation}
where $\ldim(\mathsf{D}_X)$ and $\rdim(\mathsf{D}_X)$ are the left and right dimensions defined by \eqref{eq: ldim rdim}.
\end{definition}

For the symmetry operators whose fusion rules agree exactly with those of a unitary fusion category $\mathcal{C}$, we have $\qdim_{\mathrm{lat}}(\mathsf{D}_X) = \qdim_{\mathcal{C}}(X)$ and $\ind(\mathsf{D}_X) = 1$ due to \eqref{eq: ldim=rdim=qdim}.
On the other hand, for the lattice translation operator $T$, we have $\qdim_{\text{lat}}(T) = 1$ and $\ind(T) = \dim(\mathcal{H}_{\text{o}})$ because $\ldim(T) = \dim(\mathcal{H}_{\mathrm{o}})$ and $\rdim(T) = \dim(\mathcal{H}_{\mathrm{o}})^{-1}$ as mentioned below \eqref{eq: qdim QCA}.
More generally, for any QCA, the lattice quantum dimension is equal to one due to \eqref{eq: qdim QCA} and the index agrees with the GNVW index; see Section~\ref{sec: The case of invertible symmetries} for more details.
In particular, our index generalizes the GNVW index to non-invertible symmetry operators.\footnote{The GNVW index has been generalized in several different ways in the literature. In \cite{Jones:2023imy}, the GNVW index was generalized to QCAs on abstract spin chains that do not admit a tensor product decomposition. In \cite{Ma:2024ypm}, the GNVW index was generalized to QCAs on algebras of $G$-symmetric local operators, where $G$ is a finite abelian group. These generalizations have some overlap with ours. For example, the index of the Kramers-Wannier duality operator is defined in all cases and given by $\sqrt{2}$; cf. equation~\eqref{eq: ind OKW}.}

Due to the multiplicativity~\eqref{eq: ldim rdim multiplicative} of the left and right dimensions, the lattice quantum dimension and the index are also multiplicative in the sense that
\begin{align}
\qdim_{\mathrm{lat}}(\mathsf{D}_X \mathsf{D}_Y) &= \qdim_{\mathrm{lat}}(\mathsf{D}_X) \qdim_{\mathrm{lat}}(\mathsf{D}_Y), \label{eq: qdim multiplicative} \\
\ind(\mathsf{D}_X \mathsf{D}_Y) &= \ind(\mathsf{D}_X) \ind(\mathsf{D}_Y). \label{eq: ind multiplicative}
\end{align}
On the other hand, the lattice quantum dimension and the index are not additive in general.
Indeed, as we will see below, the lattice quantum dimension is additive only for symmetry operators with the same index.
Namely, for any positive integer $n$, we have
\begin{equation}
\qdim_{\mathrm{lat}}\left(\sum_{i = 1}^{n} \mathsf{D}_{X_i}\right) = \sum_{i = 1}^{n} \qdim_{\mathrm{lat}}(\mathsf{D}_{X_i}) ~\Leftrightarrow~
\ind(\mathsf{D}_{X_i}) = \ind(\mathsf{D}_{X_j}), \quad \forall i, j = 1, 2, \cdots, n.
\label{eq: qdim additive}
\end{equation}
Furthermore, one can also show that the index is invariant under the addition of symmetry operators if and only if they have the same index, i.e.,
\begin{equation}
\ind(\mathsf{D}_X+\mathsf{D}_Y) = \ind(\mathsf{D}_X) ~\Leftrightarrow~ \ind(\mathsf{D}_X) = \ind(\mathsf{D}_Y).
\label{eq: ind invariance}
\end{equation}
In the remainder of this subsection, we will show \eqref{eq: qdim additive} and \eqref{eq: ind invariance}.

To show \eqref{eq: qdim additive}, we notice that the additivity~\eqref{eq: ldim rdim additive} of the left and right dimensions implies that the lattice quantum dimension satisfies
\begin{equation}
\begin{aligned}
&\quad \qdim_{\mathrm{lat}}\left(\sum_{i=1}^n \mathsf{D}_{X_i}\right)^2 - \left(\sum_{i=1}^n \qdim_{\mathrm{lat}}(\mathsf{D}_{X_i})\right)^2 \\
&= \frac{1}{2} \sum_{i, j} \left( \ldim(\mathsf{D}_{X_i}) \rdim(\mathsf{D}_{X_j}) + \rdim(\mathsf{D}_{X_i}) \ldim(\mathsf{D}_{X_j}) - 2 \; \qdim_{\mathrm{lat}}(\mathsf{D}_{X_i}) \qdim_{\mathrm{lat}}(\mathsf{D}_{X_j}) \right) \\
&= \frac{1}{2} \sum_{i, j} \left( \sqrt{\ldim(\mathsf{D}_{X_i}) \rdim(\mathsf{D}_{X_j})} - \sqrt{\rdim(\mathsf{D}_{X_i}) \ldim(\mathsf{D}_{X_j})} \right)^2.
\end{aligned}
\label{eq: qdim X1+X2}
\end{equation}
Therefore, $\qdim_{\mathrm{lat}}(\sum_{i=1}^n \mathsf{D}_{X_i})$ agrees with $\sum_{i=1}^n \qdim_{\mathrm{lat}}(\mathsf{D}_{X_i})$ if and only if
\begin{equation}
\sqrt{\ldim(\mathsf{D}_{X_i}) \rdim(\mathsf{D}_{X_j})} = \sqrt{\rdim(\mathsf{D}_{X_i}) \ldim(\mathsf{D}_{X_j})}, \qquad \forall i, j = 1, 2, \cdots, n.
\end{equation}
By dividing both sides by $\rdim(\mathsf{D}_{X_i}) \rdim(\mathsf{D}_{X_j})$, we obtain
\begin{equation}
\ind(\mathsf{D}_{X_i}) = \ind(\mathsf{D}_{X_j}), \qquad \forall i, j = 1, 2, \cdots, n,
\end{equation}
which shows \eqref{eq: qdim additive}.

We can also show \eqref{eq: ind invariance} by using the additivity~\eqref{eq: ldim rdim additive} of the left and right dimensions.
More specifically, the additivity~\eqref{eq: ldim rdim additive} implies that the index satisfies
\begin{equation}
\begin{aligned}
\ind(\mathsf{D}_X + \mathsf{D}_Y)^2 - \ind(\mathsf{D}_X)^2
&= \frac{\ldim(\mathsf{D}_X) + \ldim(\mathsf{D}_Y)}{\rdim(\mathsf{D}_X) + \rdim(\mathsf{D}_Y)} - \frac{\ldim(\mathsf{D}_X)}{\rdim(\mathsf{D}_X)} \\
&= \frac{\rdim(\mathsf{D}_X) \ldim(\mathsf{D}_Y) - \ldim(\mathsf{D}_X) \rdim(\mathsf{D}_Y)}{(\rdim(\mathsf{D}_X) + \rdim(\mathsf{D}_Y)) \rdim(\mathsf{D}_X)}.
\end{aligned}
\end{equation}
Therefore, $\ind(\mathsf{D}_X + \mathsf{D}_Y)$ agrees with $\ind(\mathsf{D}_X)$ if and only if
\begin{equation}
\rdim(\mathsf{D}_X) \ldim(\mathsf{D}_Y) = \ldim(\mathsf{D}_X) \rdim(\mathsf{D}_Y).
\end{equation}
By dividing both sides by $\rdim(\mathsf{D}_X)^2 \rdim(\mathsf{D}_Y)^2$ and taking the square root, we obtain
\begin{equation}
\ind(\mathsf{D}_X) = \ind(\mathsf{D}_Y),
\end{equation}
which shows \eqref{eq: ind invariance}.

\subsection{Homogeneity of the index}
\label{sec: Homogeneity of index}
In this subsection, we define the notion of homogeneity of the index, which plays a crucial role in constraining possible fusion rules of symmetry operators on a tensor product Hilbert space.

\begin{definition}
Let $\mathsf{D}_X$ and $\mathsf{D}_Y$ be indecomposable symmetry operators whose fusion rule is given by
\begin{equation}
\mathsf{D}_X \mathsf{D}_Y = \sum_{Z \in S_{XY}} \mathsf{D}_Z,
\label{eq: fusion rule 2}
\end{equation}
where $\mathsf{D}_Z$ is an indecomposable symmetry operator for every $Z \in S_{XY}$.\footnote{A symmetry operator is said to be indecomposable if it cannot be written as a sum of two symmetry operators.}
We say that the index is homogeneous for the product of $\mathsf{D}_X$ and $\mathsf{D}_Y$ if all fusion channels have the same index, i.e.,
\begin{equation}
\ind(\mathsf{D}_Z) = \ind(\mathsf{D}_W), \qquad \forall Z, W \in S_{XY}.
\label{eq: homogeneity 0}
\end{equation}
\end{definition}

We note that equation~\eqref{eq: homogeneity 0} is equivalent to
\begin{equation}
\ind(\mathsf{D}_Z) = \ind(\mathsf{D}_X) \ind(\mathsf{D}_Y), \qquad \forall Z \in S_{XY}.
\label{eq: homogeneity}
\end{equation}
Indeed, if the index satisfies \eqref{eq: homogeneity 0}, it follows that
\begin{equation}
\ind(\mathsf{D}_Z) \overset{\eqref{eq: ind invariance}}{=} \ind \left(\sum_{W \in S_{XY}} \mathsf{D}_W \right) \overset{\eqref{eq: fusion rule 2}}{=} \ind(\mathsf{D}_X \mathsf{D}_Y) \overset{\eqref{eq: ind multiplicative}}{=} \ind(\mathsf{D}_X) \ind(\mathsf{D}_Y).
\end{equation}
Conversely, if the index satisfies \eqref{eq: homogeneity}, it is clear that $\ind(\mathsf{D}_Z) = \ind(\mathsf{D}_W)$ for all $Z, W \in S_{XY}$.

For later use, let us give another equivalent characterization of the homogeneity of the index.
Specifically, we show that the index is homogeneous if and only if the lattice quantum dimension gives a one-dimensional representation of the fusion rule~\eqref{eq: fusion rule 2}, that is,
\begin{equation}
\qdim_{\mathrm{lat}}(\mathsf{D}_X) \qdim_{\mathrm{lat}}(\mathsf{D}_Y) = \sum_{Z \in S_{XY}} \qdim_{\mathrm{lat}}(\mathsf{D}_Z) ~\Leftrightarrow~
\ind(\mathsf{D}_Z) = \ind(\mathsf{D}_W), \quad \forall Z, W \in S_{XY}.
\label{eq: homogeneous index}
\end{equation}
To show this, we compute the lattice quantum dimension of $\mathsf{D}_X \mathsf{D}_Y$ in two different ways.
On the one hand, $\qdim_{\mathrm{lat}}(\mathsf{D}_X \mathsf{D}_Y)$ is equal to $\qdim_{\mathrm{lat}}(\mathsf{D}_X) \qdim_{\mathrm{lat}}(\mathsf{D}_Y)$ due to the multiplicativity~\eqref{eq: qdim multiplicative}.
On the other hand, due to \eqref{eq: qdim additive}, $\qdim_{\mathrm{lat}}(\mathsf{D}_X \mathsf{D}_Y)$ is equal to $\sum_{Z \in S_{XY}} \qdim_{\mathrm{lat}}(\mathsf{D}_Z)$ if and only if all the fusion channels have the same index.
Therefore, we find that $\qdim_{\mathrm{lat}}(\mathsf{D}_X) \qdim_{\mathrm{lat}}(\mathsf{D}_Y)$ is equal to $\sum_{Z \in S_{XY}} \qdim_{\mathrm{lat}}(\mathsf{D}_Z)$ if and only if the index is homogeneous, which shows \eqref{eq: homogeneous index}

\subsection{Homogeneity of the index implies weak integrality}
\label{sec: Homogeneity of index implies weak integrality}

The homogeneity of the index imposes a strong constraint on possible fusion rules of symmetry operators on a tensor product Hilbert space.
To see this, we suppose that there exists a finite set of indecomposable symmetry operators $\{\mathsf{D}_X \mid X \in \mathrm{Irr}(\mathcal{C})\}$ that obey the fusion rules of a unitary fusion category $\mathcal{C}$ up to QCAs, that is,
\begin{equation}
\mathsf{D}_X \mathsf{D}_Y = \sum_{Z \in X \otimes Y} \mathsf{D}_Z U_{XY}^Z,
\label{eq: fusion rule up to QCA}
\end{equation}
where $U_{XY}^Z$ is an arbitrary QCA.
In what follows, we will show that if the index is homogeneous for the product of any two symmetry operators, i.e., if the index satisfies $\ind(\mathsf{D}_ZU_{XY}^Z) = \ind(\mathsf{D}_WU_{XY}^W)$ for all $X, Y \in \mathrm{Irr}(\mathcal{C})$ and all $Z, W \in X \otimes Y$, then the unitary fusion category $\mathcal{C}$ must be weakly integral.
In other words, the conjecture in \cite{Tantivasadakarn2025KITP, Tantivasadakarn2025INI} holds if the index is homogeneous.

To show the above statement, we first recall that the homogeneity of the index implies that the lattice quantum dimension gives a one-dimensional representation of the fusion rules~\eqref{eq: fusion rule up to QCA}, i.e.,
\begin{equation}
\qdim_{\mathrm{lat}}(\mathsf{D}_X)\qdim_{\mathrm{lat}}(\mathsf{D}_Y) = \sum_{Z \in X \otimes Y} \qdim_{\mathrm{lat}}(\mathsf{D}_Z U_{XY}^Z).
\label{eq: qdim hom}
\end{equation}
Since the lattice quantum dimension is multiplicative~\eqref{eq: qdim multiplicative}, the summand on the right-hand side can be written as $\qdim_{\mathrm{lat}}(\mathsf{D}_Z) \qdim_{\mathrm{lat}}(U_{XY}^Z)$.
Furthermore, as shown in \eqref{eq: qdim QCA}, the lattice quantum dimension of any QCA is one under Assumption~\ref{assump: Compatibility with the unitary structure}.
Therefore, equation~\eqref{eq: qdim hom} reduces to
\begin{equation}
\qdim_{\mathrm{lat}}(\mathsf{D}_X) \qdim_{\mathrm{lat}}(\mathsf{D}_Y) = \sum_{Z \in X \otimes Y} \qdim_{\mathrm{lat}}(\mathsf{D}_Z).
\end{equation}
Namely, if the index is homogeneous, the lattice quantum dimension gives a one-dimensional representation of the fusion rules of $\mathcal{C}$.
Now, we recall that the lattice quantum dimension is positive because it is the square root of a positive rational number.
Since a positive one-dimensional representation of the fusion rules of $\mathcal{C}$ is uniquely given by the quantum dimension \cite[Proposition 3.3.6]{EGNO2015}, we find
\begin{equation}
\qdim_{\mathrm{lat}}(\mathsf{D}_X) = \qdim_{\mathcal{C}}(X).
\end{equation}
Furthermore, since $\qdim_{\mathrm{lat}}(\mathsf{D}_X)$ is the square root of a rational number, the above equation implies that the total dimension of $\mathcal{C}$ is rational:
\begin{equation}
\mathrm{Dim}(\mathcal{C}) \coloneq \sum_{X \in \mathrm{Irr}(\mathcal{C})} \qdim_{\mathcal{C}}(X)^2 = \sum_{X \in \mathrm{Irr}(\mathcal{C})} \qdim_{\mathrm{lat}}(\mathsf{D}_X)^2 \in \mathbb{Q}.
\end{equation}
On the other hand, the total dimension $\mathrm{Dim}(\mathcal{C})$ is an algebraic integer because $\qdim_{\mathcal{C}}(X)$ is an algebraic integer for every $X \in \mathrm{Irr}(\mathcal{C})$.\footnote{We recall that the algebraic integers form a ring. In particular, for any algebraic integers $\alpha$ and $\beta$, the sum $\alpha +\beta$ and the product $\alpha \beta$ are also algebraic integers.}
Thus, $\mathrm{Dim}(\mathcal{C})$ is a rational algebraic integer, and hence, it is an integer \cite[Chapter 2]{Marcus2018}.
This shows that $\mathcal{C}$ is weakly integral.

\vspace*{\baselineskip}
\noindent{\bf Remarks.}
The above proof also shows that the quantum dimension of every simple object of $\mathcal{C}$ must be a square root of an integer.
This is not an additional constraint.
Indeed, it is known that the quantum dimension of every simple object of any weakly integral unitary fusion category is a square root of an integer \cite[Proposition 9.6.9]{EGNO2015}.

We note that $\mathcal{C}$ must be integral if $\ind(U_{XY}^Z) = 1$ for all $X, Y \in \mathrm{Irr}(\mathcal{C})$ and $Z \in X \otimes Y$.
Indeed, in this case, the proof in Section~\ref{sec: Exact fusion rule implies integrality} is valid because the condition $\ind(U_{XY}^Z) = 1$ together with \eqref{eq: qdim QCA} implies that $\ldim(U_{XY}^Z) = \rdim(U_{XY}^Z) = 1$ and hence the logic of the proof is not affected.

However, $\mathcal{C}$ is not necessarily integral if the index of $U_{XY}^Z$ is non-trivial for some $X, Y \in \mathrm{Irr}(\mathcal{C})$ and $Z \in X \otimes Y$.
Indeed, there are examples of weakly integral fusion categories that are not integral and whose fusion rules can be realized on a tensor product Hilbert space by allowing mixing with lattice translation \cite{Seiberg:2024gek, Lu:2026rhb}.
More generally, it was shown in \cite{Wen:2026ncw} that the fusion rules of all weakly integral fusion categories can be realized on a tensor product Hilbert space by allowing mixing with non-trivial QCAs.

We also emphasize that the homogeneity of the index has not been proven.
To study the properties of the index in depth, it is desirable to have a precise formulation of non-invertible symmetry operators on a tensor product Hilbert space.
In Section~\ref{sec: Tensor network formulation}, we will provide a tensor network description of non-invertible symmetry operators on a tensor product Hilbert space.
Within this framework, we will write down sufficient conditions for the index to be homogeneous in Section~\ref{sec: Homogeneity of index TN}.

\subsection{A comment on the case without tensor product decomposition}
\label{sec: without tensor product decomposition}
The discussions so far have relied on the assumption that the Hilbert space on the lattice can be decomposed into a tensor product of local Hilbert spaces.
The conclusions are modified if we relax this assumption.
To see this, let us consider the anyon chain model based on a unitary fusion category $\mathcal{C}$ \cite{Feiguin:2006ydp, Buican:2017rxc, Aasen:2020jwb}.
The anyon chain model based on $\mathcal{C}$ is a 1+1d lattice model with a constrained Hilbert space and has an exact fusion category symmetry described by $\mathcal{C}$.
The state space $\mathcal{H}$ of the model without a defect is given by the vector space spanned by the fusion diagrams of the following form:
\begin{equation}
\adjincludegraphics[valign=c, scale=1, trim={10, 10, 10, 10}]{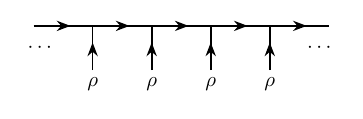}\;.
\label{eq: anyon chain untwisted}
\end{equation}
Here, $\rho \in \mathcal{C}$ is a fixed object of $\mathcal{C}$, which we can choose arbitrarily.
Similarly, the defect Hilbert spaces $\mathcal{H}_{\mathsf{D}_X}^l$ and $\mathcal{H}_{\mathsf{D}_X}^r$ associated with the symmetry operator $\mathsf{D}_X$ for $X \in \mathrm{Irr}(\mathcal{C})$ are spanned by the fusion diagrams of the following form~\cite{Buican:2017rxc, Aasen:2020jwb, Bhardwaj:2024kvy}:
\begin{equation}
\adjincludegraphics[valign=c, scale=1, trim={10, 10, 10, 10}]{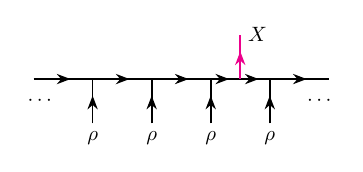}\;, \qquad
\adjincludegraphics[valign=c, scale=1, trim={10, 10, 10, 10}]{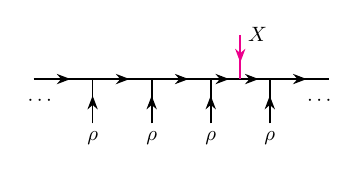}\;.
\label{eq: anyon chain twisted}
\end{equation}

For the above definition of the defect Hilbert spaces, the left and right dimensions in \eqref{eq: ldim rdim} generally depend on the system size and are not multiplicative for a fixed system size.
This motivates us to define the left and right dimensions of $\mathsf{D}_X$ in this model by taking the limit
\begin{equation}
\ldim(\mathsf{D}_X) \coloneq \lim_{N \to \infty} \frac{\dim(\mathcal{H}_{\mathsf{D}_X}^l)}{\dim(\mathcal{H})}, \qquad
\rdim(\mathsf{D}_X) \coloneq \lim_{N \to \infty} \frac{\dim(\mathcal{H}_{\mathsf{D}_X}^r)}{\dim(\mathcal{H})},
\label{eq: anyon chain ldim rdim}
\end{equation}
where $N$ is the number of vertical lines labeled by $\rho$ in \eqref{eq: anyon chain untwisted} and \eqref{eq: anyon chain twisted}.
As long as the boundary conditions are sufficiently generic, one can show that the above quantities are given by
\begin{equation}
\ldim(\mathsf{D}_X) = \rdim(\mathsf{D}_X) = \qdim_{\mathcal{C}}(X).
\label{eq: anyon chain ldim rdim 2}
\end{equation}
We will give a proof of this equation shortly.

We note that the left and right dimensions defined above are multiplicative and additive, as in the case when the state space admits a tensor product decomposition.
However, as opposed to the case of a tensor product Hilbert space, the left and right dimensions in \eqref{eq: anyon chain ldim rdim 2} are no longer rational in general.
Therefore, the discussion in the previous subsection does not apply to the anyon chain.
This is in accordance with the fact that the anyon chain model can realize an arbitrary unitary fusion category symmetry $\mathcal{C}$ without mixing with non-trivial QCAs.

\vspace*{\baselineskip}
\noindent{\bf A proof of \eqref{eq: anyon chain ldim rdim 2}.}
Now, we show \eqref{eq: anyon chain ldim rdim 2}.\footnote{The author thanks Kantaro Ohmori for explaining the proof.}
To this end, we first fix the boundary conditions so that the left and right ends of the chain in \eqref{eq: anyon chain untwisted} are labeled by fixed objects
\begin{equation}
B_L \coloneq \bigoplus_{a \in \mathrm{Irr}(\mathcal{C})} n_{a; L} a^*, \qquad
B_R \coloneq \bigoplus_{a \in \mathrm{Irr}(\mathcal{C})} n_{a; R} a^*,
\label{eq: anyon chain bc}
\end{equation}
where $n_{a; L}$ and $n_{a; R}$ are positive integers, and $a^*$ denotes the dual of $a$.
One can equally write these boundary conditions without using the dual objects as $B_L = \bigoplus_{a \in \mathrm{Irr}(\mathcal{C})} m_{a; L} a$ and $B_R = \bigoplus_{a \in \mathrm{Irr}(\mathcal{C})} m_{a; R} a$, where $m_{a; L} \coloneq n_{a^*; L}$ and $m_{a; R} \coloneq n_{a^*; R}$.
In what follows, we will write the column vectors with entries $n_{a; L}$ and $n_{a; R}$ as $\mathbf{n}_L$ and $\mathbf{n}_R$, respectively.

For the above choice of boundary conditions, the dimension of the Hilbert space $\mathcal{H}$ on an open chain with $N$ vertical legs can be computed as
\begin{equation}
\dim(\mathcal{H})
= \dim(\Hom_{\mathcal{C}}(B_L \otimes \rho^{\otimes N}, B_R))
= \mathbf{n}_L^{\mathbf{T}} \mathbf{N}_{\rho}^N \mathbf{n}_R,
\end{equation}
where $\mathbf{n}_L^{\mathbf{T}}$ is the transpose of $\mathbf{n}_L$, and $\mathbf{N}_{\rho}$ is the fusion matrix whose $(a, b)$-component is given by the fusion coefficient $N_{\rho b}^a$.
When $N$ is large, the right-hand side of the above equation is dominated by the contribution from the largest eigenvalue of $\mathbf{N}_{\rho}$.
Since the largest eigenvalue of $\mathbf{N}_{\rho}$ agrees with the quantum dimension of $\rho$ \cite{EGNO2015}, it follows that
\begin{equation}
\dim(\mathcal{H}) \sim (\mathbf{n}_L^{\mathbf{T}} \mathbf{d}) (\mathbf{d}^{\mathbf{T}} \mathbf{n}_R) \; \qdim_{\mathcal{C}}(\rho)^N, \qquad
\text{for } N \gg 1,
\label{eq: asymptotic dimH}
\end{equation}
where we neglected the subleading terms on the right-hand side.
Here, $\mathbf{d}$ is the normalized Perron-Frobenius eigenvector of $\mathbf{N}_{\rho}$, whose $a$th component is given by $\qdim_{\mathcal{C}}(a) / \sqrt{\mathrm{Dim}(\mathcal{C})}$, where $\mathrm{Dim}(\mathcal{C})$ is the total dimension of $\mathcal{C}$.\footnote{One can show that $\mathbf{d}$ defined in this way is an eigenvector of $\mathbf{N}_\rho$ with eigenvalue $\qdim_{\mathcal{C}}(\rho)$ as follows: $(\mathbf{N}_{\rho} \mathbf{d})_a = \sum_b N_{\rho b}^a \qdim_{\mathcal{C}}(b) / \sqrt{\mathrm{Dim}(\mathcal{C})} = \sum_b N_{a^* \rho}^{b^*} \qdim_{\mathcal{C}}(b) / \sqrt{\mathrm{Dim}(\mathcal{C})} = \qdim_{\mathcal{C}}(\rho) \qdim_{\mathcal{C}}(a) / \sqrt{\mathrm{Dim}(\mathcal{C})}$. Here, we used $\qdim_{\mathcal{C}}(a) = \qdim_{\mathcal{C}}(a^*)$ for all $a \in \mathrm{Irr}(\mathcal{C})$. We note that this eigenvector is unique due to the Perron-Frobenius theorem.}
We note that $\mathbf{d}$ is independent of $\rho$, and in particular it satisfies $\mathbf{N}_a \mathbf{d} = \qdim_{\mathcal{C}}(a) \mathbf{d}$ for all $a \in \mathrm{Irr}(\mathcal{C})$.
We also note that the non-universal factors $\mathbf{n}_L^{\mathbf{T}} \mathbf{d}$ and $\mathbf{d}^{\mathbf{T}} \mathbf{n}_R$ in \eqref{eq: asymptotic dimH} are non-zero as long as the boundary conditions in \eqref{eq: anyon chain bc} are sufficiently generic.

Similarly, the dimensions of the defect Hilbert spaces $\mathcal{H}_{\mathsf{D}_X}^l$ and $\mathcal{H}_{\mathsf{D}_X}^r$ on an open chain with boundary condition~\eqref{eq: anyon chain bc} can be computed as
\begin{align}
\dim(\mathcal{H}_{\mathsf{D}_X}^l)
&= \dim(\Hom_{\mathcal{C}}(B_L \otimes \rho^{\otimes N}, X \otimes B_R))
= \mathbf{n}_L^{\mathbf{T}} C \mathbf{N}_X C \mathbf{N}_{\rho}^N \mathbf{n}_R,
\\
\dim(\mathcal{H}_{\mathsf{D}_X}^r)
&= \dim(\Hom_{\mathcal{C}}(B_L \otimes \rho^{\otimes N}, X^* \otimes B_R))
= \mathbf{n}_L^{\mathbf{T}} C \mathbf{N}_{X^*} C \mathbf{N}_{\rho}^N \mathbf{n}_R,
\end{align}
where $C$ is the charge conjugation matrix whose $(a, b)$-component is given by $\delta_{a, b^*}$.
If we take $N$ to be large, the above quantities are dominated by the contributions from the largest eigenvalue of $\mathbf{N}_{\rho}$.
Since the corresponding eivenvector $\mathbf{d}$ satisfies $\mathbf{N}_X \mathbf{d} = \qdim_{\mathcal{C}}(X) \mathbf{d}$ and $C \mathbf{d} = \mathbf{d}$, it follows that
\begin{align}
\dim(\mathcal{H}_{\mathsf{D}_X}^l) &\sim (\mathbf{n}_L^{\mathbf{T}} \mathbf{d}) (\mathbf{d}^{\mathbf{T}} \mathbf{n}_R) \; \qdim_{\mathcal{C}}(X) \qdim_{\mathcal{C}}(\rho)^N,
\label{eq: asymptotic dimH left}
\\
\dim(\mathcal{H}_{\mathsf{D}_X}^r) &\sim (\mathbf{n}_L^{\mathbf{T}} \mathbf{d}) (\mathbf{d}^{\mathbf{T}} \mathbf{n}_R) \; \qdim_{\mathcal{C}}(X^*) \qdim_{\mathcal{C}}(\rho)^N,
\label{eq: asymptotic dimH right}
\end{align}
where $N \gg 1$.

By plugging the formulae~\eqref{eq: asymptotic dimH}, \eqref{eq: asymptotic dimH left}, and \eqref{eq: asymptotic dimH right} into the definition~\eqref{eq: anyon chain ldim rdim}, we find that the left and right dimensions of $\mathsf{D}_X$ are given by
\begin{equation}
\ldim(\mathsf{D}_X) = \qdim_{\mathcal{C}}(X), \qquad
\rdim(\mathsf{D}_X) = \qdim_{\mathcal{C}}(X^*) = \qdim_{\mathcal{C}}(X).
\end{equation}
Here, we used the fact that $\mathbf{n}_L^{\mathbf{T}} \mathbf{d}$ and $\mathbf{d}^{\mathbf{T}} \mathbf{n}_R$ in \eqref{eq: asymptotic dimH}, \eqref{eq: asymptotic dimH left}, and \eqref{eq: asymptotic dimH right} are non-zero for a genric choice of boundary conditions.
This concludes the proof of \eqref{eq: anyon chain ldim rdim 2}.

\section{Tensor network formulation of non-invertible symmetries}
\label{sec: Tensor network formulation}
In this section, we discuss a possible approach to establishing the index theory for non-invertible symmetry operators in 1+1d within the framework of tensor networks.
More specifically, we first propose a general class of tensor network operators that describe non-invertible symmetries on a tensor product Hilbert space in 1+1 dimensions.
We then study the corresponding defect Hilbert spaces and the associated quantities, such as the lattice quantum dimension and the index.
The properties of these quantities are not yet fully understood: for example, the homogeneity of the index has not been proven.

Before going into details, we mention that most of the computations in this section can be found in \cite{Cirac:2017vke, Sahinoglu:2017iny, Franco-Rubio:2025qss} in the context of matrix product unitaries.
Our contribution is to point out that many computations in \cite{Cirac:2017vke, Sahinoglu:2017iny, Franco-Rubio:2025qss} are valid for a more general class of tensor network operators that are generally non-invertible.
This observation enables us to define the index of non-invertible symmetry operators without relying on the physical assumptions in Section~\ref{sec: Physical assumptions}.
Throughout this section, the state space on the lattice is given by the tensor product of finite dimensional on-site Hilbert spaces.

\subsection{Preliminaries}
\label{sec: Preliminaries}
We first briefly review the basic notions of tensor networks that we will use later.
We refer the reader to \cite{Cirac:2020obd} and references therein for more details of tensor networks.

\vspace*{\baselineskip}
\noindent{\bf Injective MPOs.}
A matrix product operator (MPO) is a tensor network operator constructed from a four-leg tensor $\mathcal{O}$ as follows:
\begin{equation}
\mathsf{D}[\mathcal{O}] \coloneq 
\adjincludegraphics[valign=c, scale=1, trim={10, 10, 10, 10}]{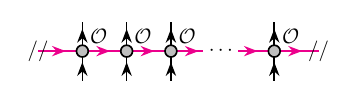} \;.
\label{eq: MPO}
\end{equation}
Here, the double slashes on both ends indicate the periodic boundary condition, and the arrows specify the source and target vector spaces of each tensor.
In what follows, we consider only translationally invariant MPOs on a periodic chain.
The physical Hilbert space on each site is denoted by $\mathcal{H}_{\mathrm{o}}$, while the bond Hilbert space is denoted by $V$.

The MPO tensor $\mathcal{O}$ in \eqref{eq: MPO} is defined as a linear map from $V \otimes \mathcal{H}_{\mathrm{o}}$ to $\mathcal{H}_{\mathrm{o}} \otimes V$.
Using the duality of finite dimensional vector spaces, one can also think of $\mathcal{O}$ as a linear map from $\mathcal{H}_{\mathrm{o}} \otimes V^* \to V^* \otimes \mathcal{H}_{\mathrm{o}}$.
Diagrammatically, the latter corresponds to the tensor viewed as a linear map from the bottom right to the top left.
On the other hand, the former corresponds to the tensor viewed as a linear map from the bottom left to the top right.
In what follows, we will interpret the four-leg tensors in the tensor network diagrams in either way depending on the context.

The MPO tensor $\mathcal{O}$ is said to be injective if it is injective as a linear map from the virtual bonds to the physical legs.
Due to the fundamental theorem of MPS/MPO \cite{Perez-Garcia:2006nqo, Cirac:2016iqe}, two injective MPO tensors $\mathcal{O}$ and $\mathcal{O}^{\prime}$ generate the same MPO $\mathsf{D}[\mathcal{O}] = \mathsf{D}[\mathcal{O}^{\prime}]$ on a periodic chain of any length if and only if they are related by a gauge transformation
\begin{equation}
\mathcal{O}^{\prime} = \adjincludegraphics[valign=c, scale=1, trim={10, 10, 10, 10}]{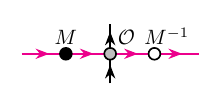} \;,
\label{eq: gauge transformation}
\end{equation}
where $M$ is an arbitrary invertible linear map.

\vspace*{\baselineskip}
\noindent{\bf Blocking.}
The MPO tensors supported on $n$ physical sites can be viewed as a single MPO tensor with physical Hilbert space $\mathcal{H}_{\mathrm{o}}^{\otimes n}$.
This operation is called blocking.
The MPO tensor obtained by blocking $n$ physical sites is denoted by $\mathcal{O}^{[n]}$.
For example, when $n = 2$, we have
\begin{equation}
\mathcal{O}^{[2]} \coloneq \;
\adjincludegraphics[valign=c, scale=1, trim={10, 10, 10, 10}]{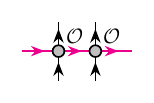} \;.
\end{equation}
We note that $\mathcal{O}^{[n]}$ is injective if $\mathcal{O}$ is injective.

\vspace*{\baselineskip}
\noindent{\bf Transfer matrix and fixed points.}
The transfer matrix of an MPO tensor $\mathcal{O}$ is the linear map defined by the following diagram:
\begin{equation}
\mathsf{T}[\mathcal{O}] \coloneq
\adjincludegraphics[valign=c, scale=1, trim={10, 10, 10, 10}]{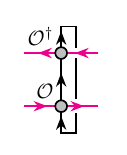} \;.
\label{eq: transfer matrix}
\end{equation}
The loop on the right-hand side represents the trace over the physical Hilbert space.
The MPO tensor $\mathcal{O}^{\dagger}$ in the above equation is the Hermitian conjugate of $\mathcal{O}$, whose components are explicitly given by
\begin{equation}
\adjincludegraphics[valign=c, scale=1, trim={10, 10, 10, 10}]{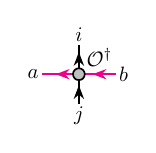} \;=\; \left(\adjincludegraphics[valign=c, scale=1, trim={10, 10, 10, 10}]{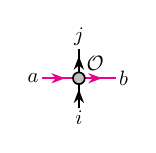}\right)^*,
\end{equation}
where $*$ denotes the complex conjugation, and the labels of the physical leg and the virtual bond take values in orthonormal bases of $\mathcal{H}_{\mathrm{o}}$ and $V$, respectively.
As indicated by the orientations of the arrows, the tensor $\mathcal{O}^{\dagger}$ is defined as a linear map from $\mathcal{H}_{\mathrm{o}} \otimes V$ to $V \otimes \mathcal{H}_{\mathrm{o}}$.

When $\mathcal{O}$ is injective, the transfer matrix $\mathsf{T}[\mathcal{O}]$ has a unique left eigenvector $\Lambda^l$ such that the corresponding eigenvalue agrees with the spectral radius of $\mathsf{T}[\mathcal{O}]$.
We denote the spectral radius of $\mathsf{T}[\mathcal{O}]$ by $\lambda$.
Similarly, $\mathsf{T}[\mathcal{O}]$ has a unique right eigenvector $\Lambda^r$ with eigenvalue $\lambda$.
The eigenvectors $\Lambda^l$ and $\Lambda^r$ are referred to as the left and right fixed points of $\mathsf{T}[\mathcal{O}]$.\footnote{In the literature, $\Lambda^l$ and $\Lambda^r$ are often referred to as the fixed points only when $\mathcal{O}$ is normalized so that $\lambda=1$. In this paper, however, we do not assume this normalization.}
Diagrammatically, the eigenvalue equations for the fixed points can be written as
\begin{equation}
\adjincludegraphics[valign=c, scale=1, trim={10, 10, 10, 10}]{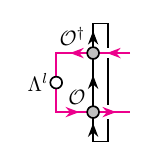}
= \lambda \adjincludegraphics[valign=c, scale=1, trim={10, 10, 10, 10}]{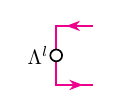} \; , 
\qquad
\adjincludegraphics[valign=c, scale=1, trim={10, 10, 10, 10}]{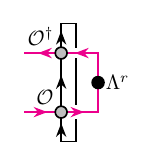}
= \lambda \adjincludegraphics[valign=c, scale=1, trim={10, 10, 10, 10}]{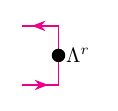} \;.
\label{eq: fixed pt eq}
\end{equation}
We note that $\Lambda^l$ and $\Lambda^r$ are also the fixed points of the transfer matrix $\mathsf{T}[\mathcal{O}^{\dagger}]$, i.e.,
\begin{equation}
\adjincludegraphics[valign=c, scale=1, trim={10, 10, 10, 10}]{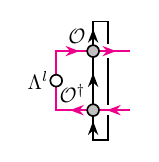} = \lambda \adjincludegraphics[valign=c, scale=1, trim={10, 10, 10, 10}]{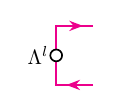} \;, \qquad
\adjincludegraphics[valign=c, scale=1, trim={10, 10, 10, 10}]{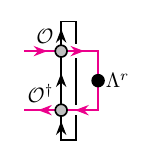} = \lambda \adjincludegraphics[valign=c, scale=1, trim={10, 10, 10, 10}]{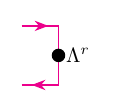} \;,
\label{eq: fixed pt eq dagger}
\end{equation}
where $\lambda$ in \eqref{eq: fixed pt eq dagger} is the same as that in \eqref{eq: fixed pt eq}.
In later subsections, we will often use the fact that the fixed points $\Lambda^l$ and $\Lambda^r$ are positive definite \cite{Perez-Garcia:2006nqo, Fannes1992} and, in particular, they have full rank and are Hermitian.

\vspace*{\baselineskip}
\noindent{\bf Left and right ranks.}
We define the left rank of an MPO tensor $\mathcal{O}$ as the rank of $\mathcal {O}$ viewed as a linear map from the bottom left to the top right in its diagrammatic representation.
Similarly, the right rank of $\mathcal{O}$ is defined as the rank of $\mathcal{O}$ viewed as a linear map from the bottom right to the top left.
Diagrammatically, these quantities are written as
\begin{equation}
\lrank(\mathcal{O}) \coloneq \rank\left(\adjincludegraphics[valign=c, scale=1, trim={10, 10, 10, 10}]{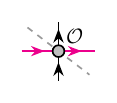}\right) \;, \qquad
\rrank(\mathcal{O}) \coloneq \rank\left(\adjincludegraphics[valign=c, scale=1, trim={10, 10, 10, 10}]{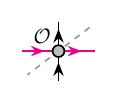}\right) \;.
\label{eq: left and right ranks}
\end{equation}
Here, the four-leg tensor in the first equation is interpreted as a linear map from $V \otimes \mathcal{H}_{\mathrm{o}}$ to $\mathcal{H}_{\mathrm{o}} \otimes V$, whereas the one in the second equation is interpreted as a linear map from $\mathcal{H}_{\mathrm{o}} \otimes V^*$ to $V^* \otimes \mathcal{H}_{\mathrm{o}}$.
We note that the left and right ranks of $\mathcal{O}^{\dagger}$ are given by the right and left ranks of $\mathcal{O}$, that is,
\begin{equation}
\lrank(\mathcal{O}^{\dagger}) = \rrank(\mathcal{O}), \qquad
\rrank(\mathcal{O}^{\dagger}) = \lrank(\mathcal{O}).
\label{eq: rank dagger}
\end{equation}
These quantities were used in \cite{Cirac:2017vke, Sahinoglu:2017iny} to define the index of matrix product unitaries.
In the next subsection, we will use these quantities to define the left and right dimensions, the lattice quantum dimension, and the index of more general MPOs that are not necessarily unitary.

\subsection{Definitions}
\label{sec: Definitions}
In Sections~\ref{sec: no mixing with QCAs} and \ref{sec: mixing with QCAs}, we defined various quantities associated with non-invertible symmetry operators in 1+1d, assuming the existence of defect Hilbert spaces.
In this subsection, we define these quantities directly using tensor network representations of the symmetry operators.

\subsubsection{General case}
We first specify the class of MPOs that we consider to be indecomposable symmetry operators, which are generally non-invertible.
\begin{definition}[Topological injective MPO] \label{def: Topological injective MPO}
An injective MPO tensor $\mathcal{O}$ is said to be topological if it satisfies
\begin{equation}
\adjincludegraphics[valign=c, scale=1, trim={10, 10, 10, 10}]{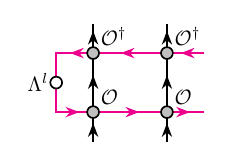}
\; = \;
\adjincludegraphics[valign=c, scale=1, trim={10, 10, 10, 10}]{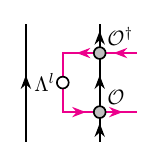},
\qquad
\adjincludegraphics[valign=c, scale=1, trim={10, 10, 10, 10}]{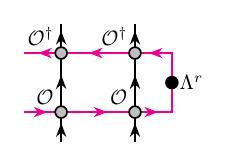}
\; = \;
\adjincludegraphics[valign=c, scale=1, trim={10, 10, 10, 10}]{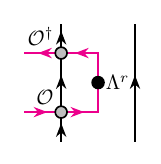} \;,
\label{eq: topological injective MPO}
\end{equation}
\begin{equation}
\adjincludegraphics[valign=c, scale=1, trim={10, 10, 10, 10}]{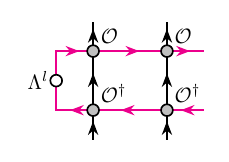}
\; = \;
\adjincludegraphics[valign=c, scale=1, trim={10, 10, 10, 10}]{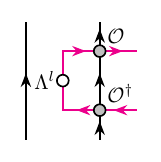},
\qquad
\adjincludegraphics[valign=c, scale=1, trim={10, 10, 10, 10}]{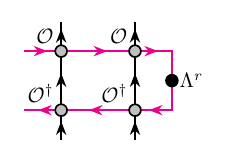}
\; = \;
\adjincludegraphics[valign=c, scale=1, trim={10, 10, 10, 10}]{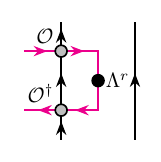} \;,
\label{eq: O dagger topological}
\end{equation}
where $\Lambda^l$ and $\Lambda^r$ are the left and right fixed points of the transfer matrix $\mathsf{T}[\mathcal{O}]$.
The MPO $\mathsf{D}[\mathcal{O}]$ generated by such a tensor $\mathcal{O}$ is called a topological injective MPO.
\end{definition}

As we will see in Section~\ref{sec: The case of invertible symmetries}, topological injective MPOs include all matrix product unitaries, which represent invertible symmetries.
More precisely, any matrix product unitary can be made into a topological injective MPO by blocking a sufficient number of physical sites.

We note that topological injective MPOs do not include the product of local projectors.
In particular, the projector onto the state space of the anyon chain model is not a topological injective MPO.
Hence, the symmetry operators of the anyon chain model are excluded from the discussion.\footnote{See \cite{Lootens:2021tet, Lootens:2022avn} for MPO representations of general fusion category symmetries of the anyon chain models.}

A topological injective MPO also has the following equivalent definition.
\begin{definition}[Equivalent definition of topological injective MPO]\label{def: Topological injective MPO 2}
An injective MPO tensor $\mathcal{O}$ is said to be topological if it satisfies \eqref{eq: topological injective MPO} and there exist non-zero complex numbers $\delta_l(\mathcal{O})$ and $\delta_r(\mathcal{O})$ such that
\begin{equation}
\adjincludegraphics[valign=c, scale=1, trim={10, 10, 10, 10}]{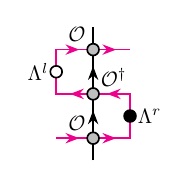} = \delta_l(\mathcal{O}) \;\; \adjincludegraphics[valign=c, scale=1, trim={10, 10, 10, 10}]{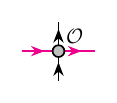}\;, \qquad
\adjincludegraphics[valign=c, scale=1, trim={10, 10, 10, 10}]{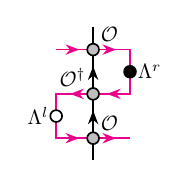} = \delta_r(\mathcal{O}) \;\; \adjincludegraphics[valign=c, scale=1, trim={10, 10, 10, 10}]{tikz/out/O.pdf}\;.
\label{eq: zigzag def}
\end{equation}
We refer to the above equation as the zigzag relations.
\end{definition}
The equivalence of the above two definitions will be shown in Appendix~\ref{sec: Equivalence of definitions}.
Furthermore, as we will show in Appendix~\ref{sec: Derivation of delta lr}, the coefficients $\delta_l(\mathcal{O})$ and $\delta_r(\mathcal{O})$ in \eqref{eq: zigzag def} are uniquely given by the real numbers \eqref{eq: delta lr}, which depend on the normalizations of $\Lambda^l$ and $\Lambda^r$.

When $\mathcal{O}$ is a topological injective MPO tensor, any other MPO tensor $\mathcal{O}^{\prime}$ related to $\mathcal{O}$ by the gauge transformation~\eqref{eq: gauge transformation} is also topological.
Thus, due to the fundamental theorem of injective MPS/MPO \cite{Perez-Garcia:2006nqo, Cirac:2016iqe}, any injective MPO tensor generating a topological injective MPO is topological.
Furthermore, it is clear from Definition~\ref{def: Topological injective MPO} that the Hermitian conjugate of a topological injective MPO is also topological.

We emphasize that we generally need at least two physical legs in the defining equations \eqref{eq: topological injective MPO} and \eqref{eq: O dagger topological}; in particular, we do not require that the virtual bond can be detached from all the physical legs simultaneously.
Indeed, in the case of invertible symmetries, the only symmetry operator that can be detached from all the physical legs simultaneously is the tensor product of on-site operators; see Section~\ref{sec: The case of invertible symmetries} for more details.

As we will show in Section~\ref{sec: Defect Hilbert spaces TN}, topological injective MPOs have the corresponding defect Hilbert spaces that satisfy Assumption~\ref{assump: defect Hilbert space}.
These defect Hilbert spaces are defined so that the left and right dimensions defined by \eqref{eq: ldim rdim} agree with those defined as follows:
\begin{definition}[Left and right dimensions] \label{def: Left and right dimensions}
The left dimension $\ldim(\mathcal{O})$ and the right dimension $\rdim(\mathcal{O})$ of a topological injective MPO tensor $\mathcal{O}$ are defined by
\begin{equation}
\ldim(\mathcal{O}) \coloneq \frac{\lrank(\mathcal{O})}{\dim(\mathcal{H}_{\mathrm{o}})}, \qquad
\rdim(\mathcal{O}) \coloneq \frac{\rrank(\mathcal{O})}{\dim(\mathcal{H}_{\mathrm{o}})},
\label{eq: ldim rdim TN}
\end{equation}
where $\lrank(\mathcal{O})$ and $\rrank(\mathcal{O})$ are the left and right ranks defined by \eqref{eq: left and right ranks}.
\end{definition}
By definition, the left and right dimensions of $\mathcal{O}$ are less than or equal to the bond dimension of $\mathcal{O}$.
We note that topological injective MPO tensors generating the same MPO have the same dimensions.
Namely, if two topological injective MPO tensors $\mathcal{O}$ and $\mathcal{O}^{\prime}$ satisfy $\mathsf{D}[\mathcal{O}] = \mathsf{D}[\mathcal{O}^{\prime}]$ on a periodic chain of any length, we have $\ldim(\mathcal{O}) = \ldim(\mathcal{O}^{\prime})$ and $\rdim(\mathcal{O}) = \rdim(\mathcal{O}^{\prime})$.
This is an immediate consequence of the fundamental theorem of injective MPS/MPO \cite{Perez-Garcia:2006nqo, Cirac:2016iqe}.
Furthermore, due to \eqref{eq: rank dagger}, the left and right dimensions of $\mathcal{O}$ and $\mathcal{O}^{\dagger}$ are related by
\begin{equation}
\ldim(\mathcal{O}^{\dagger}) = \rdim(\mathcal{O}), \qquad
\rdim(\mathcal{O}^{\dagger}) = \ldim(\mathcal{O}).
\label{eq: ldim rdim dagger}
\end{equation}
Physically, this relation means that the left and right dimensions are swapped if we flip the orientation of the symmetry operator.

By using the left and right dimensions, we can define the lattice quantum dimension and the index as in \eqref{eq: qdim ind}.
For completeness, we repeat the definition here.
\begin{definition}[Lattice quantum dimension and index] \label{def: Lattice quantum dimension and index}
The latice quantum dimension $\qdim_{\mathrm{lat}}(\mathcal{O})$ and the index $\ind(\mathcal{O})$ of a topological injective MPO tensor $\mathcal{O}$ are defined by
\begin{equation}
\qdim_{\mathrm{lat}}(\mathcal{O}) \coloneq \sqrt{\ldim(\mathcal{O}) \rdim(\mathcal{O})}, \qquad
\ind(\mathcal{O}) \coloneq \sqrt{\frac{\ldim(\mathcal{O})}{\rdim(\mathcal{O})}}.
\label{eq: qdim ind TN}
\end{equation}
\end{definition}
Since $\ldim(\mathcal{O})$ and $\rdim(\mathcal{O})$ depend only on $\mathsf{D}[\mathcal{O}]$, so do $\qdim_{\text{lat}}(\mathcal{O})$ and $\ind(\mathcal{O})$.
Furthermore, due to \eqref{eq: ldim rdim dagger}, the lattice quantum dimension and index of $\mathcal{O}$ and $\mathcal{O}^{\dagger}$ are related by
\begin{equation}
\qdim_{\text{lat}}(\mathcal{O}^{\dagger}) = \qdim_{\text{lat}}(\mathcal{O}), \qquad
 \ind(\mathcal{O}^{\dagger}) = \ind(\mathcal{O})^{-1}.
\end{equation}

\subsubsection{Example: invertible symmetries}
\label{sec: The case of invertible symmetries}
Topological injective MPOs include all invertible symmetry operators defined by translationally invariant QCAs.
To see this, let us recall the tensor network description of QCAs.
As shown in \cite{Sahinoglu:2017iny, Cirac:2017vke}, any translationally invariant QCA in 1+1 dimensions can be represented by a matrix product unitary (MPU), and conversely, any translationally invariant MPU with periodic boundary conditions is a QCA.
Here, an MPU is an MPO that is unitary.
Without loss of generality, we can take the local tensor $\mathcal{U}$ of an MPU to be injective \cite{Cirac:2017vke, Sahinoglu:2017iny}.\footnote{More precisely, the local tensor of any MPU has a single block in its canonical form \cite{Cirac:2017vke, Sahinoglu:2017iny}, and thus we can make it injective by blocking a finite number of physical sites.}
Furthermore, by blocking finitely many physical sites, one can always make $\mathcal{U}$ into a simple tensor that satisfies \cite{Cirac:2017vke, Sahinoglu:2017iny}
\begin{equation}
\adjincludegraphics[valign=c, scale=1, trim={10, 10, 10, 10}]{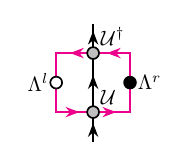}
\;\; = \;\;
\adjincludegraphics[valign=c, scale=1, trim={10, 10, 10, 10}]{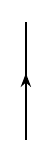}
\;\; ,
\qquad \qquad
\adjincludegraphics[valign=c, scale=1, trim={10, 10, 10, 10}]{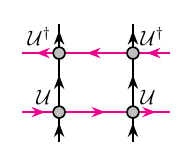}
\;\; = \;\;
\adjincludegraphics[valign=c, scale=1, trim={10, 10, 10, 10}]{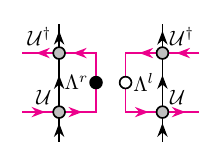}\;.
\label{eq: simple MPU}
\end{equation}
Here, the fixed points $\Lambda^l$ and $\Lambda^r$ are supposed to be properly normalized so that the above equation holds without additional scalar factors.
The above equation readily implies that the MPU tensor $\mathcal{U}$ satisfies \eqref{eq: topological injective MPO}.
Furthermore, since $\mathcal{U}^{\dagger}$ also generates an MPU, it satisfies\footnote{We can always block sufficiently many physical sites so that both \eqref{eq: simple MPU} and \eqref{eq: simple MPU dagger} are satisfied at the same time.}
\begin{equation}
\adjincludegraphics[valign=c, scale=1, trim={10, 10, 10, 10}]{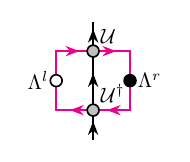}
\;\; = \;\;
\adjincludegraphics[valign=c, scale=1, trim={10, 10, 10, 10}]{tikz/out/simple_MPU2.pdf}
\;\; ,
\qquad \qquad
\adjincludegraphics[valign=c, scale=1, trim={10, 10, 10, 10}]{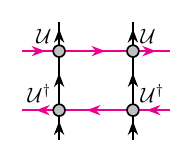}
\;\; = \;\;
\adjincludegraphics[valign=c, scale=1, trim={10, 10, 10, 10}]{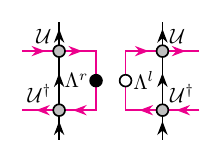}\;,
\label{eq: simple MPU dagger}
\end{equation}
which implies \eqref{eq: O dagger topological}.
Thus, any invertible symmetry operator can be represented by a topological injective MPO.

In \cite[Theorem III.8]{Cirac:2017vke}, it was shown that the left and right ranks of the MPU tensor satisfying \eqref{eq: simple MPU} obey
\begin{equation}
\lrank(\mathcal{U}) \rrank(\mathcal{U}) = \dim(\mathcal{H}_{\mathrm{o}})^2.
\label{eq: lrank U rrank U}
\end{equation}
This implies that the lattice quantum dimension of any MPU is equal to one because
\begin{equation}
\qdim_{\text{lat}}(\mathcal{U}) = \sqrt{\frac{\lrank(\mathcal{U})\rrank(\mathcal{U})}{\dim(\mathcal{H}_{\mathrm{o}})^2}} = 1.
\label{eq: qdim MPU}
\end{equation}
On the other hand, it was shown in \cite{Cirac:2017vke, Sahinoglu:2017iny} that the index of an MPU defined by \eqref{eq: qdim ind TN} agrees with the GNVW index defined in \cite{Gross:2011yvb}, i.e.,
\begin{equation}
\ind(\mathcal{U}) = \mathrm{GNVW}(\mathsf{D}[\mathcal{U}]),
\label{eq: ind MPU}
\end{equation}
where the right-hand side is the GNVW index of the MPU $\mathsf{D}[\mathcal{U}]$.
Due to \eqref{eq: qdim MPU} and \eqref{eq: ind MPU}, the left and right dimensions of $\mathcal{U}$ can be written in terms of the GNVW index as
\begin{equation}
\ldim(\mathcal{U}) = \mathrm{GNVW}(\mathsf{D}[\mathcal{U}]), \qquad
\rdim(\mathcal{U}) = \mathrm{GNVW}(\mathsf{D}[\mathcal{U}])^{-1}.
\label{eq: ldim rdim = GNVW}
\end{equation}
As we will discuss in Section~\ref{sec: Defect Hilbert spaces TN}, the left and right dimensions defined by \eqref{eq: ldim rdim TN} agree with \eqref{eq: ldim rdim} defined in terms of defect Hilbert spaces.
Thus, equation~\eqref{eq: ldim rdim = GNVW} implies that the dimension of the defect Hilbert space associated with a QCA is given by the GNVW index multiplied by the dimension of the original Hilbert space.
This fact was observed in \cite{Seifnashri:2023dpa, Seiberg:2024gek} in the case of the lattice translation operator.

\vspace*{\baselineskip}
\noindent{\bf Detachable MPU is a tensor product operator.}
Before proceeding, we demonstrate the importance of having at least two physical legs in the defining equations~\eqref{eq: topological injective MPO} and \eqref{eq: O dagger topological} of topological injective MPOs.
To this end, we consider an MPU tensor $\mathcal{U}$ that satisfies an analogue of \eqref{eq: topological injective MPO} for a single physical leg, that is,
\begin{equation}
\adjincludegraphics[valign=c, scale=1, trim={10, 10, 10, 10}]{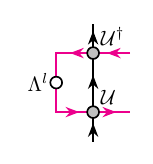}
\; = \;
\adjincludegraphics[valign=c, scale=1, trim={10, 10, 10, 10}]{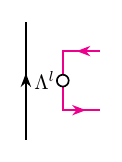},
\qquad
\adjincludegraphics[valign=c, scale=1, trim={10, 10, 10, 10}]{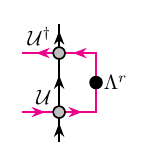}
\; = \;
\adjincludegraphics[valign=c, scale=1, trim={10, 10, 10, 10}]{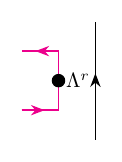} \;.
\label{eq: detachable MPU}
\end{equation}
Namely, we consider an MPU that is detachable from the physical legs.
In this case, we can show that the MPU $\mathsf{D}[\mathcal{U}]$ is the tensor product of on-site operators; in other words, the bond Hilbert space $V$ of the MPU tensor $\mathcal{U}$ is one-dimensional.
To see this, we compute the left and right ranks of $\mathcal{U}$ as
\begin{align}
\lrank(\mathcal{U}) &= \rank\left(\adjincludegraphics[valign=c, scale=1, trim={10, 10, 10, 10}]{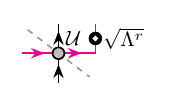}\right) = \rank\left(\adjincludegraphics[valign=c, scale=1, trim={10, 10, 10, 10}]{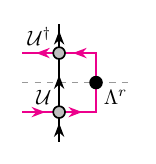}\right) \overset{\eqref{eq: detachable MPU}}{=} \dim(V) \dim(\mathcal{H}_{\mathrm{o}}),
\\
\rrank(\mathcal{U}) &= \rank\left(\adjincludegraphics[valign=c, scale=1, trim={10, 10, 10, 10}]{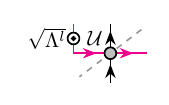}\right) = \rank\left(\adjincludegraphics[valign=c, scale=1, trim={10, 10, 10, 10}]{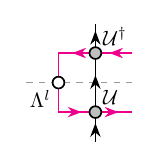}\right) \overset{\eqref{eq: detachable MPU}}{=} \dim(V) \dim(\mathcal{H}_{\mathrm{o}}),
\end{align}
where we used the fact that $\Lambda^l$ and $\Lambda^r$ are positive definite and $\rank(f) = \rank(f^{\dagger}f)$ for any linear map $f$.
The above equation implies that $\lrank(\mathcal{U}) \rrank(\mathcal{U}) = \dim(V)^2 \dim(\mathcal{H}_{\mathrm{o}})^2$.
On the other hand, as shown in \cite[Theorem III.8]{Cirac:2017vke}, the product of the left and right ranks of a general MPU tensor satisfying \eqref{eq: simple MPU} is given by \eqref{eq: lrank U rrank U}.
Therefore, it follows that $\dim(V) = 1$, i.e., the MPU satisfying the stronger condition~\eqref{eq: detachable MPU} must be the tensor product of on-site operators.

\subsection{Topological injective MPOs}
\label{sec: Properties of topological injective MPOs}
In this subsection, we derive several properties of topological injective MPO tensors.

\vspace*{\baselineskip}
\noindent{\bf Normalization of $\mathcal{O}$.}
We first point out that the normalization of a topological injective MPO tensor $\mathcal{O}$ is fixed by the condition~\eqref{eq: topological injective MPO}.
Indeed, the condition~\eqref{eq: topological injective MPO} implies that $\mathcal{O}$ is normalized so that the largest eigenvalue of the transfer matrix $\mathsf{T}[\mathcal{O}]$ is given by $\dim(\mathcal{H}_{\mathrm{o}})$, that is,
\begin{equation}
\adjincludegraphics[valign=c, scale=1, trim={10, 10, 10, 10}]{tikz/out/normalization_l1.pdf}
= \dim(\mathcal{H}_{\mathrm{o}}) \adjincludegraphics[valign=c, scale=1, trim={10, 10, 10, 10}]{tikz/out/normalization_l2.pdf} \; , 
\qquad
\adjincludegraphics[valign=c, scale=1, trim={10, 10, 10, 10}]{tikz/out/normalization_r1.pdf}
= \dim(\mathcal{H}_{\mathrm{o}}) \adjincludegraphics[valign=c, scale=1, trim={10, 10, 10, 10}]{tikz/out/normalization_r2.pdf}.
\label{eq: normalization}
\end{equation}
The first equality can be verified by taking a partial trace in the first equality of \eqref{eq: topological injective MPO}.
More specifically, by taking the trace only over the left physical leg in the first equality of \eqref{eq: topological injective MPO}, we obtain
\begin{equation}
\dim(\mathcal{H}_{\mathrm{o}}) \adjincludegraphics[valign=c, scale=1, trim={10, 10, 10, 10}]{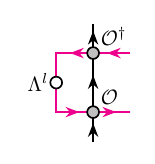}
\; \overset{\eqref{eq: topological injective MPO}}{=} \;
\adjincludegraphics[valign=c, scale=1, trim={10, 10, 10, 10}]{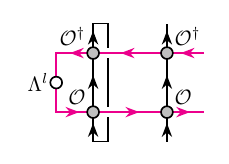}
\; \overset{\eqref{eq: fixed pt eq}}{=} \;
\lambda \adjincludegraphics[valign=c, scale=1, trim={10, 10, 10, 10}]{tikz/out/normalization_derivation1.pdf}.
\end{equation}
The above equation implies that $\lambda = \dim(\mathcal{H}_{\mathrm{o}})$, which shows the first equality of \eqref{eq: normalization}.
The second equality of \eqref{eq: normalization} can also be verified similarly.
Due to the normalization~\eqref{eq: normalization}, a scalar multiple of a topological injective MPO tensor is no longer topological in general.
We note that the same normalization of an MPO tensor is employed in \cite{Tantivasadakarn:2025txn}.

\vspace*{\baselineskip}
\noindent{\bf Bubble removal.}
The condition~\eqref{eq: topological injective MPO} also implies that a small loop of a topological injective MPO acting on a single physical leg can be removed at the cost of a scalar multiplication.
More specifically, one can show that
\begin{equation}
\adjincludegraphics[valign=c, scale=1, trim={10, 10, 10, 10}]{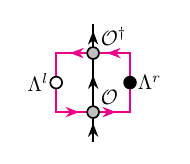} = \; \gamma \;\;
\adjincludegraphics[valign=c, scale=1, trim={10, 10, 10, 10}]{tikz/out/simple_MPU2.pdf}\;,
\label{eq: bubble removal}
\end{equation}
where the proportionality constant $\gamma$ is given by
\begin{equation}
\gamma \coloneq
\adjincludegraphics[valign=c, scale=1, trim={10, 10, 10, 10}]{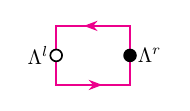} = \tr{\Lambda^l \Lambda^r}.
\label{eq: loop}
\end{equation}
We note that $\gamma$ is a positive real number because $\Lambda^l$ and $\Lambda^r$ are positive definite.\footnote{Since $\sqrt{\Lambda^l} \Lambda^r \sqrt{\Lambda^l}$ is positive definite, we have $\tr{\Lambda^l \Lambda^r} = \tr{\sqrt{\Lambda^l} \Lambda^r \sqrt{\Lambda^l}} > 0$.}
By definition, $\gamma$ depends on the normalization of $\Lambda^l$ and $\Lambda^r$.
To show \eqref{eq: bubble removal}, we again consider a partial trace in the first equation of \eqref{eq: topological injective MPO}.
By taking the trace only for the right physical leg in the first equality and capping the virtual leg with $\Lambda^r$, we obtain
\begin{equation*}
\adjincludegraphics[valign=c, scale=1, trim={10, 10, 10, 10}]{tikz/out/bubble_removal1.pdf}
\overset{\eqref{eq: normalization}}{=}
\frac{1}{\dim(\mathcal{H}_{\mathrm{o}})} \adjincludegraphics[valign=c, scale=1, trim={10, 10, 10, 10}]{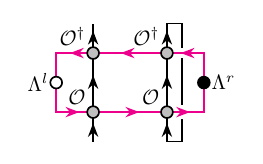}
\overset{\eqref{eq: topological injective MPO}}{=}
\frac{1}{\dim(\mathcal{H}_{\mathrm{o}})} \; \adjincludegraphics[valign=c, scale=1, trim={10, 10, 10, 10}]{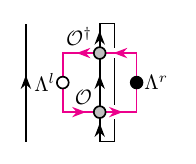}
\overset{\eqref{eq: normalization}}{=}
\gamma \;\;\adjincludegraphics[valign=c, scale=1, trim={10, 10, 10, 10}]{tikz/out/simple_MPU2.pdf}\;,
\end{equation*}
which shows \eqref{eq: bubble removal}.
We note that a loop of a topological injective MPO acting on multiple physical legs can also be removed similarly.
This is because the condition~\eqref{eq: topological injective MPO} allows us to shrink the loop into the one acting only on a single physical leg, which can be removed by \eqref{eq: bubble removal}.

\vspace*{\baselineskip}
\noindent{\bf Zigzag relations.}
Using the conditions~\eqref{eq: topological injective MPO} and \eqref{eq: O dagger topological} together with the injectivity, one can show that a topological injective MPO tensor $\mathcal{O}$ satisfies the following zigzag relations:\footnote{In other words, a topological injective MPO in the sense of Definition~\ref{def: Topological injective MPO} is a topological injective MPO in the sense of Definition~\ref{def: Topological injective MPO 2}. The converse is also true; we refer the reader to Appendix~\ref{sec: Equivalence of definitions} for a proof.}
\begin{equation}
\adjincludegraphics[valign=c, scale=1, trim={10, 10, 10, 10}]{tikz/out/zigzag1.pdf} = \delta_l(\mathcal{O}) \;\; \adjincludegraphics[valign=c, scale=1, trim={10, 10, 10, 10}]{tikz/out/O.pdf}\;, \qquad
\adjincludegraphics[valign=c, scale=1, trim={10, 10, 10, 10}]{tikz/out/zigzag2.pdf} = \delta_r(\mathcal{O}) \;\; \adjincludegraphics[valign=c, scale=1, trim={10, 10, 10, 10}]{tikz/out/O.pdf}\;.
\label{eq: zigzag}
\end{equation}
Here, $\delta_l(\mathcal{O})$ and $\delta_r(\mathcal{O})$ are the positive real numbers defined by
\begin{equation}
\delta_l(\mathcal{O}) \coloneq \frac{\gamma}{\ldim(\mathcal{O})}, \qquad
\delta_r(\mathcal{O}) \coloneq \frac{\gamma}{\rdim(\mathcal{O})},
\label{eq: delta lr}
\end{equation}
where $\gamma$ is given by \eqref{eq: loop}.
We note that $\delta_l(\mathcal{O})$ and $\delta_r(\mathcal{O})$ depend on the normalization of $\Lambda^l$ and $\Lambda^r$ just as $\gamma$ does.
We also emphasize that when $\ind(\mathcal{O}) \neq 1$, we cannot set $\delta_l(\mathcal{O})$ and $\delta_r(\mathcal{O})$ to one simultaneously because $\delta_l(\mathcal{O})/\delta_r(\mathcal{O}) = \ind(\mathcal{O})^{-2}$.
Equation~\eqref{eq: zigzag} can be verified by following the proof of \cite[Proposition 1]{Franco-Rubio:2025qss}.
For completeness, we give a proof below.

To show the first equality of \eqref{eq: zigzag}, we notice that the linear map defined by the left-hand side, which we denoted by $s$, commutes with $\mathcal{O}$ in the following sense:
\begin{equation}
\adjincludegraphics[valign=c, scale=1, trim={10, 10, 10, 10}]{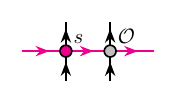}
\;=\;
\adjincludegraphics[valign=c, scale=1, trim={10, 10, 10, 10}]{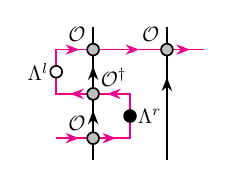}
\;=\;
\adjincludegraphics[valign=c, scale=1, trim={10, 10, 10, 10}]{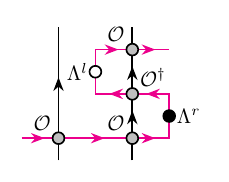}
\;=\; 
\adjincludegraphics[valign=c, scale=1, trim={10, 10, 10, 10}]{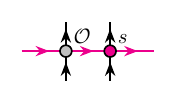}\;.
\end{equation}
Here, the second equality follows from \eqref{eq: topological injective MPO} and \eqref{eq: O dagger topological}.
By applying the above equation repeatedly, we obtain
\begin{equation}
\adjincludegraphics[valign=c, scale=1, trim={10, 10, 10, 10}]{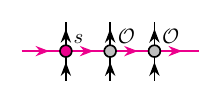} \;=\; \adjincludegraphics[valign=c, scale=1, trim={10, 10, 10, 10}]{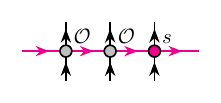}\;.
\label{eq: soo = oos}
\end{equation}
Now, since $\mathcal{O}$ is injective, it has a left inverse $\widetilde{O}$ that satisfies
\begin{equation}
\adjincludegraphics[valign=c, scale=1, trim={10, 10, 10, 10}]{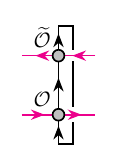} \;=\; \adjincludegraphics[valign=c, scale=1, trim={10, 10, 10, 10}]{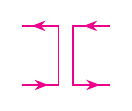}\;.
\end{equation}
By applying $\widetilde{\mathcal{O}}$ to the middle physical leg of \eqref{eq: soo = oos} and taking the trace over this physical leg, we obtain
\begin{equation}
\adjincludegraphics[valign=c, scale=1, trim={10, 10, 10, 10}]{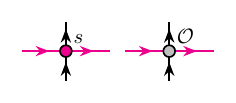} \;=\; \adjincludegraphics[valign=c, scale=1, trim={10, 10, 10, 10}]{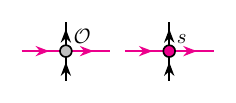}\;.
\end{equation}
Therefore, there exists a constant $\delta_l(\mathcal{O})$ such that
\begin{equation}
\adjincludegraphics[valign=c, scale=1, trim={10, 10, 10, 10}]{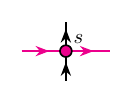} \;=\; \delta_l(\mathcal{O}) \;\; \adjincludegraphics[valign=c, scale=1, trim={10, 10, 10, 10}]{tikz/out/O.pdf}\;.
\end{equation}
This shows the first equality of \eqref{eq: zigzag}.
A similar computation shows that there also exists another constant $\delta_r(\mathcal{O})$ that satisfies the second equation of \eqref{eq: zigzag}.
In Appendix~\ref{sec: Derivation of delta lr}, we will show that $\delta_l(\mathcal{O})$ and $\delta_r(\mathcal{O})$ are given by \eqref{eq: delta lr}.

\subsection{Defect Hilbert spaces and movement operators}
\label{sec: Defect Hilbert spaces TN}
In this subsection, we define the defect Hilbert spaces associated with a topological injective MPO $\mathsf{D}[\mathcal{O}]$ and describe the unitary operator that moves the defect.

\vspace*{\baselineskip}
\noindent{\bf Defect Hilbert spaces.}
The defect Hilbert spaces $\mathcal{H}_{\mathsf{D}[\mathcal{O}]}^l$ and $\mathcal{H}_{\mathsf{D}[\mathcal{O}]}^r$ are defined by replacing the on-site Hilbert space $\mathcal{H}_{\mathrm{o}}$ on a single site with another finite dimensional Hilbert space $L$ or $R$.
More specifically, we define
\begin{equation}
\mathcal{H}_{\mathsf{D}[\mathcal{O}]}^l \coloneq L \otimes \mathcal{H}_{\mathrm{o}}^{\otimes N-1}, \qquad
\mathcal{H}_{\mathsf{D}[\mathcal{O}]}^r \coloneq R \otimes \mathcal{H}_{\mathrm{o}}^{\otimes N-1},
\label{eq: defect Hilbert spaces def}
\end{equation}
where $N$ is the total number of sites.
Here, the vector spaces $L$ and $R$ are given by
\begin{equation}
L \coloneq \mathrm{Im}(P_l), \qquad
R \coloneq \mathrm{Im}(P_r),
\label{eq: LR}
\end{equation}
where $P_l$ and $P_r$ are the projectors defined by
\begin{equation}
P_l \coloneq \frac{1}{\delta_l(\mathcal{O})} \adjincludegraphics[valign=c, scale=1, trim={10, 10, 10, 10}]{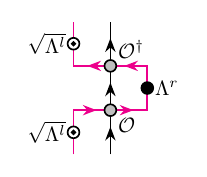}\;, \qquad
P_r \coloneq \frac{1}{\delta_r(\mathcal{O})} \adjincludegraphics[valign=c, scale=1, trim={10, 10, 10, 10}]{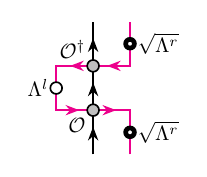}\;.
\label{eq: Pl Pr}
\end{equation}
In particular, $L$ and $R$ are subspaces of $V \otimes \mathcal{H}_{\text{o}}$ and $\mathcal{H}_{\text{o}} \otimes V^*$, where $V$ is the bond Hilbert space of $\mathcal{O}$ and $V^*$ is its dual vector space.
We note that $P_l$ is an orthogonal projector, i.e., it satisfies $P_l^2 = P_l$ and $P_l^{\dagger} = P_l$.
The first equality follows from the zigzag equation~\eqref{eq: zigzag} and the second equality follows from the Hermiticity of $\sqrt{\Lambda^l}$ and $\Lambda^r$.
Similarly, $P_r$ is also an orthogonal projector.

For the above definition of the defect Hilbert spaces, the left and right dimensions defined by \eqref{eq: ldim rdim TN} agree with those in \eqref{eq: ldim rdim}, that is,
\begin{equation}
\ldim(\mathcal{O}) \coloneq \frac{\lrank(\mathcal{O})}{\dim(\mathcal{H}_{\mathrm{o}})} = \frac{\dim(\mathcal{H}_{\mathsf{D}[\mathcal{O}]}^l)}{\dim(\mathcal{H})}, \qquad
\rdim(\mathcal{O}) \coloneq \frac{\rrank(\mathcal{O})}{\dim(\mathcal{H}_{\mathrm{o}})} = \frac{\dim(\mathcal{H}_{\mathsf{D}[\mathcal{O}]}^r)}{\dim(\mathcal{H})}.
\label{eq: dim TN = dim defect}
\end{equation}
This can be verified by a direct computation as follows:
\begin{equation*}
\frac{\dim(\mathcal{H}_{\mathsf{D}[\mathcal{O}]}^l)}{\dim(\mathcal{H})} = \frac{\tr{P_l}}{\dim(\mathcal{H}_{\mathrm{o}})} = \frac{\gamma}{\delta_l(\mathcal{O})} = \ldim(\mathcal{O}), \quad
\frac{\dim(\mathcal{H}_{\mathsf{D}[\mathcal{O}]}^r)}{\dim(\mathcal{H})} = \frac{\tr{P_r}}{\dim(\mathcal{H}_{\mathrm{o}})} = \frac{\gamma}{\delta_r(\mathcal{O})} = \rdim(\mathcal{O}).
\end{equation*}
Here, we used \eqref{eq: bubble removal} and \eqref{eq: delta lr}.

\vspace*{\baselineskip}
\noindent{\bf Movement operators.}
On the defect Hilbert spaces $\mathcal{H}_{\mathsf{D}[\mathcal{O}]}^l$ and $\mathcal{H}_{\mathsf{D}[\mathcal{O}]}^r$, one can define unitary operators that move the defect by one site.
We refer to such unitary operators as movement operators following \cite{Seifnashri:2023dpa, Seiberg:2024gek, Seifnashri:2025fgd}.
Concretely, the movement operators on $\mathcal{H}_{\mathsf{D}[\mathcal{O}]}^l$ and $\mathcal{H}_{\mathsf{D}[\mathcal{O}]}^r$ are given by
\begin{equation}
u_l \coloneq \frac{1}{\delta_l(\mathcal{O})} \adjincludegraphics[valign=c, scale=1, trim={10, 10, 10, 10}]{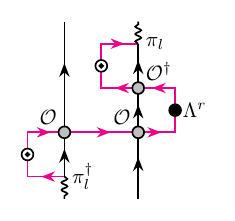}, \qquad
u_r \coloneq \frac{1}{\delta_r(\mathcal{O})} \adjincludegraphics[valign=c, scale=1, trim={10, 10, 10, 10}]{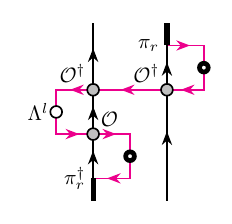}.
\label{eq: movement ops}
\end{equation}
Here, the wiggly and thick lines represent the new physical legs $L$ and $R$ at the defect locus, and $\pi_l: V \otimes \mathcal{H}_{\mathrm{o}} \to L$ and $\pi_r: \mathcal{H}_{\mathrm{o}} \otimes V^* \to R$ are the linear maps obtained by restricting the target vector spaces of $P_l$ and $P_r$ to their images $L$ and $R$, respectively.
The unlabeled dots in the first and second equations of \eqref{eq: movement ops} represent $\sqrt{\Lambda^l}$ and $\sqrt{\Lambda^r}$.
For later use, we mention that $\pi_l$ and $\pi_r$ satisfy
\begin{equation}
\pi_l^{\dagger} \pi_l = P_l, \qquad
\pi_l \pi_l^{\dagger} = \id_L, \qquad
\pi_r^{\dagger} \pi_r = P_r, \qquad
\pi_r \pi_r^{\dagger} = \id_R.
\label{eq: pi identities}
\end{equation}
The operators $u_l$ and $u_r$ defined above move the defect to the right by one site.
On the other hand, the operators that move the defect to the left by one site are given by 
\begin{equation}
u_l^{\dagger} = \frac{1}{\delta_l(\mathcal{O})} \adjincludegraphics[valign=c, scale=1, trim={10, 10, 10, 10}]{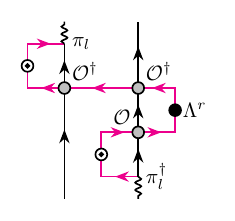}, \qquad
u_r^{\dagger} = \frac{1}{\delta_r(\mathcal{O})} \adjincludegraphics[valign=c, scale=1, trim={10, 10, 10, 10}]{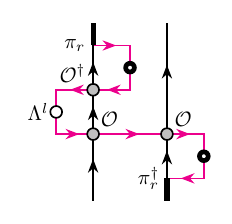}.
\label{eq: movement ops inverse}
\end{equation}
The operators in \eqref{eq: movement ops inverse} are the inverses of those in \eqref{eq: movement ops}.
In particular, the movement operators $u_l$ and $u_r$ are unitary.
The unitarity of $u_l$ can be verified by a direct computation as follows:
\begin{equation*}
u_l^{\dagger} u_l =
\frac{1}{\delta_l(\mathcal{O})^3} \adjincludegraphics[valign=c, scale=1, trim={10, 10, 10, 10}]{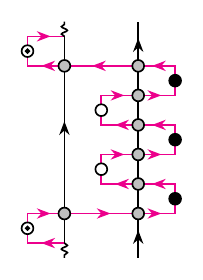} \;=\;
\frac{1}{\delta_l(\mathcal{O})} \adjincludegraphics[valign=c, scale=1, trim={10, 10, 10, 10}]{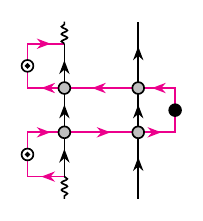} \;=\;
\frac{1}{\delta_l(\mathcal{O})} \adjincludegraphics[valign=c, scale=1, trim={10, 10, 10, 10}]{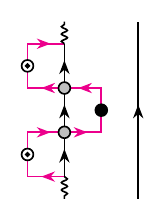} \;=\;
\id_{L \otimes \mathcal{H}_{\mathrm{o}}}.
\end{equation*}
Here, the first equality follows from $\pi_l^{\dagger} \pi_l = P_l$, the second equality follows from the zigzag relation~\eqref{eq: zigzag}, the third equality follows from the condition~\eqref{eq: topological injective MPO}, and the last equality follows from $\pi_l P_l = \pi_l$ and $\pi_l \pi_l^{\dagger} = \id_L$.
A similar computation also shows $u_l u_l^{\dagger} = \id_{\mathcal{H}_{\mathrm{o}} \otimes L}$, and hence $u_l$ is unitary.
One can also show the unitarity of $u_r$ in the same way.

\vspace*{\baselineskip}
\noindent{\bf Compatibility with the unitary structure.}
As an immediate consequence of \eqref{eq: dim TN = dim defect} and \eqref{eq: ldim rdim dagger},  the dimensions of the defect Hilbert spaces associated with topological injective MPOs $\mathsf{D}[\mathcal{O}]$ and $\mathsf{D}[\mathcal{O}]^{\dagger} = \mathsf{D}[\mathcal{O}^{\dagger}]$ are related by
\begin{equation}
\dim(\mathcal{H}^l_{\mathsf{D}[\mathcal{O}]^{\dagger}}) = \dim(\mathcal{H}^r_{\mathsf{D}[\mathcal{O}]}), \qquad
\dim(\mathcal{H}^r_{\mathsf{D}[\mathcal{O}]^{\dagger}}) = \dim(\mathcal{H}^l_{\mathsf{D}[\mathcal{O}]}).
\end{equation}
This implies that there exist isomorphisms between these defect Hilbert spaces, i.e.,
\begin{equation}
\mathcal{H}^l_{\mathsf{D}[\mathcal{O}]^{\dagger}} \cong \mathcal{H}^r_{\mathsf{D}[\mathcal{O}]}, \qquad
\mathcal{H}^r_{\mathsf{D}[\mathcal{O}]^{\dagger}} \cong \mathcal{H}^l_{\mathsf{D}[\mathcal{O}]}.
\label{eq: HO isom HOdagger}
\end{equation}
In other words, the defect Hilbert spaces defined above satisfy Assumption~\ref{assump: Compatibility with the unitary structure}.

Let us now construct these isomorphisms in a canonical way.
To this end, we notice that the projectors $P_l$ and $P_r$ defined for $\mathcal{O}$ and $\mathcal{O}^{\dagger}$ are related by\footnote{To avoid confusion, the projectors $P_l$ and $P_r$ defined by~\eqref{eq: Pl Pr} are denoted by $P_l[\mathcal{O}]$ and $P_r[\mathcal{O}]$ for now. The same convention applies to $\pi_l$ and $\pi_r$, as well as the vector spaces $L$ and $R$ defined by \eqref{eq: LR}.}
\begin{equation}
P_l[\mathcal{O}^{\dagger}] = f_{lr} P_r[\mathcal{O}] f_{lr}^{\dagger}, \qquad
P_r[\mathcal{O}^{\dagger}] = f_{rl} P_l[\mathcal{O}] f_{rl}^{\dagger},
\label{eq: POdagger}
\end{equation}
where $f_{lr}$ and $f_{rl}$ are defined by
\begin{equation}
f_{lr} \coloneq \frac{1}{\sqrt{\delta_r(\mathcal{O})}} \adjincludegraphics[valign=c, scale=1, trim={10, 10, 10, 10}]{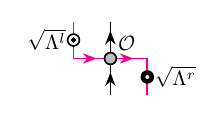}, \qquad
f_{rl} \coloneq \frac{1}{\sqrt{\delta_l(\mathcal{O})}} \adjincludegraphics[valign=c, scale=1, trim={10, 10, 10, 10}]{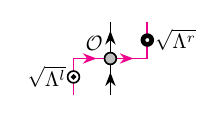}.
\end{equation}
Equation~\eqref{eq: POdagger} readily follows from the zigzag relations~\eqref{eq: zigzag}.
Similarly, one can also show that
\begin{equation}
P_l[\mathcal{O}] = f_{rl}^{\dagger} P_r[\mathcal{O}^{\dagger}] f_{rl}, \qquad
P_r[\mathcal{O}] = f_{lr}^{\dagger} P_l[\mathcal{O}^{\dagger}] f_{lr}.
\label{eq: POdagger2}
\end{equation}
Using the linear maps $f_{lr}$ and $f_{rl}$, we can construct canonical isomorphisms~\eqref{eq: HO isom HOdagger} between the defect Hilbert spaces.
More specifically, we can define unitary maps between the local Hilbert spaces at the defect locus as follows:
\begin{equation}
g_{lr} \coloneq \pi_l[\mathcal{O}^{\dagger}] f_{lr} \pi_r[\mathcal{O}]^{\dagger}: R[\mathcal{O}] \xrightarrow{\cong} L[\mathcal{O}^{\dagger}], \qquad
g_{rl} \coloneq \pi_r[\mathcal{O}^{\dagger}] f_{rl} \pi_l[\mathcal{O}]^{\dagger}: L[\mathcal{O}] \xrightarrow{\cong} R[\mathcal{O}^{\dagger}].
\label{eq: glr grl}
\end{equation}
The unitarity of $g_{lr}$ can be verified by a direct computation as
\begin{equation}
\begin{aligned}
g_{lr}^{\dagger} g_{lr} &= \pi_r[\mathcal{O}] f_{lr}^{\dagger} P_l[\mathcal{O}^{\dagger}] f_{lr} \pi_r[\mathcal{O}]^{\dagger} = \pi_r[\mathcal{O}] P_r[\mathcal{O}] \pi_r[\mathcal{O}]^{\dagger} = \id_{R[\mathcal{O}]}, \\
g_{lr} g_{lr}^{\dagger} &= \pi_l[\mathcal{O}^{\dagger}] f_{lr} P_r[\mathcal{O}] f_{lr}^{\dagger} \pi_l[\mathcal{O}^{\dagger}]^{\dagger} = \pi_l[\mathcal{O}^{\dagger}] P_l[\mathcal{O}]^{\dagger} \pi_l[\mathcal{O}^{\dagger}]^{\dagger} = \id_{L[\mathcal{O}^{\dagger}]}.
\end{aligned}
\end{equation}
Here, we used equations \eqref{eq: POdagger}, \eqref{eq: POdagger2}, and \eqref{eq: pi identities}.
A similar computation shows that $g_{rl}$ is also unitary.
The unitary maps~\eqref{eq: glr grl} give isomorphisms~\eqref{eq: HO isom HOdagger} between the defect Hilbert spaces.

\subsection{Relation to sequential quantum circuits}
\label{sec: Topological injective MPOs as sequential circuits}
It was recently argued in \cite{Tantivasadakarn:2025txn} that non-invertible symmetry operators on the lattice can be generally implemented by using sequential quantum circuits \cite{Chen:2023qst}.
In this subsection, we discuss the relation between sequential quantum circuits and topological injective MPOs.
In particular, we will derive the sequential circuit representations of topological injective MPOs and show that each local unitary gate comprising the circuit can be identified with the movement operator defined in the previous subsection.
We will also write down the defect creation and annihilation operators, which complement the unitary circuit part of the symmetry operator and make the full symmetry action non-invertible.

\vspace*{\baselineskip}
\noindent{\bf Decompositions of topological injective MPO tensors.}
To derive the sequential circuit representation, we first decompose a topological injective MPO tensor $\mathcal{O}$ into two three-leg tensors as was done in the case of MPUs in \cite{Cirac:2017vke}.
To this end, we begin with the singular value decomposition of $\mathcal{O}$, which can be depicted as follows:
\begin{equation}
\adjincludegraphics[valign=c, scale=1, trim={10, 10, 10, 10}]{tikz/out/O.pdf}
\;=\;
\adjincludegraphics[valign=c, scale=1, trim={10, 10, 10, 10}]{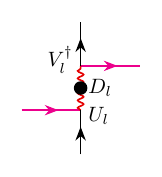}
\;=\;
\adjincludegraphics[valign=c, scale=1, trim={10, 10, 10, 10}]{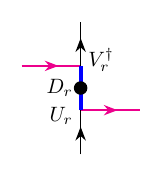} \;.
\end{equation}
Here, the bond dimensions of the red wiggly and blue thick lines are given by $\lrank(\mathcal{O})$ and $\rrank(\mathcal{O})$, respectively.
The two-leg tensors $D_i$ for $i = l, r$ are diagonal matrices with positive diagonal entries, and (the Hermitian conjugates of) the three-leg tensors $V_i$ and $U_i$ are isometries, which satisfy
\begin{equation}
\adjincludegraphics[valign=c, scale=1, trim={10, 10, 10, 10}]{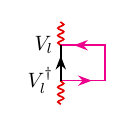}
\;=\;
\adjincludegraphics[valign=c, scale=1, trim={10, 10, 10, 10}]{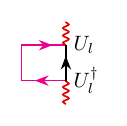}
\;=\;
\adjincludegraphics[valign=c, scale=1, trim={10, 10, 10, 10}]{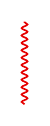}
\;, \qquad
\adjincludegraphics[valign=c, scale=1, trim={10, 10, 10, 10}]{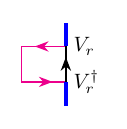}
\;=\;
\adjincludegraphics[valign=c, scale=1, trim={10, 10, 10, 10}]{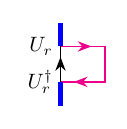}
\;=\;
\adjincludegraphics[valign=c, scale=1, trim={10, 10, 10, 10}]{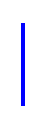} \;.
\end{equation}
Using the above decomposition, we can also write the four-leg tensor $\mathcal{O}$ as
\begin{equation}
\adjincludegraphics[valign=c, scale=1, trim={10, 10, 10, 10}]{tikz/out/O.pdf}
\;=\;
\adjincludegraphics[valign=c, scale=1, trim={10, 10, 10, 10}]{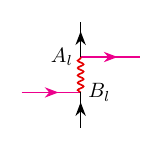}
\;=\;
\adjincludegraphics[valign=c, scale=1, trim={10, 10, 10, 10}]{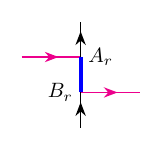} \;,
\label{eq: AB decomposition}
\end{equation}
where $A_l$, $B_l$, $A_r$, and $B_r$ are defined by
\begin{align}
A_l & \coloneq \;
\adjincludegraphics[valign=c, scale=1, trim={10, 10, 10, 10}]{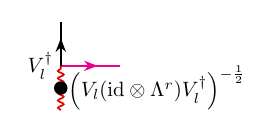} \;, &
B_l &\coloneq
\adjincludegraphics[valign=c, scale=1, trim={10, 10, 10, 10}]{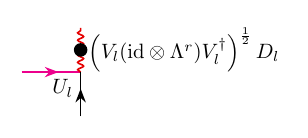} \;, \label{eq: Al Bl}
\\
A_r & \coloneq \;
\adjincludegraphics[valign=c, scale=1, trim={10, 10, 10, 10}]{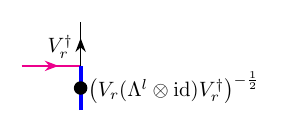} \;, &
B_r & \coloneq
\adjincludegraphics[valign=c, scale=1, trim={10, 10, 10, 10}]{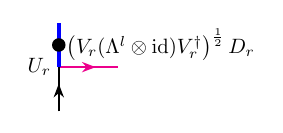} \;. \label{eq: Ar Br}
\end{align}
We note that the two-leg tensors $\left( V_l (\id \otimes \Lambda^r)V_l^{\dagger} \right)^{\pm 1/2}$ and $\left( V_r (\Lambda^l \otimes \id) V_r^{\dagger} \right)^{\pm 1/2}$ make sense because $\Lambda^l$ and $\Lambda^r$ are positive definite.
By construction, $A_l$ and $A_r$ satisfy
\begin{equation}
\adjincludegraphics[valign=c, scale=1, trim={10, 10, 10, 10}]{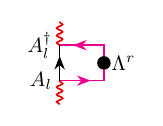}
\;=\; 
\adjincludegraphics[valign=c, scale=1, trim={10, 10, 10, 10}]{tikz/out/isometry3.pdf} \;, \qquad
\adjincludegraphics[valign=c, scale=1, trim={10, 10, 10, 10}]{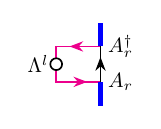}
\;=\; 
\adjincludegraphics[valign=c, scale=1, trim={10, 10, 10, 10}]{tikz/out/isometry6.pdf} \;.
\label{eq: A Lambda A}
\end{equation}
We remark that the same decomposition was used in \cite{Cirac:2017vke} to derive the standard form of MPUs.
We also note that the above decomposition is always possible for any MPO tensor $\mathcal{O}$ and any positive definite linear maps $\Lambda^l$ and $\Lambda^r$.
In what follows, we will be interested in the case where $\mathcal{O}$ is a topological injective MPO tensor and $\Lambda^l$ and $\Lambda^r$ are the fixed points of the transfer matrix $\mathsf{T}[\mathcal{O}]$.

\vspace*{\baselineskip}
\noindent{\bf Sequential circuit representation.}
Using the decomposition~\eqref{eq: AB decomposition}, we can express the topological injective MPO $\mathsf{D}[\mathcal{O}]$ on a periodic chain as
\begin{equation}
\mathsf{D}[\mathcal{O}] =\;
\adjincludegraphics[valign=c, scale=1, trim={10, 10, 10, 10}]{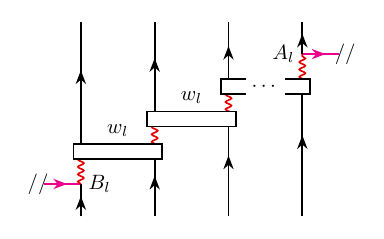}
\;=\;
\adjincludegraphics[valign=c, scale=1, trim={10, 10, 10, 10}]{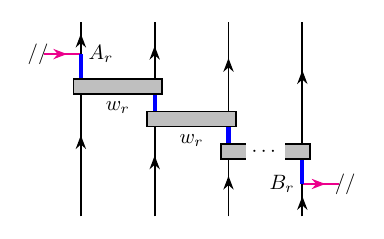} \;,
\label{eq: sequential circuit rep}
\end{equation}
where the local operators $w_l$ and $w_r$ are defined by
\begin{equation}
w_l \coloneq \; \adjincludegraphics[valign=c, scale=1, trim={10, 10, 10, 10}]{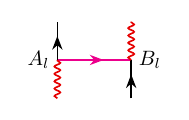} \;, \qquad
w_r \coloneq \; \adjincludegraphics[valign=c, scale=1, trim={10, 10, 10, 10}]{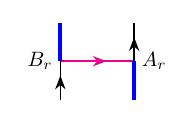} \;.
\label{eq: wl wr}
\end{equation}
Let us show that $w_l$ and $w_r$ are unitary when $\mathcal{O}$ is topological.
To show the unitarity of $w_l$, we express the second equation in \eqref{eq: topological injective MPO} using $A_l$ and $B_l$ as follows:
\begin{equation}
\adjincludegraphics[valign=c, scale=1, trim={10, 10, 10, 10}]{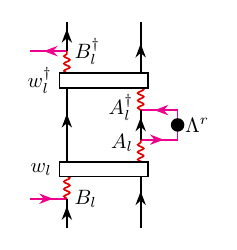} \;=\;
\adjincludegraphics[valign=c, scale=1, trim={10, 10, 10, 10}]{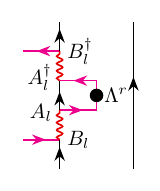} \;.
\label{eq: AB topological}
\end{equation}
The small loops involving $\Lambda^r$ on both sides can be removed by using \eqref{eq: A Lambda A}.
Furthermore, we can remove $B_l$ from both sides by multiplying its right inverse
\begin{equation}
\widetilde{B}_l \coloneq U_l^{\dagger} D_l^{-1} \left( V_l (\id \otimes \Lambda^r) V_l^{\dagger} \right)^{-\frac{1}{2}},
\end{equation}
which satisfies $B_l \widetilde{B}_l = \id$.
Similarly, we can also remove $B_l^{\dagger}$ from both sides of \eqref{eq: AB topological} by multiplying its left inverse $\widetilde{B}_l^{\dagger}$.
Thus, after multiplying $\widetilde{B}_l$ from the bottom and $\widetilde{B}^{\dagger}_l$ from the top, equation~\eqref{eq: AB topological} reduces to
\begin{equation}
w_l^{\dagger} w_l = \id.
\end{equation}
Since $w_l$ is a square matrix of finite size, the above equation also implies
\begin{equation}
w_l w_l^{\dagger} = \id,
\label{eq: wl unitary}
\end{equation}
which shows that $w_l$ is unitary.
The unitarity of $w_r$ can also be shown similarly.
The sequence of local unitary operators $w_l$ or $w_r$ in \eqref{eq: sequential circuit rep} is referred to as a sequential quantum circuit \cite{Chen:2023qst}.
The entire MPO of the form \eqref{eq: sequential circuit rep} is called a sequential MPO in \cite{Tantivasadakarn:2025txn}.

We note that the derivation of the sequential circuit representation~\eqref{eq: sequential circuit rep} does not directly use the injectivity of $\mathcal{O}$.
More specifically, even if the MPO tensor $\mathcal{O}$ is not injective, the MPO $\mathsf{D}[\mathcal{O}]$ has a sequential circuit representation as long as there exist positive definite matrices $\Lambda^l$ and $\Lambda^r$ that satisfy \eqref{eq: topological injective MPO}.
When $\mathcal{O}$ is injective, such $\Lambda^l$ and $\Lambda^r$, if they exist, can always be chosen to be the fixed points of the transfer matrix as shown in Appendix~\ref{sec: Choice of Lambda}.

\vspace*{\baselineskip}
\noindent{\bf Movement operators revisited.}
The sequential circuit representation~\eqref{eq: sequential circuit rep} suggests that $w_l$ and $w_r$ can be regarded as movement operators.
Indeed, due to the decomposition~\eqref{eq: AB decomposition}, the movement operators $u_l$ and $u_r$ defined by~\eqref{eq: movement ops} can be expressed in terms of $w_l$ and $w_r$ as follows:
\begin{equation}
u_l = \adjincludegraphics[valign=c, scale=1, trim={10, 10, 10, 10}]{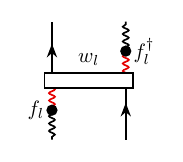} \;, \qquad
u_r = \adjincludegraphics[valign=c, scale=1, trim={10, 10, 10, 10}]{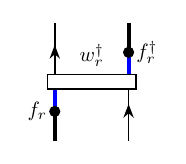} \;.
\label{eq: ul=wl}
\end{equation}
Here, the linear maps $f_l$ and $f_r$ are defined by
\begin{equation}
f_l \coloneq \frac{1}{\sqrt{\delta_l(\mathcal{O})}} \adjincludegraphics[valign=c, scale=1, trim={10, 10, 10, 10}]{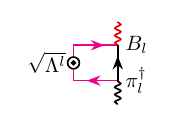}, \qquad
f_r \coloneq \frac{1}{\sqrt{\delta_r(\mathcal{O})}} \adjincludegraphics[valign=c, scale=1, trim={10, 10, 10, 10}]{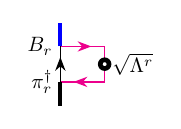}.
\end{equation}
We note that $f_l$ and $f_r$ are unitary because
\begin{equation}
f_l^{\dagger} f_l = \adjincludegraphics[valign=c, scale=1, trim={10, 10, 10, 10}]{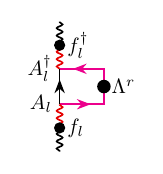} = \pi_l P_l \pi_l^{\dagger} = \id_L, \qquad
f_r^{\dagger} f_r = \adjincludegraphics[valign=c, scale=1, trim={10, 10, 10, 10}]{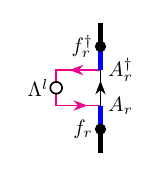} = \pi_rP_r\pi_r^{\dagger} = \id_R,
\end{equation}
where we used \eqref{eq: A Lambda A} and \eqref{eq: pi identities}.
Since $f_l$ and $f_r$ are linear maps between finite dimensional vector spaces of the same dimension, the above equation guarantees that $f_l f_l^{\dagger}$ and $f_r f_r^{\dagger}$ are also the identity maps, which implies that $f_l$ and $f_r$ are unitary.
Thus, equation~\eqref{eq: ul=wl} shows that $w_l$ and $w_r$ agree with the movement operators $u_l$ and $u_r^{\dagger}$ up to on-site unitary acting on the physical Hilbert space at the defect locus.
In other words, $w_l$ and $w_r$ can be regarded as movement operators on the defect Hilbert spaces isomorphic to \eqref{eq: defect Hilbert spaces def}.
More specifically, $w_l$ and $w_r$ are the movement operators on
\begin{equation}
\widetilde{\mathcal{H}}_{\mathsf{D}[\mathcal{O}]}^l \coloneq \widetilde{L} \otimes \mathcal{H}_{\mathrm{o}}^{\otimes N-1}, \qquad
\widetilde{\mathcal{H}}_{\mathsf{D}[\mathcal{O}]}^r \coloneq \widetilde{R} \otimes \mathcal{H}_{\mathrm{o}}^{\otimes N-1},
\label{eq: defect Hilbert spaces def2}
\end{equation}
where $\widetilde{L}$ and $\widetilde{R}$ are the $\lrank(\mathcal{O})$-dimensional and $\rrank(\mathcal{O})$-dimensional vector spaces associated with the red wiggly line and the blue thick line, respectively.
Equation~\eqref{eq: ul=wl} then shows that the movement operators on the isomorphic defect Hilbert spaces~\eqref{eq: defect Hilbert spaces def} and \eqref{eq: defect Hilbert spaces def2} are intertwined by unitary operators $f_l$ and $f_r$.
We mention that the movement operators $w_l$ and $w_r$ for invertible symmetries are studied in \cite{Franco-Rubio:2025qss}.

\vspace*{\baselineskip}
\noindent{\bf Defect creation and annihilation operators.}
The sequential circuit representation~\eqref{eq: sequential circuit rep} was obtained by using the same decomposition of $\mathcal{O}$ on all physical sites.
We can obtain other variants of the sequential circuit representation by using different decompositions of $\mathcal{O}$ depending on the sites.
For example, if we use a different decomposition of $\mathcal{O}$ only on the first site, we obtain
\begin{equation}
\mathsf{D}[\mathcal{O}] =
\adjincludegraphics[valign=c, scale=0.95, trim={10, 10, 10, 10}]{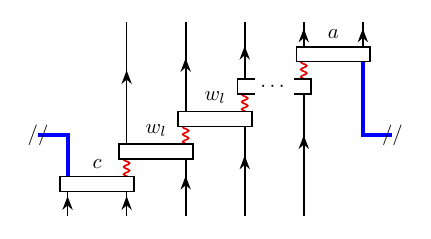} \;=\;
\adjincludegraphics[valign=c, scale=0.95, trim={10, 10, 10, 10}]{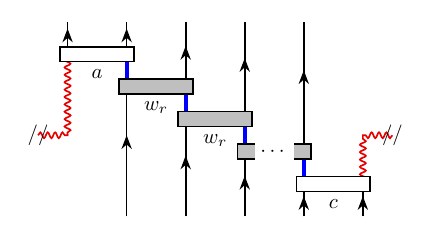} \;.
\label{eq: another sequential rep}
\end{equation}
Here, the local operators $c$ and $a$ are defined by
\begin{equation}
c \coloneq \adjincludegraphics[valign=c, scale=1, trim={10, 10, 10, 10}]{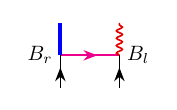} \;, \qquad
a \coloneq \adjincludegraphics[valign=c, scale=1, trim={10, 10, 10, 10}]{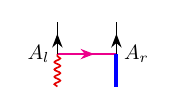} \;.
\label{eq: creation and annihilation}
\end{equation}
Physically, $c$ and $a$ are regarded as operators that create and annihilate a pair of defects.
We refer to these operators as defect creation and annihilation operators.
These operators were originally introduced in \cite{Cirac:2017vke} for MPUs.
We note that the defect creation and annihilation operators are not unitary when the topological injective MPO $\mathsf{D}[\mathcal{O}]$ is non-invertible.
Indeed, as is clear from \eqref{eq: another sequential rep}, these operators can be unitary only when $\mathsf{D}[\mathcal{O}]$ is unitary.
Conversely, when $\mathsf{D}[\mathcal{O}]$ is unitary, these operators are indeed unitary as long as $\Lambda^l$ and $\Lambda^r$ are normalized properly \cite{Cirac:2017vke}.
In general, $c$ and $a^{\dagger}$ are isometries up to scalar, i.e., they satisfy
\begin{align}
c^{\dagger} c &= \adjincludegraphics[valign=c, scale=1, trim={10, 10, 10, 10}]{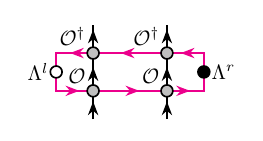} = \adjincludegraphics[valign=c, scale=1, trim={10, 10, 10, 10}]{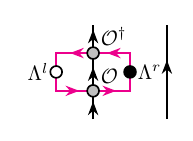} = \gamma \; \id_{\mathcal{H}_{\mathrm{o}} \otimes \mathcal{H}_{\mathrm{o}}} \label{eq: c dagger c}, \\
aa^{\dagger} & = \frac{1}{\delta_l(\mathcal{O})\delta_r(\mathcal{O})} \adjincludegraphics[valign=c, scale=1, trim={10, 10, 10, 10}]{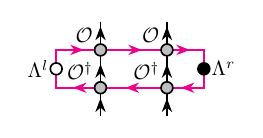} = \frac{\ldim(\mathcal{O}) \rdim(\mathcal{O})}{\gamma} \; \id_{\mathcal{H}_{\mathrm{o}} \otimes \mathcal{H}_{\mathrm{o}}}. \label{eq: a a dagger}
\end{align}
Here, the first equality of \eqref{eq: c dagger c} follows from \eqref{eq: AB decomposition} and \eqref{eq: A Lambda A}.
Similarly, the first equality of \eqref{eq: a a dagger} follows from \eqref{eq: AB decomposition} and \eqref{eq: B Lambda B}.
We note that the product of $c^{\dagger}c$ and $aa^{\dagger}$ is an invariant scalar $\ldim(\mathcal{O}) \rdim(\mathcal{O}) = \qdim_{\mathrm{lat}}(\mathcal{O})^2$.
In particular, one cannot rescale $c^{\dagger}c$ and $aa^{\dagger}$ to one simultaneously by changing the normalization of $\Lambda^l$ and $\Lambda^r$.
Using the defect creation and annihilation operators, we can also write $\mathsf{D}[\mathcal{O}]$ as
\begin{equation}
\mathsf{D}[\mathcal{O}] = \;
\adjincludegraphics[valign=c, scale=1, trim={10, 10, 10, 10}]{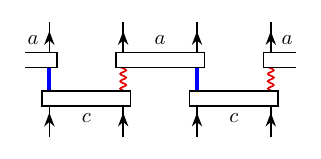} \;.
\end{equation}

\subsection{Properties of dimensions}
\label{sec: Properties}
In this subsection, we show several basic properties of the left and right dimensions~\eqref{eq: ldim rdim TN} and the lattice quantum dimension~\eqref{eq: qdim ind TN} of a topological injective MPO tensor $\mathcal{O}$.

\vspace*{\baselineskip}
\noindent{\bf Invariance under blocking.}
We first show that the left and right dimensions are invariant under blocking a finite number of physical sites.
To this end, we consider the local tensor $\mathcal{O}^{[2]}$ obtained by blocking two physical sites as follows:
\begin{equation}
\mathcal{O}^{[2]} = \;
\adjincludegraphics[valign=c, scale=1, trim={10, 10, 10, 10}]{tikz/out/O_blocked1.pdf} \;.
\end{equation}
Since $\mathcal{O}$ is topological and injective,  $\mathcal{O}^{[2]}$ is also topological and injective.
By definition, the left dimension of $\mathcal{O}^{[2]}$ is given by
\begin{equation}
\ldim(\mathcal{O}^{[2]}) = \frac{\lrank(\mathcal{O}^{[2]})}{\dim(\mathcal{H}_{\mathrm{o}})^2}.
\label{eq: ldim blocking}
\end{equation}
Following the proof of \cite[Theorem 3]{Sahinoglu:2017iny}, one can compute the left rank of $\mathcal{O}^{[2]}$ as
\begin{equation}
\lrank(\mathcal{O}^{[2]}) = 
\rank\left( \adjincludegraphics[valign=c, scale=0.95, trim={10, 10, 10, 10}]{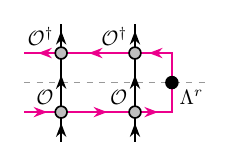} \right)
= \rank\left( \adjincludegraphics[valign=c, scale=0.95, trim={10, 10, 10, 10}]{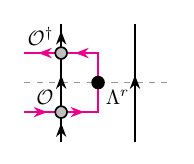} \right)
= \lrank(\mathcal{O}) \dim(\mathcal{H}_{\mathrm{o}}).
\label{eq: rank O blocked}
\end{equation}
In the first and third equalities, we used the fact that $\Lambda^r$ is positive definite and $\rank(f) = \rank(f^{\dagger}f)$ for an arbitrary linear map $f$.
The second equality follows from~\eqref{eq: topological injective MPO}.
Plugging the above equation into \eqref{eq: ldim blocking}, we find
\begin{equation}
\ldim(\mathcal{O}^{[2]}) = \frac{\lrank(\mathcal{O})}{\dim(\mathcal{H}_{\mathrm{o}})} = \ldim(\mathcal{O}).
\label{eq: ldim O2 = ldim O}
\end{equation}
The same computation shows that $\ldim(\mathcal{O}^{[n+1]}) = \ldim(\mathcal{O}^{[n]})$ for any positive integer $n$, where $\mathcal{O}^{[n]}$ is the local tensor obtained by blocking $n$ physical sites.
Thus, combining this with \eqref{eq: ldim O2 = ldim O}, we find
\begin{equation}
\ldim(\mathcal{O}^{[n]}) = \ldim(\mathcal{O}).
\end{equation}
This shows that the left dimension is invariant under blocking any number of physical sites.
A similar computation shows that the right dimension is also invariant under blocking.
We note that the lattice quantum dimension and the index defined by \eqref{eq: qdim ind TN} are also invariant under blocking because they consist only of the left and right dimensions.

\vspace*{\baselineskip}
\noindent{\bf Upper and lower bounds of the lattice quantum dimension.}
Next, we show that the lattice quantum dimension defined in \eqref{eq: qdim ind TN} satisfies
\begin{equation}
1 \leq \qdim_{\text{lat}}(\mathcal{O}) \leq \dim(V),
\label{eq: bounds of qdim}
\end{equation}
where $V$ is the bond Hilbert space of $\mathcal{O}$.
We note that the upper bound in \eqref{eq: bounds of qdim} is invariant under the gauge transformation of the MPO tensor $\mathcal{O}$.
Thus, it depends only on the operator $\mathsf{D}[\mathcal{O}]$ due to the fundamental theorem of injective MPS/MPO \cite{Perez-Garcia:2006nqo, Cirac:2016iqe}.
The upper bound in \eqref{eq: bounds of qdim} immediately follows from the fact that the left and right dimensions of $\mathcal{O}$ are less than or equal to $\dim(V)$ by definition~\eqref{eq: ldim rdim TN}.
On the other hand, the lower bound in \eqref{eq: bounds of qdim} can be obtained by following the proof of \cite[Lemma III.7]{Cirac:2017vke}.
More specifically, to derive the lower bound of $\qdim_{\text{lat}}(\mathcal{O})$, we recall that the defect creation operator $c$ satisfies \eqref{eq: c dagger c}, i.e.,
\begin{equation}
c^{\dagger} c = \gamma \; \id_{\mathcal{H}_{\mathrm{o}} \otimes \mathcal{H}_{\mathrm{o}}}.
\end{equation}
Since $\gamma$ is non-zero,  the above equation implies that $\rank(c) = \rank(c^{\dagger}c) = \dim(\mathcal{H}_{\mathrm{o}})^2$.
On the other hand, since the codomain of $c$ has dimension $\lrank(\mathcal{O}) \rrank(\mathcal{O})$, the rank of $c$ must be less than or equal to $\lrank(\mathcal{O}) \rrank(\mathcal{O})$.
Therefore, it follows that
\begin{equation}
\qdim_{\text{lat}}(\mathcal{O}) = \frac{\lrank(\mathcal{O}) \rrank(\mathcal{O})}{\dim(\mathcal{H}_{\mathrm{o}})^2} \geq 1.
\end{equation}
We note that $\qdim_{\text{lat}}(\mathcal{O})=1$ if and only if the MPO $\mathsf{D}[\mathcal{O}]$ is unitary \cite[Theorem III.8]{Cirac:2017vke}.

\vspace*{\baselineskip}
\noindent{\bf Multiplicativity.}
The left and right dimensions are multiplicative under the composition of topological injective MPO tensors.
More specifically, after blocking a sufficient number of physical sites, we have
\begin{equation}
\ldim(\mathcal{O}_X \mathcal{O}_Y) = \ldim(\mathcal{O}_X) \ldim(\mathcal{O}_Y), \qquad
\rdim(\mathcal{O}_X \mathcal{O}_Y) = \rdim(\mathcal{O}_X) \rdim(\mathcal{O}_Y),
\label{eq: ldim rdim multiplicative TN}
\end{equation}
where $\mathcal{O}_X$ and $\mathcal{O}_Y$ are topological injective MPO tensors, and $\mathcal{O}_X \mathcal{O}_Y$ is the composite tensor defined by
\begin{equation}
\mathcal{O}_X \mathcal{O}_Y \coloneq 
\adjincludegraphics[valign=c, scale=1, trim={10, 10, 10, 10}]{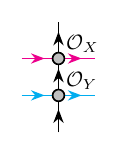} \;.
\label{eq: OX OY}
\end{equation}
In~\eqref{eq: ldim rdim multiplicative TN}, the left and right dimensions of a generally non-injective MPO tensor $\mathcal{O}_X \mathcal{O}_Y$ are defined in the same way as~\eqref{eq: ldim rdim TN}, that is, $\ldim(\mathcal{O}_X \mathcal{O}_Y) \coloneq \lrank(\mathcal{O}_X \mathcal{O}_Y) / \dim(\mathcal{H}_{\mathrm{o}})$ and $\rdim(\mathcal{O}_X \mathcal{O}_Y) \coloneq \rrank(\mathcal{O}_X \mathcal{O}_Y) / \dim(\mathcal{H}_{\mathrm{o}})$.
Following the proof of \cite[Theorem 3]{Sahinoglu:2017iny}, we can show the multiplicativity of the left dimension by computing the left rank of $\mathcal{O}_X^{[2]} \mathcal{O}_Y^{[2]}$ as follows: 
\begin{equation*}
\begin{aligned}
&\quad \lrank(\mathcal{O}_X^{[2]} \mathcal{O}_Y^{[2]})
= \rank\left( \adjincludegraphics[valign=c, scale=0.9, trim={10, 10, 10, 10}]{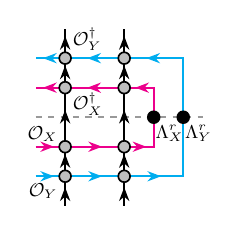} \right)
= \rank\left( \adjincludegraphics[valign=c, scale=0.9, trim={10, 10, 10, 10}]{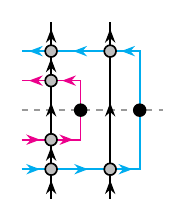} \right)
= \rank\left( \adjincludegraphics[valign=c, scale=0.9, trim={10, 10, 10, 10}]{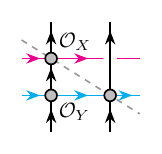} \right)
\\
&= \rank\left( \adjincludegraphics[valign=c, scale=0.9, trim={10, 10, 10, 10}]{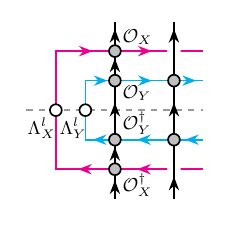} \right)
= \rank\left( \adjincludegraphics[valign=c, scale=0.9, trim={10, 10, 10, 10}]{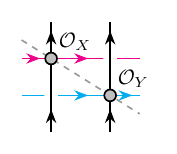} \right)
= \lrank(\mathcal{O}_X) \lrank(\mathcal{O}_Y).
\end{aligned}
\end{equation*}
By dividing both sides by $\dim(\mathcal{H}_{\mathrm{o}})^2$, we find
\begin{equation}
\ldim(\mathcal{O}_X^{[2]} \mathcal{O}_Y^{[2]}) = \ldim(\mathcal{O}_X) \ldim(\mathcal{O}_Y) = \ldim(\mathcal{O}_X^{[2]}) \ldim(\mathcal{O}_Y^{[2]}),
\end{equation}
where the last equality follows from the fact that $\ldim(\mathcal{O}_X)$ and $\ldim(\mathcal{O}_Y)$ are invariant under blocking.
Thus, by blocking two physical sites in the above equation, we obtain the first equality of \eqref{eq: ldim rdim multiplicative TN}.\footnote{More precisely, in \eqref{eq: ldim rdim multiplicative TN}, we have redefined $\mathcal{O}_X$ and $\mathcal{O}_Y$ by $\mathcal{O}_X^{[2]}$ and $\mathcal{O}_Y^{[2]}$, respectively.}
By a similar computation, we can also show the multiplicativity of the right dimension.
The multiplicativity~\eqref{eq: ldim rdim multiplicative TN} implies that the lattice quantum dimension and the index are also multiplicative in the sense that
\begin{equation}
\qdim_{\mathrm{lat}}(\mathcal{O}_X \mathcal{O}_Y) = \qdim_{\mathrm{lat}}(\mathcal{O}_X) \qdim_{\mathrm{lat}}(\mathcal{O}_Y), \qquad
\ind(\mathcal{O}_X \mathcal{O}_Y) = \ind(\mathcal{O}_X) \ind(\mathcal{O}_Y).
\end{equation}

\vspace*{\baselineskip}
\noindent{\bf Additivity.}
It immediately follows from the definition~\eqref{eq: ldim rdim TN} that the left and right dimensions are additive under the direct sum of MPO tensors, i.e., 
\begin{equation}
\ldim(\mathcal{O}_X \oplus \mathcal{O}_Y) = \ldim(\mathcal{O}_X) + \ldim(\mathcal{O}_Y), \quad
\rdim(\mathcal{O}_X \oplus \mathcal{O}_Y) = \rdim(\mathcal{O}_X) + \rdim(\mathcal{O}_Y).
\label{eq: ldim rdim additive TN}
\end{equation}
Here, $\mathcal{O}_X \oplus \mathcal{O}_Y$ is the block injective MPO tensor whose injective blocks are given by $\mathcal{O}_X$ and $\mathcal{O}_Y$.
In other words, $\mathcal{O}_X \oplus \mathcal{O}_Y$ is the direct sum of the linear maps $\mathcal{O}_X: V_X \otimes \mathcal{H}_{\mathrm{o}} \to \mathcal{H}_{\mathrm{o}} \otimes V_X$ and $\mathcal{O}_Y: V_Y \otimes \mathcal{H}_{\mathrm{o}} \to \mathcal{H}_{\mathrm{o}} \otimes V_Y$, where $V_X$ and $V_Y$ are the bond Hilbert spaces of $\mathcal{O}_X$ and $\mathcal{O}_Y$.
In \eqref{eq: ldim rdim additive TN}, the left and right dimensions of a non-injective MPO tensor $\mathcal{O}_X \oplus \mathcal{O}_Y$ are defined in the same way as \eqref{eq: ldim rdim TN}, that is, $\ldim(\mathcal{O}_X \oplus \mathcal{O}_Y) \coloneq \lrank(\mathcal{O}_X \oplus \mathcal{O}_Y) / \dim(\mathcal{H}_{\mathrm{o}})$ and $\rdim(\mathcal{O}_X \oplus \mathcal{O}_Y) \coloneq \rrank(\mathcal{O}_X \oplus \mathcal{O}_Y) / \dim(\mathcal{H}_{\mathrm{o}})$.
The additivity~\eqref{eq: ldim rdim additive TN} does not imply that the lattice quantum dimension and the index are additive.

\vspace*{\baselineskip}
\noindent{\bf Remark.}
The multiplicativity~\eqref{eq: ldim rdim multiplicative TN} and the additivity~\eqref{eq: ldim rdim additive TN} do not necessarily imply that the left and right dimensions are one-dimensional representations of the fusion rules.
This is because the rank of the non-injective MPO tensor $\mathcal{O}_X \mathcal{O}_Y$ may not be equal to the sum of the ranks of its injective blocks due to the off-diagonal components of the tensor.
Nevertheless, under the conditions spelled out in Section~\ref{sec: Homogeneity of index TN}, the left and right dimensions of topological injective MPO tensors become one-dimensional representations of the fusion rules; see Section~\ref{sec: The left and right dimensions obey the fusion rules} for more details.

\section{Sufficient conditions for the homogeneity of the index}
\label{sec: Homogeneity of index TN}
In this section, we discuss sufficient conditions for the homogeneity of the index of topological injective MPOs.
To this end, we first define fusion and splitting tensors for the fusion channels of topological injective MPOs and introduce two conditions on them, which we refer to as the broken zipper condition and the two-sided zipper condition.
Assuming that these conditions are satisfied, we will show that the fusion channels of topological injective MPOs are topological, the left and right dimensions are one-dimensional representations of the fusion rules, and the index is homogeneous.
As a result, as long as these conditions hold, the fusion category symmetry $\mathcal{C}$ realized by topological injective MPOs must be weakly integral even if we allow mixing with non-trivial QCAs.
The broken zipper condition and the two-sided zipper condition for general topological injective MPOs are not proven.\footnote{Nevertheless, all examples of topological injective MPOs discussed in Section~\ref{sec: Examples} have fusion and splitting tensors that satisfy both the broken zipper condition and the two-sided zipper condition.}

\subsection{Fusion and splitting tensors}
\label{sec: Fusion and splitting tensors}
Let us first define fusion tensors and splitting tensors for the fusion channels of topological injective MPOs.
\begin{definition}[Fusion and splitting tensors]\label{def: Fusion and splitting tensors}
Let $\mathcal{O}_X$ and $\mathcal{O}_Y$ be topological injective MPO tensors.
We suppose that MPOs $\mathsf{D}[\mathcal{O}_X]$ and $\mathsf{D}[\mathcal{O}_Y]$ on a periodic chain obey the fusion rule
\begin{equation}
\mathsf{D}[\mathcal{O}_X] \mathsf{D}[\mathcal{O}_Y] = \sum_{Z \in S_{XY}} \mathsf{D}[\mathcal{O}_Z],
\label{eq: fusion rules MPO}
\end{equation}
where $\mathcal{O}_Z$ is an injective MPO tensor for every $Z \in S_{XY}$.
We denote the bond Hilbert spaces of $\mathcal{O}_X$, $\mathcal{O}_Y$, and $\mathcal{O}_Z$ by $V_X$, $V_Y$, and $V_Z$, respectively.
The fusion tensor $\phi_Z: V_X \otimes V_Y \to V_Z$ and the splitting tensor $\overline{\phi}_Z: V_Z \to V_X \otimes V_Y$ for fusion channel $Z$ are three-leg tensors that satisfy
\begin{align}
\adjincludegraphics[valign=c, scale=1, trim={10, 10, 10, 10}]{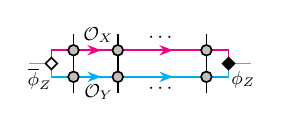} &\;=\; \adjincludegraphics[valign=c, scale=1, trim={10, 10, 10, 10}]{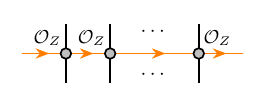}\;,
\label{eq: fusion and splitting 1}
\\
\adjincludegraphics[valign=c, scale=1, trim={10, 10, 10, 10}]{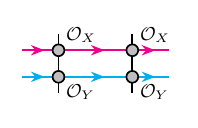}  &\;=\; \sum_{Z \in S_{XY}} \adjincludegraphics[valign=c, scale=1, trim={10, 10, 10, 10}]{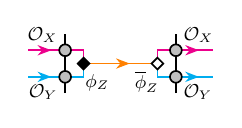}\;,
\label{eq: fusion and splitting 2}
\end{align}
where the first equality is required to be satisfied for any number of physical legs.
\end{definition}

As we will see in Section~\ref{sec: Invertible symmetries}, the fusion and splitting tensors exist when both $\mathsf{D}[\mathcal{O}_X]$ and $\mathsf{D}[\mathcal{O}_Y]$ are MPUs.
For more general topological injective MPOs, the existence of the fusion and splitting tensors has not been proven except for the concrete examples discussed in Section~\ref{sec: Examples}.
In what follows, we assume that these tensors exist for general topological injective MPOs.

\vspace*{\baselineskip}
\noindent{\bf Remark.}
As long as we block a sufficient number of physical legs, there always exist three-leg tensors satisfying \eqref{eq: fusion and splitting 1} according to the general theory of MPS/MPO \cite{Perez-Garcia:2006nqo, Bultinck:2015bot}.
For example, the pair of the projection and inclusion map for each injective block satisfies \eqref{eq: fusion and splitting 1}.
However, it is non-trivial whether such tensors also satisfy \eqref{eq: fusion and splitting 2}.
We note that the condition~\eqref{eq: fusion and splitting 2} was not imposed in the original definition of fusion tensors in \cite{Bultinck:2015bot}.
We also mention that the fusion and splitting tensors defined above are more general than those discussed in, e.g., \cite{Lootens:2020mso, Garre-Rubio:2022uum, Inamura:2024jke}.
More specifically, the fusion and splitting tensors in \cite{Lootens:2020mso, Garre-Rubio:2022uum, Inamura:2024jke} are supposed to satisfy the zipper condition in Definition~\ref{def: Zipper condition}, which does not hold in general as we will see later.

\subsection{Broken zipper condition and two-sided zipper condition}
\label{sec: Broken zipper condition and two-sided zipper condition}
Let us introduce the broken zipper condition and the two-sided zipper condition for the fusion and splitting tensors.

\begin{definition}[Broken zipper condition]\label{def: Broken zipper condition}
Let $\mathcal{O}_X$ and $\mathcal{O}_Y$ be topological injective MPO tensors whose fusion rule is given by \eqref{eq: fusion rules MPO}.
We say that the fusion tensor $\phi_Z$ and the splitting tensor $\overline{\phi}_Z$ for fusion channel $Z \in S_{XY}$ satisfy the broken zipper condition if the following equations hold:
\begin{equation}
\adjincludegraphics[valign=c, scale=1, trim={10, 10, 10, 10}]{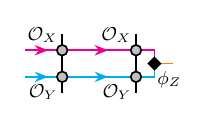} \;=\; \adjincludegraphics[valign=c, scale=1, trim={10, 10, 10, 10}]{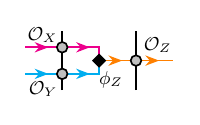}\;, \qquad
\adjincludegraphics[valign=c, scale=1, trim={10, 10, 10, 10}]{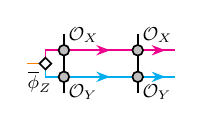} \;=\; \adjincludegraphics[valign=c, scale=1, trim={10, 10, 10, 10}]{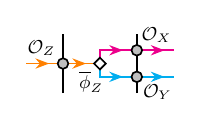}\;.
\label{eq: broken zipper}
\end{equation}
\end{definition}

We note that the above condition is called a weaker zipping condition in \cite{Lu:2025rwd}.
The term ``broken zipper condition" was introduced in \cite{Ohyama:2024ytt} in the case where the fusion channel is unique.

\begin{definition}[Two-sided zipper condition]\label{def: Two-sided zipper condition}
Let $\mathcal{O}_X$ and $\mathcal{O}_Y$ be topological injective MPO tensors whose fusion rule is given by \eqref{eq: fusion rules MPO}.
We say that the fusion tensor $\phi_Z$ and the splitting tensor $\overline{\phi}_Z$ for fusion channel $Z \in S_{XY}$ satisfy the two-sided zipper condition if there exist non-zero complex numbers $a_Z^{l}$, $a_Z^r$, $b_Z^{l}$, and $b_Z^r$ such that
\begin{alignat}{2}
&\adjincludegraphics[valign=c, scale=1, trim={10, 10, 10, 10}]{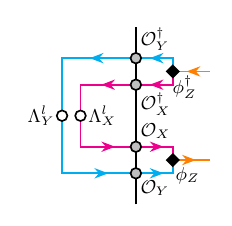} \;= a_Z^{l} \;\; \adjincludegraphics[valign=c, scale=1, trim={10, 10, 10, 10}]{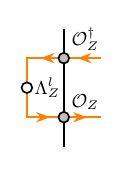}\;,
\qquad
&&\adjincludegraphics[valign=c, scale=1, trim={10, 10, 10, 10}]{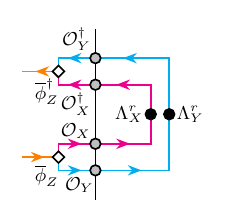} \;= a_Z^r \;\; \adjincludegraphics[valign=c, scale=1, trim={10, 10, 10, 10}]{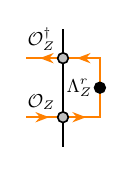}\;,
\label{eq: two-sided zipper 1}
\\
&\adjincludegraphics[valign=c, scale=1, trim={10, 10, 10, 10}]{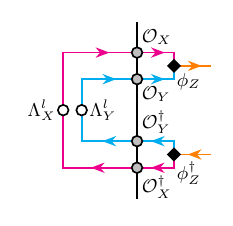} \;= b_Z^{l} \;\; \adjincludegraphics[valign=c, scale=1, trim={10, 10, 10, 10}]{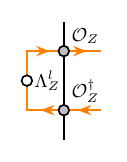}\;,
\qquad
&&\adjincludegraphics[valign=c, scale=1, trim={10, 10, 10, 10}]{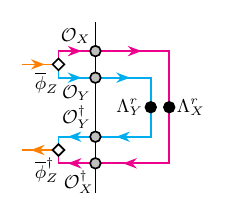} \;= b_Z^r \;\; \adjincludegraphics[valign=c, scale=1, trim={10, 10, 10, 10}]{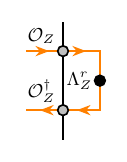} \;,
\label{eq: two-sided zipper 2}
\end{alignat}
where $\Lambda_X^{l/r}$, $\Lambda_Y^{l/r}$, and $\Lambda_Z^{l/r}$ are the left/right fixed points of the transfer matrices $\mathsf{T}[\mathcal{O}_X]$, $\mathsf{T}[\mathcal{O}_Y]$, and $\mathsf{T}[\mathcal{O}_Z]$.
\end{definition}

As we will see in Section~\ref{sec: Invertible symmetries}, the fusion and splitting tensors satisfy both the broken zipper condition and the two-sided zipper condition when $\mathsf{D}[\mathcal{O}_X]$ and $\mathsf{D}[\mathcal{O}_Y]$ are MPUs.
For more general topological injective MPOs, these conditions have not been proven except for the concrete examples discussed in Section~\ref{sec: Examples}.

Before proceeding, we mention that the broken zipper condition defined above is weaker than the following zipper condition \cite{Bultinck:2015bot}:

\begin{definition}[Zipper condition]\label{def: Zipper condition}
Let $\mathcal{O}_X$ and $\mathcal{O}_Y$ be topological injective MPO tensors whose fusion rule is given by \eqref{eq: fusion rules MPO}.
We say that the fusion tensor $\phi_Z$ and the splitting tensor $\overline{\phi}_Z$ satisfy the zipper condition if the following equations hold:
\begin{equation}
\adjincludegraphics[valign=c, scale=1, trim={10, 10, 10, 10}]{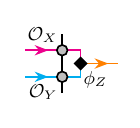} \;=\; \adjincludegraphics[valign=c, scale=1, trim={10, 10, 10, 10}]{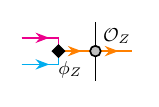}\;, \qquad
\adjincludegraphics[valign=c, scale=1, trim={10, 10, 10, 10}]{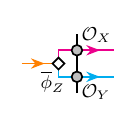} \;=\; \adjincludegraphics[valign=c, scale=1, trim={10, 10, 10, 10}]{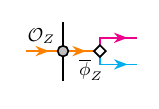}\;.
\label{eq: zipper condition}
\end{equation}
\end{definition}
The zipper condition~\eqref{eq: zipper condition} does not hold in general.
To see this, let us consider the case when $\mathsf{D}[\mathcal{O}_X]$ and $\mathsf{D}[\mathcal{O}_Y]$ are MPUs.
If we choose $\mathcal{O}_X = \mathcal{O}_Y^{\dagger}$, the fusion and splitting tensors for the unique fusion channel (i.e., the identity channel) are given by the normalized right and left fixed points of the transfer matrix $\mathsf{T}[\mathcal{O}_Y]$.\footnote{Indeed, the normalized fixed points of the transfer matrix satisfy the defining equations~\eqref{eq: fusion and splitting 1} and \eqref{eq: fusion and splitting 2} of fusion and splitting tensors due to \eqref{eq: simple MPU}.}
Thus, the zipper condition for this fusion channel reduces to \eqref{eq: detachable MPU}.
However, as discussed in Section~\ref{sec: The case of invertible symmetries}, an MPU tensor satisfying \eqref{eq: detachable MPU} must be an on-site operator.
Therefore, the zipper condition does not hold in general.

\subsection{Consequences}
\label{sec: Consequences}
We now discuss some consequences of the broken zipper condition and the two-sided zipper condition.
In what follows, unless otherwise specified, the black and white dots in tensor network diagrams represent the right and left fixed points of the transfer matrices, respectively.

\subsubsection{Fusion channels are topological}
We first show that the fusion channel $Z \in S_{XY}$ is topological if the fusion tensor $\phi_Z$ and the splitting tensor $\overline{\phi}_Z$ satisfy both the broken zipper condition and the two-sided zipper condition.
More specifically, the broken zipper condition and the two-sided zipper condition imply
\begin{alignat}{2}
&\adjincludegraphics[valign=c, scale=1, trim={10, 10, 10, 10}]{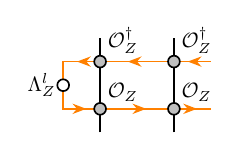} \;=\; \adjincludegraphics[valign=c, scale=1, trim={10, 10, 10, 10}]{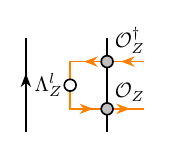}\;, \qquad
&&\adjincludegraphics[valign=c, scale=1, trim={10, 10, 10, 10}]{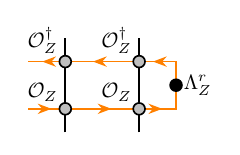} \;=\; \adjincludegraphics[valign=c, scale=1, trim={10, 10, 10, 10}]{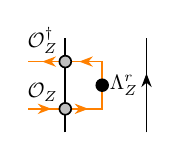}\;,
\label{eq: topological zipper 1}
\\
&\adjincludegraphics[valign=c, scale=1, trim={10, 10, 10, 10}]{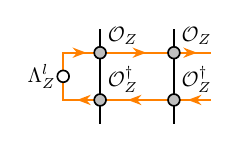} \;=\; \adjincludegraphics[valign=c, scale=1, trim={10, 10, 10, 10}]{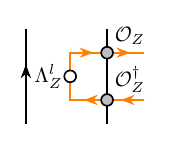}\;, \qquad
&&\adjincludegraphics[valign=c, scale=1, trim={10, 10, 10, 10}]{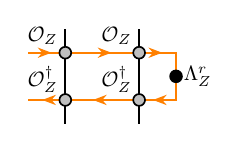} \;=\; \adjincludegraphics[valign=c, scale=1, trim={10, 10, 10, 10}]{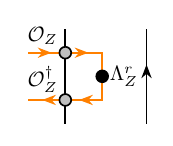}\;,
\label{eq: topological zipper 2}
\end{alignat}
as long as the physical legs are blocked sufficiently,

To show the first equality of \eqref{eq: topological zipper 1}, we note that the broken zipper condition and the two-sided zipper condition lead to the following equality:
\begin{equation}
\adjincludegraphics[valign=c, scale=1, trim={10, 10, 10, 10}]{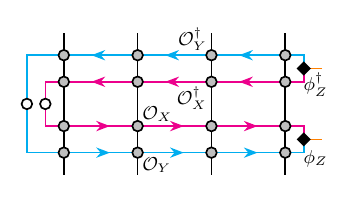} = a_Z^{l} \;\; \adjincludegraphics[valign=c, scale=1, trim={10, 10, 10, 10}]{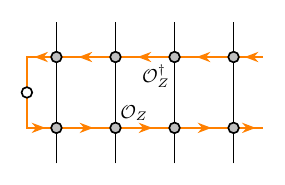} \;.
\end{equation}
The left-hand side of the above equation can also be computed as
\begin{equation}
\adjincludegraphics[valign=c, scale=1, trim={10, 10, 10, 10}]{tikz/out/topological_zipper_der1.pdf} =\; \adjincludegraphics[valign=c, scale=1, trim={10, 10, 10, 10}]{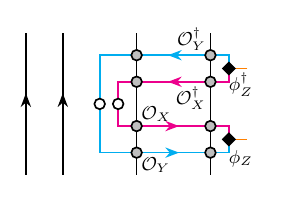} \;= a_Z^{l} \;\; \adjincludegraphics[valign=c, scale=1, trim={10, 10, 10, 10}]{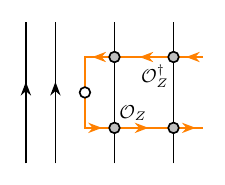} \;,
\end{equation}
where the first equality follows from the topological property~\eqref{eq: topological injective MPO} of $\mathcal{O}_X$ and $\mathcal{O}_Y$.
By comparing the above two equations, we find
\begin{equation}
\adjincludegraphics[valign=c, scale=1, trim={10, 10, 10, 10}]{tikz/out/topological_zipper_der2.pdf}  \;=\; \adjincludegraphics[valign=c, scale=1, trim={10, 10, 10, 10}]{tikz/out/topological_zipper_der4.pdf} \;.
\end{equation}
By blocking two physical legs in the above equation, we obtain the first equality of~\eqref{eq: topological zipper 1}.
The other equalities in~\eqref{eq: topological zipper 1} and \eqref{eq: topological zipper 2} can also be obtained similarly.

\subsubsection{The left and right dimensions obey the fusion rules}
\label{sec: The left and right dimensions obey the fusion rules}
We next show that the left and right dimensions are one-dimensional representations of the fusion rules~\eqref{eq: fusion rules MPO} if the fusion and splitting tensors for all fusion channels satisfy both the broken zipper condition and the two-sided zipper condition.
More specifically, if \eqref{eq: broken zipper}, \eqref{eq: two-sided zipper 1}, and \eqref{eq: two-sided zipper 2} hold for all $Z \in S_{XY}$, it follows that
\begin{align}
\ldim(\mathcal{O}_X) \ldim(\mathcal{O}_Y) &= \sum_{Z \in S_{XY}} \ldim(\mathcal{O}_Z), \label{eq: ldim zipper} \\
\rdim(\mathcal{O}_X) \rdim(\mathcal{O}_Y) &= \sum_{Z \in S_{XY}} \rdim(\mathcal{O}_Z), \label{eq: rdim zipper}
\end{align}
as long as the physical sites are blocked sufficiently.

To show \eqref{eq: ldim zipper}, we first recall that the left dimension is multiplicative~\eqref{eq: ldim rdim multiplicative TN}, i.e., it satisfies $\ldim(\mathcal{O}_X) \ldim(\mathcal{O}_Y) = \ldim(\mathcal{O}_X \mathcal{O}_Y)$, where $\mathcal{O}_X \mathcal{O}_Y$ is the composite tensor defined by \eqref{eq: OX OY}.
Thus, equation~\eqref{eq: ldim zipper} reduces to $\ldim(\mathcal{O}_X \mathcal{O}_Y) = \sum_{Z \in S_{XY}} \ldim(\mathcal{O}_Z)$, or equivalently,
\begin{equation}
\lrank(\mathcal{O}_X \mathcal{O}_Y) = \sum_{Z \in S_{XY}} \lrank(\mathcal{O}_Z).
\label{eq: rank fusion rule}
\end{equation}
To show the above equality, we note that the MPO matrices of the composite tensor $\mathcal{O}_X \mathcal{O}_Y$ can be made into a block upper triangular form, where the non-zero block diagonal components are given by the injective blocks $\{\mathcal{O}_Z \mid Z \in S_{XY}\}$ \cite{Perez-Garcia:2006nqo, Cirac:2016iqe}.
Since the rank of an upper triangular matrix is greater than or equal to the sum of the ranks of its block diagonal components, it follows that
\begin{equation}
\sum_{Z \in S_{XY}} \lrank(\mathcal{O}_Z) \leq \lrank(\mathcal{O}_X \mathcal{O}_Y).
\label{eq: rank inequality 1}
\end{equation}
On the other hand, due to the defining equation~\eqref{eq: fusion and splitting 2} of the fusion and splitting tensors, the rank of $\mathcal{O}_X \mathcal{O}_Y$ also satisfies
\begin{equation}
\lrank(\mathcal{O}_X \mathcal{O}_Y) = \frac{\lrank(\mathcal{O}_X^{[2]} \mathcal{O}_Y^{[2]})}{\dim(\mathcal{H}_{\mathrm{o}})} \leq \frac{1}{\dim(\mathcal{H}_{\mathrm{o}})} \sum_{Z \in S_{XY}} \lrank \left( \adjincludegraphics[valign=c, scale=1, trim={10, 10, 10, 10}]{tikz/out/fusion_split_def2.pdf} \right),
\label{eq: rank inequality 2}
\end{equation}
where the inequality follows from $\rank(f_1 + f_2) \leq \rank(f_1) + \rank(f_2)$ for any linear maps $f_1$ and $f_2$.
The first equality in \eqref{eq: rank inequality 2} holds as long as we block a sufficient number of physical sites.\footnote{More precisely, for any pair of topological injective MPO tensors $\mathcal{O}_X$ and $\mathcal{O}_Y$, it follows that $\lrank(\mathcal{O}_X^{[4]} \mathcal{O}_Y^{[4]}) = \lrank(\mathcal{O}_X^{[2]} \mathcal{O}_Y^{[2]}) \dim(\mathcal{H}_{\mathrm{o}})^2$ due to \eqref{eq: topological injective MPO}. Thus, by blocking two physical sites, we obtain the first equality of \eqref{eq: rank inequality 2}.}
Using the broken zipper condition and the two-sided zipper condition, we can compute the summand on the right-hand side of \eqref{eq: rank inequality 2} as follows:
\begin{equation}
\begin{aligned}
& \quad \lrank \left( \adjincludegraphics[valign=c, scale=1, trim={10, 10, 10, 10}]{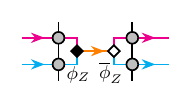} \right)
=
\rank \left( \adjincludegraphics[valign=c, scale=1, trim={10, 10, 10, 10}]{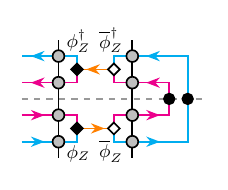} \right)
=
\rank \left( \adjincludegraphics[valign=c, scale=1, trim={10, 10, 10, 10}]{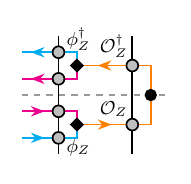} \right) \\
&=
\lrank \left( \adjincludegraphics[valign=c, scale=1, trim={10, 10, 10, 10}]{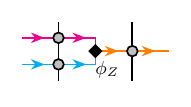} \right)
=
\rank \left( \adjincludegraphics[valign=c, scale=1, trim={10, 10, 10, 10}]{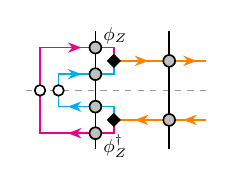} \right)
=
\lrank \left( \adjincludegraphics[valign=c, scale=1, trim={10, 10, 10, 10}]{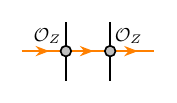} \right) \\
&= \dim(\mathcal{H}_{\mathrm{o}}) \lrank(\mathcal{O}_Z).
\end{aligned}
\end{equation}
Thus, equation~\eqref{eq: rank inequality 2} reduces to
\begin{equation}
\lrank(\mathcal{O}_X \mathcal{O}_Y) \leq \sum_{Z \in S_{XY}} \lrank(\mathcal{O}_Z).
\label{eq: rank inequality 3}
\end{equation}
Equations~\eqref{eq: rank inequality 1} and \eqref{eq: rank inequality 3} imply \eqref{eq: rank fusion rule}, which in turn implies \eqref{eq: ldim zipper}.
Equation~\eqref{eq: rdim zipper} can also be obtained in the same way.

\subsubsection{The index is homogeneous}
Finally, we show that the index is homogeneous if the broken zipper condition and the two-sided zipper condition hold for all fusion channels.
More specifically, if \eqref{eq: broken zipper}, \eqref{eq: two-sided zipper 1}, and \eqref{eq: two-sided zipper 2} hold for fusion channel $Z \in S_{XY}$, we have
\begin{equation}
\ind(\mathcal{O}_Z) = \ind(\mathcal{O}_X) \ind(\mathcal{O}_Y).
\label{eq: ind zipper}
\end{equation}
In particular, if this holds for all fusion channels, it follows that $\ind(\mathcal{O}_Z) = \ind(\mathcal{O}_W)$ for any pair of $Z, W \in S_{XY}$, and hence the index is homogeneous.

To show \eqref{eq: ind zipper}, we consider the following equalities:
\begin{align}
\adjincludegraphics[valign=c, scale=1, trim={10, 10, 10, 10}]{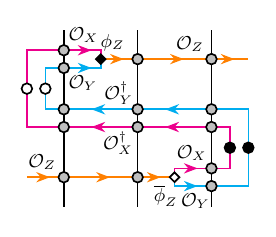} &\;=\; \adjincludegraphics[valign=c, scale=1, trim={10, 10, 10, 10}]{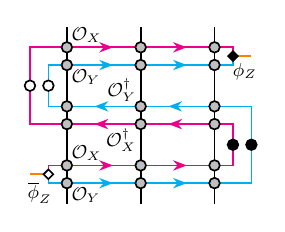} \;= c_Z^{l} \; \adjincludegraphics[valign=c, scale=1, trim={10, 10, 10, 10}]{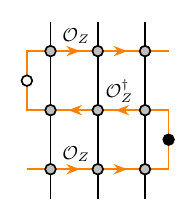}\;,
\label{eq: ind zipper derivation 1}
\\
\adjincludegraphics[valign=c, scale=1, trim={10, 10, 10, 10}]{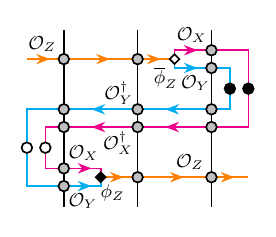} &\;=\; \adjincludegraphics[valign=c, scale=1, trim={10, 10, 10, 10}]{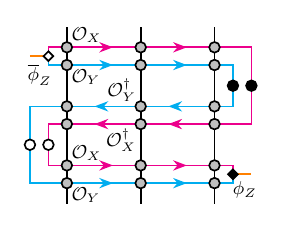} \;= c_Z^r \; \adjincludegraphics[valign=c, scale=1, trim={10, 10, 10, 10}]{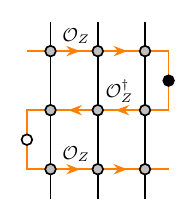}\;.
\label{eq: ind zipper derivation 2}
\end{align}
The left equalities of \eqref{eq: ind zipper derivation 1} and \eqref{eq: ind zipper derivation 2} follow from the broken zipper condition~\eqref{eq: broken zipper}, and the right equalities of \eqref{eq: ind zipper derivation 1} and \eqref{eq: ind zipper derivation 2} follow from the fact that $\mathcal{O}_X$, $\mathcal{O}_Y$, and $\mathcal{O}_Z$ are topological.
Due to the zigzag relations~\eqref{eq: zigzag}, the complex numbers $c_Z^{l}$ and $c_Z^r$ can be written as
\begin{equation}
c_Z^{l} = \frac{\gamma_X \gamma_Y}{\gamma_Z} \frac{\ldim(\mathcal{O}_Z)}{\ldim(\mathcal{O}_X) \ldim(\mathcal{O}_Y)}, \qquad
c_Z^r = \frac{\gamma_X \gamma_Y}{\gamma_Z} \frac{\rdim(\mathcal{O}_Z)}{\rdim(\mathcal{O}_X) \rdim(\mathcal{O}_Y)},
\label{eq: cZl cZr}
\end{equation}
where we defined $\gamma_i \coloneq \tr{\Lambda_i^l \Lambda_i^r}$ for $i = X, Y, Z$.
The above equation implies that equation~\eqref{eq: ind zipper} is equivalent to $c_Z^{l} = c_Z^r$.
In what follows, we will show that $c_Z^{l} = c_Z^r$.

To show that $c_Z^{l} = c_Z^r$, we first show the following equality:
\begin{equation}
\tr{\adjincludegraphics[valign=c, scale=0.89, trim={10, 10, 10, 10}]{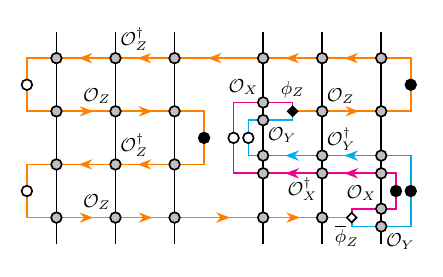}} \;=\; \tr{\adjincludegraphics[valign=c, scale=0.89, trim={10, 10, 10, 10}]{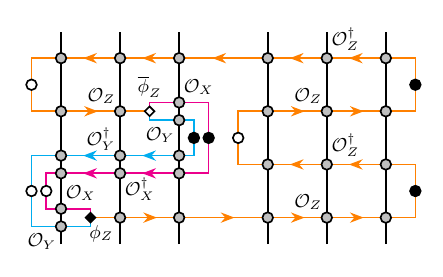}}.
\label{eq: cZl=cZr derivation}
\end{equation}
Using the cyclicity of the trace and the zigzag relations for $\mathcal{O}_Z$, one can compute the left-hand side and the right-hand side of the above equation as
\begin{equation}
\text{LHS} = \delta_l(\mathcal{O}_Z) \delta_r(\mathcal{O}_Z) \dim(\mathcal{H}_{\mathrm{o}})^3 \tr{\adjincludegraphics[valign=c, scale=1, trim={10, 10, 10, 10}]{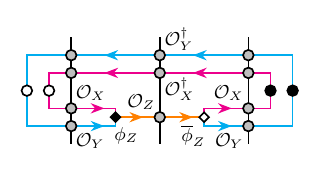}} = \text{RHS},
\end{equation}
where $\delta_l$ and $\delta_r$ are defined by \eqref{eq: delta lr}.
The above equation shows \eqref{eq: cZl=cZr derivation}.
On the other hand, due to \eqref{eq: ind zipper derivation 1} and \eqref{eq: ind zipper derivation 2}, the left-hand side and the right-hand side of~\eqref{eq: cZl=cZr derivation} can also be written as
\begin{equation}
\text{LHS} = c_Z^{l} \delta_l(\mathcal{O}_Z) \delta_r(\mathcal{O}_Z) \gamma_Z \dim(\mathcal{H}_{\mathrm{o}})^6, \qquad
\text{RHS} = c_Z^r \delta_l(\mathcal{O}_Z) \delta_r(\mathcal{O}_Z) \gamma_Z \dim(\mathcal{H}_{\mathrm{o}})^6.
\end{equation}
Therefore, equation~\eqref{eq: cZl=cZr derivation} implies that $c_Z^{l} = c_Z^r$, which in turn implies~\eqref{eq: ind zipper}.

\section{Examples}
\label{sec: Examples}
In this section, we discuss various examples of topological injective MPOs and show that the fusion and splitting tensors for their fusion channels satisfy both the broken zipper condition and the two-sided zipper condition.
We also compute the lattice quantum dimension and the index of these topological injective MPOs.

\subsection{Invertible symmetries}
\label{sec: Invertible symmetries}
Invertible symmetry operators defined by QCAs in 1+1d can be represented by injective MPUs \cite{Cirac:2017vke, Sahinoglu:2017iny}.
As shown in Section~\ref{sec: The case of invertible symmetries}, injective MPUs are examples of topological injective MPOs.
In this subsection, we show the existence of the fusion and splitting tensors for the unique fusion channel of any pair of injective MPUs.
We will also show that the fusion and splitting tensors satisfy both the broken zipper condition and the two-sided zipper condition.

\vspace*{\baselineskip}
\noindent{\bf Fusion and splitting tensors.}
Let $\mathcal{O}_X$ and $\mathcal{O}_Y$ be injective MPU tensors with bond Hilbert spaces $V_X$ and $V_Y$.
Since the product of two MPUs is an MPU, there exists an MPU tensor $\mathcal{O}_Z$ such that
\begin{equation}
\mathsf{D}[\mathcal{O}_X] \mathsf{D}[\mathcal{O}_Y] = \mathsf{D}[\mathcal{O}_Z]
\label{eq: MPU fusion rule}
\end{equation}
on a periodic chain of any length.
We can take $\mathcal{O}_Z$ to be injective by blocking a sufficient number of physical sites.
We denote the bond Hilbert space of $\mathcal{O}_Z$ by $V_Z$.
The above equation implies that the composite tensor $\mathcal{O}_X \mathcal{O}_Y$ defined by~\eqref{eq: OX OY} generates the same MPO as an injective MPO tensor $\mathcal{O}_Z$.
Therefore, due to \cite[Proposition 20]{Molnar:2018hls}, there exists a pair of three leg tensors $\phi_Z: V_X \otimes V_Y \to V_Z$ and $\overline{\phi}_Z: V_Z \to V_X \otimes V_Y$ that satisfies~\eqref{eq: fusion and splitting 1} for any number of physical legs.
In \cite[Definition 7]{Molnar:2018hls}, the pair $(\phi_Z, \overline{\phi}_Z)$ satisfying \eqref{eq: fusion and splitting 1} is called a reduction from $\mathcal{O}_X \mathcal{O}_Y$ to $\mathcal{O}_Z$.
In what follows, we will show that the reduction $(\phi_Z, \overline{\phi}_Z)$ also satisfies~\eqref{eq: fusion and splitting 2}, that is, $\phi_Z$ and $\overline{\phi}_Z$ are the fusion and splitting tensors for the unique fusion channel $Z$.

To show \eqref{eq: fusion and splitting 2}, we first show the following equality:
\begin{equation}
\adjincludegraphics[valign=c, scale=1, trim={10, 10, 10, 10}]{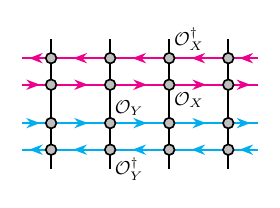} \;=\; \adjincludegraphics[valign=c, scale=1, trim={10, 10, 10, 10}]{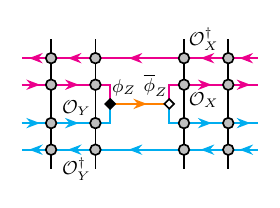}\;.
\label{eq: MPU fusion and splitting derivation 1}
\end{equation}
Using~\eqref{eq: simple MPU}, one can compute the left-hand side and the right-hand side of the above equation as
\begin{equation}
\text{LHS} = \adjincludegraphics[valign=c, scale=1, trim={10, 10, 10, 10}]{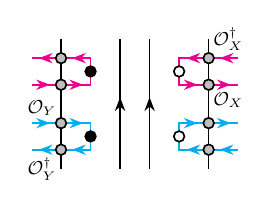} \;, \qquad
\text{RHS} = \adjincludegraphics[valign=c, scale=1, trim={10, 10, 10, 10}]{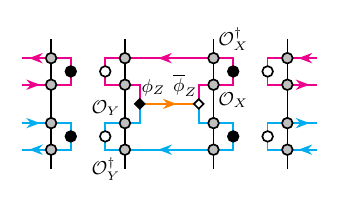} \;,
\label{eq: MPU fusion and splitting derivation 2}
\end{equation}
where the black and white dots represent the normalized fixed points of the transfer matrices $\mathsf{T}[\mathcal{O}_X]$ and $\mathsf{T}[\mathcal{O}_Y]$.
The middle diagram on the right-hand side of the second equation can further be computed as
\begin{equation}
\adjincludegraphics[valign=c, scale=1, trim={10, 10, 10, 10}]{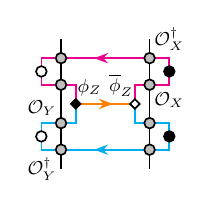} = \adjincludegraphics[valign=c, scale=1, trim={10, 10, 10, 10}]{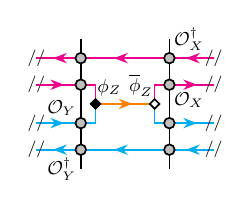} = \mathsf{D}[\mathcal{O}_X]^{\dagger} \mathsf{D}[\mathcal{O}_Z] \mathsf{D}[\mathcal{O}_Y]^{\dagger} = 1,
\label{eq: MPU fusion and splitting derivation 3}
\end{equation}
where the double slashes indicate the periodic boundary conditions.
The first equality follows from \eqref{eq: simple MPU},\footnote{More specifically, we used the second equality of \eqref{eq: simple MPU}. The left and right physical legs in \eqref{eq: simple MPU} correspond to the right and left physical legs in \eqref{eq: MPU fusion and splitting derivation 3}, respectively.} the second equality follows from \eqref{eq: fusion and splitting 1}, and the third equality follows from \eqref{eq: MPU fusion rule}.
Equations~\eqref{eq: MPU fusion and splitting derivation 2} and \eqref{eq: MPU fusion and splitting derivation 3} show that \eqref{eq: MPU fusion and splitting derivation 1} holds.

Equation~\eqref{eq: MPU fusion and splitting derivation 1} implies
\begin{equation}
\adjincludegraphics[valign=c, scale=1, trim={10, 10, 10, 10}]{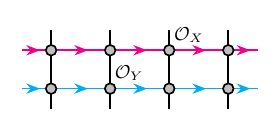} \;=\; \adjincludegraphics[valign=c, scale=1, trim={10, 10, 10, 10}]{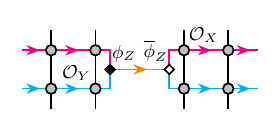} \;.
\label{eq: MPU fusion and splitting derivation 4}
\end{equation}
Furthermore, due to the zigzag relations~\eqref{eq: zigzag} for $\mathcal{O}_X$ and $\mathcal{O}_Y$, the above equation implies
\begin{equation}
\begin{aligned}
\adjincludegraphics[valign=c, scale=0.9, trim={10, 10, 10, 10}]{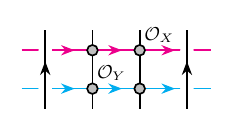}
= \alpha_{XY} \adjincludegraphics[valign=c, scale=0.9, trim={10, 10, 10, 10}]{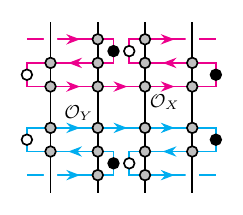}
= \alpha_{XY} \adjincludegraphics[valign=c, scale=0.9, trim={10, 10, 10, 10}]{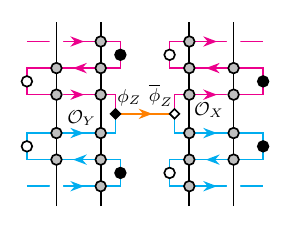}
= \adjincludegraphics[valign=c, scale=0.9, trim={10, 10, 10, 10}]{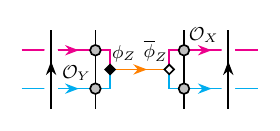}\;,
\end{aligned}
\label{eq: MPU fusion and splitting derivation 5}
\end{equation}
where $\alpha_{XY} = (\delta_l(\mathcal{O}_X) \delta_r(\mathcal{O}_X) \delta_l(\mathcal{O}_Y) \delta_r(\mathcal{O}_Y))^{-1}$.
Equation~\eqref{eq: MPU fusion and splitting derivation 5} shows that the reduction $(\phi_Z, \overline{\phi}_Z)$ satisfies~\eqref{eq: fusion and splitting 2}.

\vspace*{\baselineskip}
\noindent{\bf Broken zipper condition.}
It was shown in \cite[Proposition 5]{Ohyama:2024ytt} that the reduction $(\phi_Z, \overline{\phi}_Z)$ satisfies
\begin{equation}
\adjincludegraphics[valign=c, scale=1, trim={10, 10, 10, 10}]{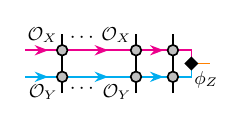} \;=\; \adjincludegraphics[valign=c, scale=1, trim={10, 10, 10, 10}]{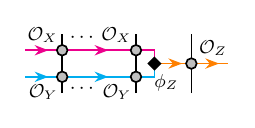}\;,
\label{eq: broken zipper original}
\end{equation}
where the number of physical legs is supposed to be greater than or equal to a positive integer called the nilpotency length of the reduction \cite[Definition 8]{Molnar:2018hls}.
We note that the nilpotency length of the reduction is independent of the system size \cite{Molnar:2018hls}.
Therefore, after blocking a finite number of physical legs, the reduction $(\phi_Z, \overline{\phi}_Z)$ satisfies the first equality of the broken zipper condition~\eqref{eq: broken zipper}.
The second equality of~\eqref{eq: broken zipper} also follows similarly.

\vspace*{\baselineskip}
\noindent{\bf Two-sided zipper condition.}
To show the two-sided zipper condition~\eqref{eq: two-sided zipper 1}, we first show the following equality:
\begin{equation}
\adjincludegraphics[valign=c, scale=1, trim={10, 10, 10, 10}]{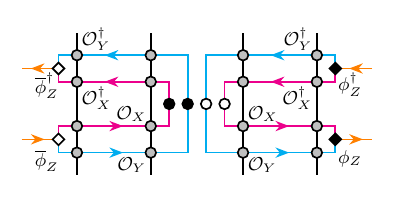} \;=\; \adjincludegraphics[valign=c, scale=1, trim={10, 10, 10, 10}]{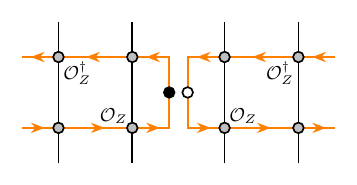} \;.
\label{eq: MPU two-sided zipper derivation 1}
\end{equation}
One can show the above equation by a direct computation as
\begin{equation}
\text{LHS} =\;
\adjincludegraphics[valign=c, scale=1, trim={10, 10, 10, 10}]{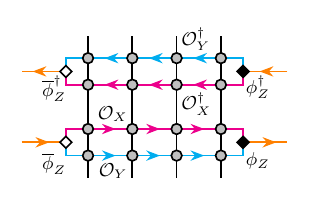}
\;=\;
\adjincludegraphics[valign=c, scale=1, trim={10, 10, 10, 10}]{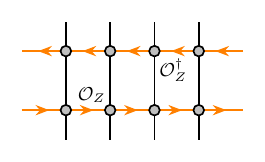}
\;= \text{RHS},
\end{equation}
where the first equality follows from \eqref{eq: simple MPU} and \eqref{eq: topological injective MPO}, and the second equality follows from \eqref{eq: fusion and splitting 1}.
Equation~\eqref{eq: MPU two-sided zipper derivation 1} implies that there exist complex numbers $a_Z^{l}$ and $a_Z^r$ such that 
\begin{equation}
\adjincludegraphics[valign=c, scale=1, trim={10, 10, 10, 10}]{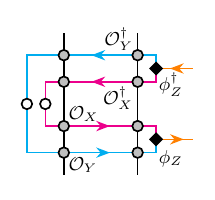} = a_Z^{l} \;\; \adjincludegraphics[valign=c, scale=1, trim={10, 10, 10, 10}]{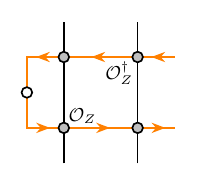} \;, \qquad
\adjincludegraphics[valign=c, scale=1, trim={10, 10, 10, 10}]{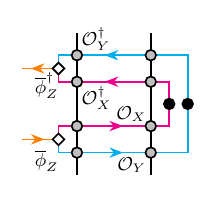} = a_Z^r \adjincludegraphics[valign=c, scale=1, trim={10, 10, 10, 10}]{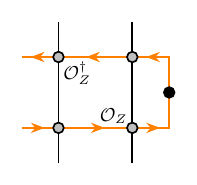} \;.
\label{eq: MPU two-sided zipper derivation 2}
\end{equation}
We note that $a_Z^{l}$ and $a_Z^r$ are non-zero because the diagrams on the left-hand side of the above two equations are non-zero, which also follows from \eqref{eq: MPU two-sided zipper derivation 1}.
Equation~\eqref{eq: MPU two-sided zipper derivation 2} shows that the reduction $(\phi_Z, \overline{\phi}_Z)$ satisfies the two-sided zipper condition~\eqref{eq: two-sided zipper 1} after blocking two physical sites.
One can also show the other half~\eqref{eq: two-sided zipper 2} of the two-sided zipper condition similarly.

\vspace*{\baselineskip}
\noindent{\bf Lattice quantum dimension and index.}
As discussed in Section~\ref{sec: The case of invertible symmetries}, the lattice quantum dimension and the index of an injective MPU tensor $\mathcal{O}$ are given by
\begin{equation}
\qdim_{\mathrm{lat}}(\mathcal{O}) = 1, \qquad
\ind(\mathcal{O}) = \text{GNVW}(\mathsf{D}[\mathcal{O}]),
\end{equation}
where $\text{GNVW}(\mathsf{D}[\mathcal{O}])$ is the GNVW index of the MPU $\mathsf{D}[\mathcal{O}]$.
Equivalently, the left and right dimensions of $\mathcal{O}$ are given by
\begin{equation}
\ldim(\mathcal{O}) = \text{GNVW}(\mathsf{D}[\mathcal{O}]), \qquad
\rdim(\mathcal{O}) = \text{GNVW}(\mathsf{D}[\mathcal{O}])^{-1}.
\end{equation}

\vspace*{\baselineskip}
\noindent{\bf Example: lattice translation.}
As an example of an injective MPU, let us consider the lattice translation operator $T$.
The MPO tensor of $T$ and its Hermitian conjugate can be written as
\begin{equation}
\mathcal{O}_T = \;\adjincludegraphics[valign=c, scale=1, trim={10, 10, 10, 10}]{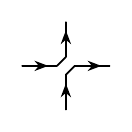}\;, \qquad
\mathcal{O}_T^{\dagger} = \;\adjincludegraphics[valign=c, scale=1, trim={10, 10, 10, 10}]{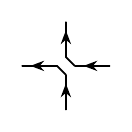}\;.
\end{equation}
We note that the bond Hilbert space and the physical Hilbert space are both given by $\mathcal{H}_{\mathrm{o}}$.
The left and right fixed points of the transfer matrix $\mathsf{T}[\mathcal{O}_T]$ are given by
\begin{equation}
\Lambda_T^l = \frac{1}{\sqrt{\dim(\mathcal{H}_{\mathrm{o}})}} \; \adjincludegraphics[valign=c, scale=1, trim={10, 10, 10, 10}]{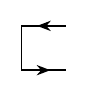} \;, \qquad
\Lambda_T^r = \frac{1}{\sqrt{\dim(\mathcal{H}_{\mathrm{o}})}} \; \adjincludegraphics[valign=c, scale=1, trim={10, 10, 10, 10}]{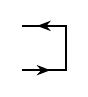} \;.
\end{equation}
Here, we normalized $\Lambda_T^l$ and $\Lambda_T^r$ so that they satisfy \eqref{eq: simple MPU}.
Since $T^{\dagger}T=1$, the product $\mathcal{O}_T^{\dagger} \mathcal{O}_T$ has a unique fusion channel, which is the identity channel denoted by $\mathds{1}$.
The corresponding fusion tensor $\phi_{\mathds{1}}$ and the splitting tensor $\overline{\phi}_{\mathds{1}}$ satisfying~\eqref{eq: fusion and splitting 1} and \eqref{eq: fusion and splitting 2} are given by
\begin{equation}
\phi_{\mathds{1}} = \Lambda_T^r, \qquad
\overline{\phi}_{\mathds{1}} = \Lambda_T^l.
\end{equation}
A direct computation shows that the above fusion and splitting tensors satisfy the broken zipper condition and the two-sided zipper condition.
The complex numbers in the defining equations \eqref{eq: two-sided zipper 1} and \eqref{eq: two-sided zipper 2} of the two-sided zipper condition are given by
\begin{equation}
a_{\mathds{1}}^{l} = \dim(\mathcal{H}_\mathrm{o})^{-1}, \qquad
a_{\mathds{1}}^r = \dim(\mathcal{H}_\mathrm{o}), \qquad 
b_{\mathds{1}}^{l} = \dim(\mathcal{H}_\mathrm{o}), \qquad
b_{\mathds{1}}^r = \dim(\mathcal{H}_\mathrm{o})^{-1}.
\end{equation}

\subsection{Non-anomalous fusion category symmetries}
\label{sec: Non-anomalous fusion category symmetries}
As an example of non-invertible topological injective MPOs, we consider the symmetry operators of a non-anomalous fusion category symmetry $\Rep(A)$, where $A$ is a finite dimensional semisimple Hopf $*$-algebra.
We will write down the fusion and splitting tensors for any fusion channel of these symmetry operators and show that they satisfy both the broken zipper condition and the two-sided zipper condition.

\vspace*{\baselineskip}
\noindent{\bf Hopf $*$-algebras and their representations.}
Let us begin with the definitions of a Hopf $*$-algebra and its $*$-representations.
We refer the reader to \cite{Timmermann2008} for a review of this subject.

\begin{definition}[Hopf algebra]
A Hopf algebra is a vector space $A$ equipped with multiplication $m: A \otimes A \to A$, unit $\eta: \mathbb{C} \to A$, comultiplication $\Delta: A \to A \otimes A$, counit $\epsilon: A \to \mathbb{C}$, and antipode $S: A \to A$, which satisfy the following axioms:
\begin{enumerate}
\item the triple $(A, m, \eta)$ is a unital algebra, i.e.,
\begin{equation}
m \circ (m \otimes \id_A) = m \circ (\id_A \otimes m), \qquad
m \circ (\eta \otimes \id_A) = m \circ (\id_A \otimes \eta) = \id_A;
\end{equation}
\item the triple $(A, \Delta, \epsilon)$ is a counital coalgebra, i.e.,
\begin{equation}
(\Delta \otimes \id_A) \circ \Delta = (\id_A \otimes \Delta) \circ \Delta, \qquad
(\epsilon \otimes \id_A) \circ \Delta = (\id_A \otimes \epsilon) \circ \Delta = \id_A;
\end{equation}
\item the compultiplication $\Delta$ and the counit $\epsilon$ are unital algebra homomorphisms;
\item the antipode $S$ satisfies
\begin{equation}
m \circ (S \otimes \id_A) \circ \Delta = m \circ (\id_A \otimes S) \circ \Delta = \eta \circ \epsilon.
\label{eq: antipode axiom}
\end{equation}
\end{enumerate}
\end{definition}
We note that the antipode $S$ is both a unital algebra antihomomorphism and a counital coalgebra antihomomorphism \cite[Proposition 1.3.12]{Timmermann2008}.
When $A$ is finite dimensional, $S$ is also invertible \cite{Larson1969, Radford1976}.
Furthermore, when $A$ is semisimple, we have $S^2 = \id_A$ \cite{LR1988}.

\begin{definition}[Hopf $*$-algebra]
A Hopf $*$-algebra is a Hopf algebra $A$ equipped with a conjugate-linear involution $*: A \to A$ that satisfies
\begin{equation}
m (a \otimes b)^* = m (b^* \otimes a^*), \qquad
\Delta(a)^* = \Delta(a^*), \qquad
\forall a, b \in A,
\end{equation}
where $\Delta(a)^*$ is defined by $\Delta(a)^* \coloneq ((* \otimes *) \circ \Delta)(a)$.
\end{definition}
We note that the antipode of a Hopf $*$-algebra satisfies $S \circ * \circ S \circ * = \id_A$ \cite[Proposition 1.3.28]{Timmermann2008}.
In particular, when $A$ is finite dimensional and semisimple, we have $S \circ * = * \circ S$ due to $S^2 = \id_A$.

In what follows, we will restrict our attention to finite dimensional semisimple Hopf $*$-algebras.\footnote{We note that every finite dimensional semisimple (Hopf) $*$-algebra is a finite dimensional (Hopf) $C^*$-algebra, and vice versa.}
For later use, we mention that a finite dimensional semisimple Hopf $*$-algebra $A$ can be equipped with the structure of a Hilbert space with the inner product defined by \cite[Example 3.1.6]{Timmermann2008}
\begin{equation}
(a, b) \coloneq \Lambda(a^*b), \qquad
\forall a, b \in A,
\label{eq: inner product on A}
\end{equation}
where $\Lambda: A \to \mathbb{C}$ is the Haar measure of $A$.
The above inner product will be used to define the Hermitian conjugate of MPO representations of the symmetry operators.

\begin{definition}[$*$-representation]
A finite dimensional $*$-representation of a finite dimensional semisimple Hopf $*$-algebra $A$ is a pair $(V, \rho)$, where $V$ is a finite dimensional Hilbert space and $\rho: A \to \End(V)$ is a unital $*$-algebra homomorphism.
Here, $\End(V)$ denotes the $*$-algebra of endomorphisms of $V$ with the $*$-operation given by the Hermitian conjugation.
\end{definition}

It is known that finite dimensional $*$-representations of a finite dimensional semisimple Hopf $*$-algebra $A$ form a unitary fusion category denoted by $\Rep(A)$ \cite{Neshveyev2013, EGNO2015}.
More concretely, the tensor product of $*$-representations $(V_X, \rho_X)$ and $(V_Y, \rho_Y)$ is given by $(V_X \otimes V_Y, \rho_{XY})$, where the representation map $\rho_{XY}: A \to \End(V_X \otimes V_Y)$ is given by
\begin{equation}
\rho_{XY} \coloneq (\rho_X \otimes \rho_Y) \circ \Delta.
\end{equation}
We note that $(V_X \otimes V_Y, \rho_{XY})$ is a $*$-representation.
The unit of the above tensor product is given by the trivial $*$-representation $(\mathbb{C}, \rho_{\epsilon})$, where $\mathbb{C}$ is equipped with the canonical Hilbert space structure and the representation map $\rho_{\epsilon}: A \to \End(\mathbb{C}) \cong \mathbb{C}$ is given by
\begin{equation}
\rho_{\epsilon} (a) \coloneq \epsilon(a), \qquad
\forall a \in A.
\end{equation}
Furthermore, for any finite dimensional $*$-representation $(V, \rho)$, one can define the dual $*$-representation $(V^*, \overline{\rho})$, where $V^*$ is the dual vector space equipped with the inner product induced by that of $V$, and the representation map $\overline{\rho}: A \to \End(V^*)$ is given by
\begin{equation}
\overline{\rho}(a) \coloneq \rho(S(a))^{\mathbf{T}}, \qquad
\forall a \in A.
\label{eq: dual rep map}
\end{equation}
Here, the superscript $\mathbf{T}$ denotes the transpose.

An isomorphism between $*$-representations $(V, \rho)$ and $(W, \sigma)$ is a unitary $u: V \to W$ that intertwines the representaion maps $\rho$ and $\sigma$ in the sense that $u \circ \rho(a) = \sigma(a) \circ u$ for all $a \in A$.
Two $*$-representations are said to be isomorphic if there exists an isomorphism between them.

\vspace*{\baselineskip}
\noindent{\bf MPO representation of $\Rep(A)$ symmetry.}
Let us now introduce MPO representations of the symmetry operators of $\Rep(A)$ symmetry following \cite{Inamura:2021szw, Molnar:2022nmh, Jia:2024bng}.
For concreteness, we take the physical Hilbert space on each site to be the Hopf algebra $A$ equipped with the inner product defined by~\eqref{eq: inner product on A}.
We then define the MPO tensor of the symmetry operator labeled by a $*$-representation $(V, \rho) \in \Rep(A)$ as follows:
\begin{equation}
\mathcal{O}_{(V, \rho)} \coloneq \;\adjincludegraphics[valign=c, scale=1, trim={10, 10, 10, 10}]{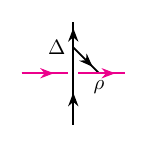} \;.
\label{eq: RepA MPO}
\end{equation}
Here, the Hilbert space on the virtual bond is the representation space $V$.
The above MPO representation is called an ``on-site" representation of $\Rep(A)$ symmetry according to the terminology introduced in \cite{Meng:2024nxx}.

By definition, it follows that the periodic MPOs generated by the above MPO tensors obey
\begin{equation}
\mathsf{D}[\mathcal{O}_{(V_X, \rho_X)}] \mathsf{D}[\mathcal{O}_{(V_Y, \rho_Y)}] = \mathsf{D}[\mathcal{O}_{(V_X \otimes V_Y, \rho_{XY})}],
\end{equation}
where $(V_X \otimes V_Y, \rho_{XY})$ is the tensor product representation defined above.
In addition, one can easily see that isomorphic $*$-representations give rise to the same MPO on a periodic chain because the corresponding MPO tensors differ only by a gauge transformation.
As a result, the MPOs on a periodic chain obey the same fusion rules as those of $\Rep(A)$, i.e.,
\begin{equation}
(V_X, \rho_X) \otimes (V_Y, \rho_Y) \cong \bigoplus_{Z \in S_{XY}} (V_Z, \rho_Z)
\; \Rightarrow \;
\mathsf{D}[\mathcal{O}_{(V_X, \rho_X)}] \mathsf{D}[\mathcal{O}_{(V_Y, \rho_Y)}] = \sum_{Z \in S_{XY}} \mathsf{D}[\mathcal{O}_{(V_Z, \rho_Z)}],
\label{eq: RepA MPO fusion rules}
\end{equation}
where $S_{XY}$ is the set of fusion channels of $(V_X, \rho_X)$ and $(V_Y, \rho_Y)$, including fusion multiplicities.

The set of MPO tensors of the form~\eqref{eq: RepA MPO} is closed under Hermitian conjugation.
Indeed, as shown in Appendix~\ref{sec: RepA MPO dagger}, the Hermitian conjugate of $\mathcal{O}_{(V, \rho)}$ agrees with the MPO tensor $\mathcal{O}_{(V^*, \overline{\rho})}$ labeled by the dual $*$-representation $(V^*, \overline{\rho})$.
More specifically, we have
\begin{equation}
\mathcal{O}_{(V, \rho)}^{\dagger}
=\; \adjincludegraphics[valign=c, scale=1, trim={10, 10, 10, 10}]{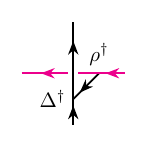}
\;=\; \adjincludegraphics[valign=c, scale=1, trim={10, 10, 10, 10}]{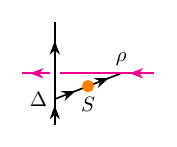}
\;= \mathcal{O}_{(V^*, \overline{\rho})},
\label{eq: RepA MPO dagger}
\end{equation}
where the last equality follows from the definition of the dual representation map~\eqref{eq: dual rep map}.

The MPO tensor $\mathcal{O}_{(V, \rho)}$ is injective when $(V, \rho)$ is an irreducible $*$-representation of $A$.
Indeed, when $(V, \rho)$ is irreducible, Schur's lemma implies that an operator acting on the virtual bond commutes with $\mathcal{O}_{(V, \rho)}$ if and only if it is proportional to the identity, and hence $\mathcal{O}_{(V, \rho)}$ is injective.
When $\mathcal{O}_{(V, \rho)}$ is injective, the left and right fixed points of the transfer matrix $\mathsf{T}[\mathcal{O}_{(V, \rho)}]$ are both given by the identity, i.e.,
\begin{equation}
\Lambda_{(V, \rho)}^l = \adjincludegraphics[valign=c, scale=1, trim={10, 10, 10, 10}]{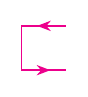} \;, \qquad
\Lambda_{(V, \rho)}^r = \adjincludegraphics[valign=c, scale=1, trim={10, 10, 10, 10}]{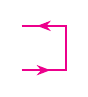} \;.
\end{equation}
A direct computation shows that $\Lambda_{(V, \rho)}^l$ and $\Lambda_{(V, \rho)}^r$ defined above satisfy\footnote{Equation~\eqref{eq: RepA fixed points} immediately follows from \eqref{eq: RepA MPO topological 1} that we will show shortly.}
\begin{equation}
\Lambda_{(V, \rho)}^l \mathsf{T}[\mathcal{O}_{(V, \rho)}] = \dim(A) \Lambda_{(V, \rho)}^l, \qquad
\mathsf{T}[\mathcal{O}_{(V, \rho)}] \Lambda_{(V, \rho)}^r = \dim(A) \Lambda_{(V, \rho)}^r.
\label{eq: RepA fixed points}
\end{equation}
Since $\Lambda_{(V, \rho)}^l$ and $\Lambda_{(V, \rho)}^r$ are positive definite, the above equation implies that they are the fixed points of $\mathsf{T}[\mathcal{O}_{(V, \rho)}]$.\footnote{We recall that a positive definite eigenvector of the transfer matrix of an injective MPO tensor is uniquely given by the fixed point due to \cite[Theorem 2.4]{Evans1978}. \label{fn: fixed points}}

\vspace*{\baselineskip}
\noindent{\bf The $\Rep(A)$ MPOs are topological.}
The MPO tensor $\mathcal{O}_{(V, \rho)}$ labeled by an irreducible $*$-representation $(V, \rho) \in \Rep(A)$ is a topological injective MPO tensor, i.e., it satisfies \eqref{eq: topological injective MPO} and \eqref{eq: O dagger topological}.
More specifically, $\mathcal{O}_{(V, \rho)}$ satisfies the following stronger conditions:
\begin{alignat}{2}
&\adjincludegraphics[valign=c, scale=1, trim={10, 10, 10, 10}]{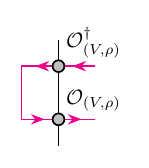} \;=\; \adjincludegraphics[valign=c, scale=1, trim={10, 10, 10, 10}]{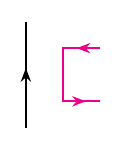} \;, \qquad
&&\adjincludegraphics[valign=c, scale=1, trim={10, 10, 10, 10}]{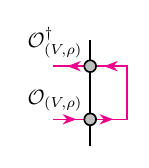} \;=\; \adjincludegraphics[valign=c, scale=1, trim={10, 10, 10, 10}]{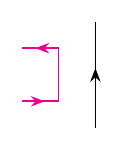} \;,
\label{eq: RepA MPO topological 1}
\\
&\adjincludegraphics[valign=c, scale=1, trim={10, 10, 10, 10}]{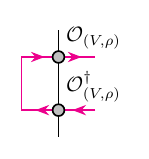} \;=\; \adjincludegraphics[valign=c, scale=1, trim={10, 10, 10, 10}]{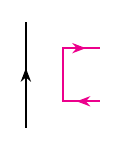} \;, \qquad
&&\adjincludegraphics[valign=c, scale=1, trim={10, 10, 10, 10}]{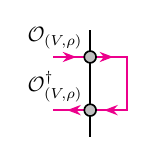} \;=\; \adjincludegraphics[valign=c, scale=1, trim={10, 10, 10, 10}]{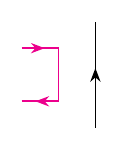} \;.
\label{eq: RepA MPO topological 2}
\end{alignat}
One can show the above equations by a direct computation.
For example, the left equation in \eqref{eq: RepA MPO topological 1} can be obtained as
\begin{equation}
\text{LHS}
=\; \adjincludegraphics[valign=c, scale=1, trim={10, 10, 10, 10}]{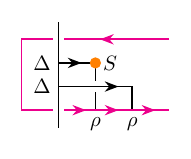}
\;=\; \adjincludegraphics[valign=c, scale=1, trim={10, 10, 10, 10}]{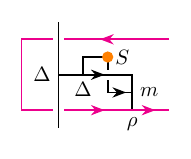}
\;=\; \adjincludegraphics[valign=c, scale=1, trim={10, 10, 10, 10}]{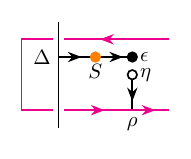}
\;= \text{RHS},
\end{equation}
where the first equality follows from \eqref{eq: RepA MPO dagger} and the third equality follows from the antipode axiom \eqref{eq: antipode axiom} together with the fact that $S$ is a coalgebra antihomomorphism.
The other equations in \eqref{eq: RepA MPO topological 1} and \eqref{eq: RepA MPO topological 2} can also be obtained similarly.

\vspace*{\baselineskip}
\noindent{\bf Fusion and splitting tensors.}
As shown in \eqref{eq: RepA MPO fusion rules}, the MPOs $\{\mathsf{D}[\mathcal{O}_{(V, \rho)}] \mid (V, \rho) \in \Rep(A)\}$ on a periodic chain obey the same fusion rules as $\Rep(A)$.
For each fusion channel $Z \in S_{XY}$, the fusion tensor $\phi_Z: V_X \otimes V_Y \to V_Z$ and the splitting tensor $\overline{\phi}_Z: V_Z \to V_X \otimes V_Y$ are given by
\begin{equation}
\phi_Z = \pi_Z \circ u_{XY}, \qquad
\overline{\phi}_Z = u_{XY}^{\dagger} \circ \iota_Z,
\label{eq: RepA phi}
\end{equation}
where $u_{XY}: V_X \otimes V_Y \to \bigoplus_{Z \in S_{XY}} V_Z$ is an isomorphism of $*$-representations, $\pi_Z: \bigoplus_{Z^{\prime} \in S_{XY}} V_{Z^{\prime}} \to V_Z$ is the projection, and $\iota_Z: V_Z \to \bigoplus_{Z^{\prime} \in S_{XY}} V_{Z^{\prime}}$ is the inclusion.
By definition, the above fusion and splitting tensors satisfy
\begin{equation}
\adjincludegraphics[valign=c, scale=1, trim={10, 10, 10, 10}]{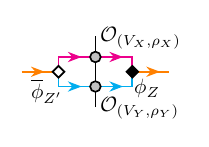} = \delta_{Z, Z^{\prime}} \;\adjincludegraphics[valign=c, scale=1, trim={10, 10, 10, 10}]{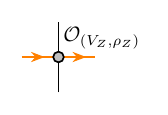}\;, \qquad
\sum_{Z \in S_{XY}} \adjincludegraphics[valign=c, scale=1, trim={10, 10, 10, 10}]{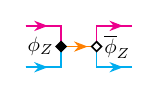} = \adjincludegraphics[valign=c, scale=1, trim={10, 10, 10, 10}]{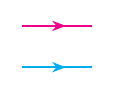}\;.
\label{eq: RepA completeness}
\end{equation}
This readily implies the defining equations~\eqref{eq: fusion and splitting 1} and \eqref{eq: fusion and splitting 2} of the fusion and splitting tensors.

\vspace*{\baselineskip}
\noindent{\bf Broken zipper condition.}
The fusion and splitting tensors defined by \eqref{eq: RepA phi} satisfy the broken zipper condition.
More specifically, these tensors satisfy the zipper condition
\begin{equation}
\adjincludegraphics[valign=c, scale=1, trim={10, 10, 10, 10}]{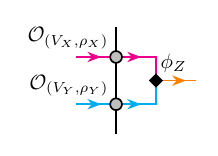} = \adjincludegraphics[valign=c, scale=1, trim={10, 10, 10, 10}]{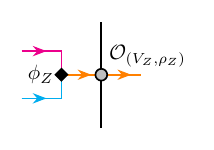}, \qquad
\adjincludegraphics[valign=c, scale=1, trim={10, 10, 10, 10}]{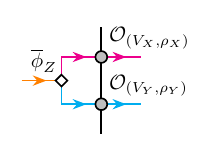} = \adjincludegraphics[valign=c, scale=1, trim={10, 10, 10, 10}]{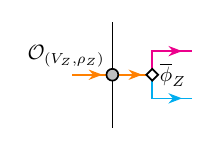}\;,
\label{eq: RepA zipper}
\end{equation}
which is stronger than the broken zipper condition.
The above equation immediately follows from \eqref{eq: RepA completeness}.
For example, the first equality of \eqref{eq: RepA zipper} can be obtained as
\begin{equation}
\adjincludegraphics[valign=c, scale=1, trim={10, 10, 10, 10}]{tikz/out/RepA_unbroken_zipper1.pdf}
= \sum_{Z^{\prime} \in S_{XY}} \adjincludegraphics[valign=c, scale=1, trim={10, 10, 10, 10}]{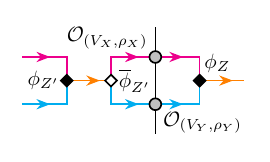}
= \adjincludegraphics[valign=c, scale=1, trim={10, 10, 10, 10}]{tikz/out/RepA_unbroken_zipper2.pdf}.
\end{equation}
The second equality can also be obtained in the same way.

\vspace*{\baselineskip}
\noindent{\bf Two-sided zipper condition.}
The fusion and splitting tensors in \eqref{eq: RepA phi} also satisfy the two-sided zipper conditions \eqref{eq: two-sided zipper 1} and \eqref{eq: two-sided zipper 2}, where $a_Z^{l} = a_Z^r = b_Z^{l} = b_Z^r = 1$ for all fusion channels $Z \in S_{XY}$.
This immediately follows from the zipper condition~\eqref{eq: RepA zipper} together with the following equalities:
\begin{equation}
\adjincludegraphics[valign=c, scale=1, trim={10, 10, 10, 10}]{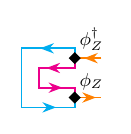} =\; \adjincludegraphics[valign=c, scale=1, trim={10, 10, 10, 10}]{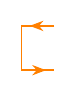}\;,
\qquad
\adjincludegraphics[valign=c, scale=1, trim={10, 10, 10, 10}]{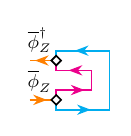} \;= \adjincludegraphics[valign=c, scale=1, trim={10, 10, 10, 10}]{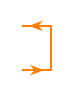}\;,
\qquad
\adjincludegraphics[valign=c, scale=1, trim={10, 10, 10, 10}]{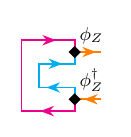} =\; \adjincludegraphics[valign=c, scale=1, trim={10, 10, 10, 10}]{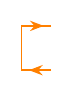}\;,
\qquad
\adjincludegraphics[valign=c, scale=1, trim={10, 10, 10, 10}]{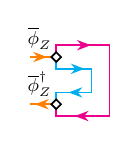} \;= \adjincludegraphics[valign=c, scale=1, trim={10, 10, 10, 10}]{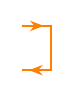}\;.
\end{equation}

\vspace*{\baselineskip}
\noindent{\bf Lattice quantum dimension and index.}
Finally, let us compute the lattice quantum dimension and the index of $\mathcal{O}_{(V, \rho)}$.
Equation~\eqref{eq: RepA MPO topological 1} implies that the left and right dimensions of $\mathcal{O}_{(V, \rho)}$ are given by
\begin{equation}
\ldim(\mathcal{O}_{(V, \rho)}) = \dim(V), \qquad
\rdim(\mathcal{O}_{(V, \rho)}) = \dim(V).
\end{equation}
Therefore, the lattice quantum dimension and the index of $\mathcal{O}_{(V, \rho)}$ can be computed as
\begin{equation}
\qdim_{\mathrm{lat}}(\mathcal{O}_{(V, \rho)}) = \dim(V), \qquad
\ind(\mathcal{O}_{(V, \rho)}) = 1.
\end{equation}
This result is consistent with the fact that the $\Rep(A)$ symmetry is realized exactly on a tensor product Hilbert space without mixing with QCAs.

\subsection{Kramers-Wannier symmetry for $\mathbb{Z}_2$ gauging}
\label{sec: Kramers-Wannier symmetry}
As an example of a non-invertible symmetry that mixes with a non-trivial QCA, we consider the Kramers-Wannier duality symmetry \cite{KW1941}, i.e., the self-duality under gauging a $\mathbb{Z}_2$ symmetry.
We will write down the fusion and splitting tensors for the fusion channels of two copies of the Kramers-Wannier duality operator and show that they satisfy both the broken zipper condition and the two-sided zipper condition.
We refer the reader to \cite[Appendix A]{Lu:2025rwd} for a similar analysis.
A generalization to the case of $\mathbb{Z}_N$ gauging will be discussed in the next subsection.

\vspace*{\baselineskip}
\noindent{\bf The Kramers-Wannier duality operator.}
Let us first recall the definition of the Kramers-Wannier duality operator.
We consider a one-dimensional chain of qubits on a periodic lattice of $L$ sites.
The Pauli operators on each site are denoted by $\{X_i, Y_i, Z_i\}$ for $i = 1, 2, \cdots, L$.
We take a basis of the Hilbert space on the entire lattice as $\{\ket{a_1, a_2, \cdots, a_L} \coloneq \bigotimes_{i = 1, 2, \cdots, L} \ket{a_i}_i \mid a_i = 0, 1\}$, where $\ket{a_i}_i$ is the eigenstate of $Z_i$ with eigenvalue $(-1)^{a_i}$.
The action of the Kramers-Wannier duality operator $\mathsf{D}_{\mathrm{KW}}$ is then given by \cite{Aasen:2016dop, Tantivasadakarn:2021vel, Li:2023ani}
\begin{equation}
\mathsf{D}_{\mathrm{KW}} \ket{a_1, a_2, \cdots, a_L} = \left(\bigotimes_{i = 1, \cdots, L} H_i \right) \ket{a_1-a_L, a_2-a_1, \cdots, a_{L}-a_{L-1}},
\label{eq: KW}
\end{equation}
where $H_i = \frac{1}{\sqrt{2}}(X_i + Z_i)$ is the Hadamard gate on site $i$, and $a_i - a_{i-1}$ is defined modulo 2.
The above operator $\mathsf{D}_{\mathrm{KW}}$ obeys the fusion rule \cite{Seiberg:2023cdc, Seiberg:2024gek}
\begin{equation}
\mathsf{D}_{\mathrm{KW}} \mathsf{D}_{\mathrm{KW}} = T (1 + U),
\label{eq: KW fusion rule}
\end{equation}
where $T$ is the lattice translation operator defined by $T \ket{a_1, a_2, \cdots, a_L} = \ket{a_L, a_1, \cdots, a_{L-1}}$, and $U = \bigotimes_{i = 1, \cdots, L} X_i$ is the $\mathbb{Z}_2$ symmetry operator.

\vspace*{\baselineskip}
\noindent{\bf MPO representation.}
The Kramers-Wannier duality operator~\eqref{eq: KW} can be written as an MPO as follows \cite{Tantivasadakarn:2021vel, Gorantla:2024ocs}:\footnote{The orientations of the physical legs and the virtual bonds do not matter in this example. Nevertheless, we keep the orientations in view of the generalization to the $\mathbb{Z}_N$ case, which we will discuss in the next subsection.}
\begin{equation}
\mathsf{D}_{\mathrm{KW}} =\; \adjincludegraphics[valign=c, scale=1, trim={10, 10, 10, 10}]{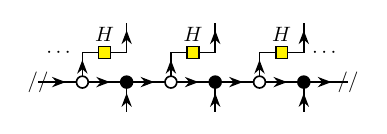} \;.
\label{eq: KW MPO}
\end{equation}
Here, the bond Hilbert space is $\mathbb{C}^2$ whose basis is again labeled by $0$ and $1$, the yellow square represents the Hadamard gate, and the three-leg tensors represented by the black and white dots are defined by
\begin{equation}
\adjincludegraphics[valign=c, scale=1, trim={10, 10, 10, 10}]{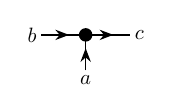} = \delta_{a, b} \delta_{a, c}, \qquad
\adjincludegraphics[valign=c, scale=1, trim={10, 10, 10, 10}]{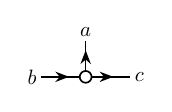} = \delta_{a+c, b}.
\label{eq: copy and multiplication}
\end{equation}
The above tensors are referred to as the copy tensor and the multiplication tensor, respectively.
We note that these tensors are intertwined by the Hadamard gate up to a scalar, that is, 
\begin{equation}
\adjincludegraphics[valign=c, scale=1, trim={10, 10, 10, 10}]{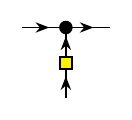}  = \frac{1}{\sqrt{2}} \; \adjincludegraphics[valign=c, scale=1, trim={10, 10, 10, 10}]{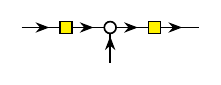}\;, \qquad
\adjincludegraphics[valign=c, scale=1, trim={10, 10, 10, 10}]{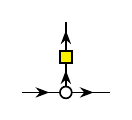} = \sqrt{2} \; \adjincludegraphics[valign=c, scale=1, trim={10, 10, 10, 10}]{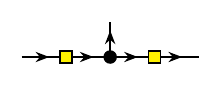}\;.
\label{eq: H intertwiner}
\end{equation}
The MPO representation~\eqref{eq: KW MPO} shows that the local MPO tensor of the Kramers-Wannier duality operator is given by
\begin{equation}
\mathcal{O}_{\mathrm{KW}} = \; \adjincludegraphics[valign=c, scale=1, trim={10, 10, 10, 10}]{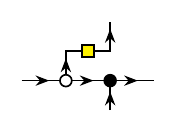}\;.
\label{eq: OKW}
\end{equation}
This MPO tensor is injective.
Indeed, as a linear map from the virtual bonds to the physical legs, the above MPO tensor can be viewed as the controlled-$X$ operator followed by the Hadamard gate, which is invertible and hence injective.

For later use, let us write down the left and right fixed points of the transfer matrix $\mathsf{T}[\mathcal{O}_{\mathrm{KW}}]$.
To this end, we first write down the Hermitian conjugate of the MPO tensor~\eqref{eq: OKW} as follows:\footnote{The MPO tensor $\mathcal{O}_{\mathrm{KW}}^{\dagger}$ generates the Hermitian conjugate of $\mathsf{D}_{\mathrm{KW}}$. Using \eqref{eq: H intertwiner}, one can show that $\mathsf{D}_{\mathrm{KW}}^{\dagger}$ and $\mathsf{D}_{\mathrm{KW}}$ differ by the lattice translation, i.e., $\mathsf{D}_{\mathrm{KW}}^{\dagger} = T^{-1} \mathsf{D}_{\mathrm{KW}}$ \cite{Seiberg:2024gek}. The same relation holds for the Kramers-Wannier duality operator for the $\mathbb{Z}_N$ gauging \cite{Cao:2024qjj}.}
\begin{equation}
\mathcal{O}_{\mathrm{KW}}^{\dagger} = \; \adjincludegraphics[valign=c, scale=1, trim={10, 10, 10, 10}]{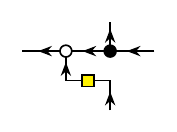}\;.
\label{eq: OKW dagger}
\end{equation}
Using \eqref{eq: OKW} and \eqref{eq: OKW dagger}, the transfer matrix $\mathsf{T}[\mathcal{O}_{\mathrm{KW}}]$ can be written as
\begin{equation}
\mathsf{T}[\mathcal{O}_{\mathrm{KW}}] = \; \adjincludegraphics[valign=c, scale=1, trim={10, 10, 10, 10}]{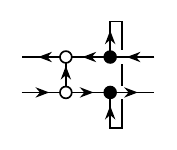} \;.
\end{equation}
The left and right fixed points $\Lambda_{\mathrm{KW}}^l$ and $\Lambda_{\mathrm{KW}}^r$ of this transfer matrix are both given by the identity, that is,
\begin{equation}
\Lambda_{\mathrm{KW}}^l = \adjincludegraphics[valign=c, scale=1, trim={10, 10, 10, 10}]{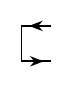}\;, \qquad
\Lambda_{\mathrm{KW}}^r = \adjincludegraphics[valign=c, scale=1, trim={10, 10, 10, 10}]{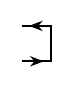}\;.
\label{eq: Lambda KW}
\end{equation}
A direct computation shows that $\Lambda_{\mathrm{KW}}^l$ and $\Lambda_{\mathrm{KW}}^r$ are eivenvectors of $\mathsf{T}[\mathcal{O}_{\mathrm{KW}}]$ with eigenvalue $2$:
\begin{equation}
\Lambda_{\mathrm{KW}}^l \mathsf{T}[\mathcal{O}_{\mathrm{KW}}] = \adjincludegraphics[valign=c, scale=1, trim={10, 10, 10, 10}]{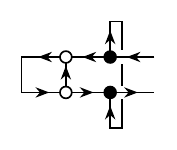} = 2 \Lambda_{\mathrm{KW}}^l, \qquad
\mathsf{T}[\mathcal{O}_{\mathrm{KW}}] \Lambda_{\mathrm{KW}}^r = \adjincludegraphics[valign=c, scale=1, trim={10, 10, 10, 10}]{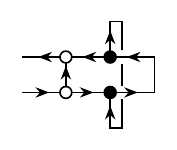} = 2 \Lambda_{\mathrm{KW}}^r.
\label{eq: KW fixed point eq}
\end{equation}
Here, we used 
\begin{equation}
\adjincludegraphics[valign=c, scale=1, trim={10, 10, 10, 10}]{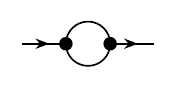} = \adjincludegraphics[valign=c, scale=1, trim={10, 10, 10, 10}]{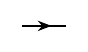}\;, \qquad
\adjincludegraphics[valign=c, scale=1, trim={10, 10, 10, 10}]{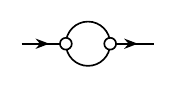} = 2 \; \adjincludegraphics[valign=c, scale=1, trim={10, 10, 10, 10}]{tikz/out/qubit_bubble3.pdf}\;.
\label{eq: qubit bubble}
\end{equation}
Since $\Lambda_{\mathrm{KW}}^l$ and $\Lambda_{\mathrm{KW}}^r$ in \eqref{eq: Lambda KW} are positive definite, equation~\eqref{eq: KW fixed point eq} implies that they are the left and right fixed points of $\mathsf{T}[\mathcal{O}_{\mathrm{KW}}]$; cf. footnote~\ref{fn: fixed points}.

\vspace*{\baselineskip}
\noindent{\bf The Kramers-Wannier MPO is topological.}
Now, we show that the injective MPO tensor $\mathcal{O}_{\mathrm{KW}}$ is topological.
Specifically, we show that $\mathcal{O}_{\mathrm{KW}}$ satisfies \eqref{eq: topological injective MPO} and \eqref{eq: O dagger topological}, i.e.,
\begin{equation}
\adjincludegraphics[valign=c, scale=1, trim={10, 10, 10, 10}]{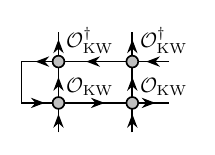}
\; = \;
\adjincludegraphics[valign=c, scale=1, trim={10, 10, 10, 10}]{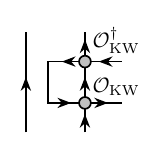},
\qquad
\adjincludegraphics[valign=c, scale=1, trim={10, 10, 10, 10}]{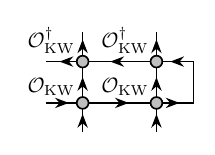}
\; = \;
\adjincludegraphics[valign=c, scale=1, trim={10, 10, 10, 10}]{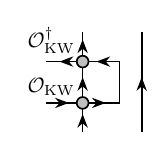}
\;,
\label{eq: KW topological 1}
\end{equation}
\begin{equation}
\adjincludegraphics[valign=c, scale=1, trim={10, 10, 10, 10}]{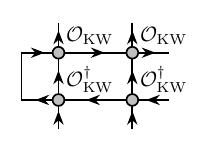}
\; = \;
\adjincludegraphics[valign=c, scale=1, trim={10, 10, 10, 10}]{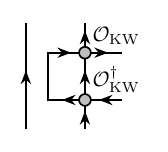},
\qquad
\adjincludegraphics[valign=c, scale=1, trim={10, 10, 10, 10}]{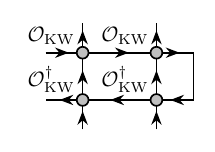}
\; = \;
\adjincludegraphics[valign=c, scale=1, trim={10, 10, 10, 10}]{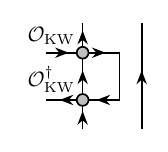}
\;.
\label{eq: KW topological 2}
\end{equation}
To show the first equality of \eqref{eq: KW topological 1}, we compute the diagram on the left-hand side as follows:
\begin{equation}
\text{LHS} 
=\;
\adjincludegraphics[valign=c, scale=1, trim={10, 10, 10, 10}]{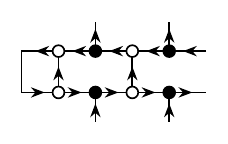}
\;=
2 \: \adjincludegraphics[valign=c, scale=1, trim={10, 10, 10, 10}]{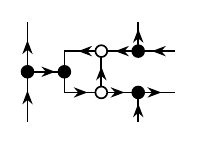}
\;=
2 \; \adjincludegraphics[valign=c, scale=1, trim={10, 10, 10, 10}]{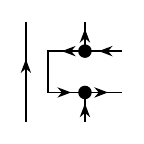}
\;= \text{RHS}.
\label{eq: KW topological derivation}
\end{equation}
In the third equality, we used
\begin{equation}
\adjincludegraphics[valign=c, scale=1, trim={10, 10, 10, 10}]{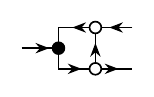} \;=\; \adjincludegraphics[valign=c, scale=1, trim={10, 10, 10, 10}]{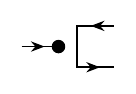}\;, \qquad
\text{where } \adjincludegraphics[valign=c, scale=1, trim={10, 10, 10, 10}]{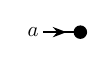} \coloneq 1 \quad \text{for $a = 0, 1$}.
\end{equation}
Equation~\eqref{eq: KW topological derivation} shows the first equality of \eqref{eq: KW topological 1}.
A similar computation shows the other equalities in \eqref{eq: KW topological 1} and \eqref{eq: KW topological 2}.

\vspace*{\baselineskip}
\noindent{\bf Fusion and splitting tensors.}
As shown in \eqref{eq: KW fusion rule}, the square of the Kramers-Wannier duality operator has two fusion channels $T$ and $TU$.
The fusion and splitting tensors for these fusion channels can be written explicitly as \cite{Lu:2025rwd}
\begin{alignat}{2}
&\phi_T = \adjincludegraphics[valign=c, scale=1, trim={10, 10, 10, 10}]{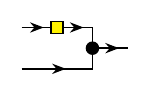}\;, \qquad
&&\overline{\phi}_T  = \adjincludegraphics[valign=c, scale=1, trim={10, 10, 10, 10}]{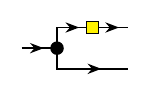}\;,
\label{eq: phi T}
\\
&\phi_{TU} = \adjincludegraphics[valign=c, scale=1, trim={10, 10, 10, 10}]{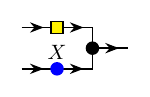}\;, \qquad
&&\overline{\phi}_{TU} = \adjincludegraphics[valign=c, scale=1, trim={10, 10, 10, 10}]{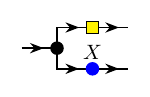}\;.
\label{eq: phi TU}
\end{alignat}
Here, we recall that the yellow square represents the Hadamard gate and the black dot represents the copy tensor.
We note that the fusion and splitting tensors in \eqref{eq: phi T} and \eqref{eq: phi TU} are related by Hermitian conjugation as $\overline{\phi}_T = \phi_T^{\dagger}$ and $\overline{\phi}_{TU} = \phi_{TU}^{\dagger}$.
As we will see below, the above tensors satisfy the defining equations of the fusion and splitting tensors, that is,
\begin{align}
\adjincludegraphics[valign=c, scale=1, trim={10, 10, 10, 10}]{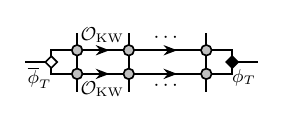} &= \adjincludegraphics[valign=c, scale=1, trim={10, 10, 10, 10}]{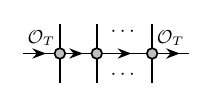}\;, \qquad
\adjincludegraphics[valign=c, scale=1, trim={10, 10, 10, 10}]{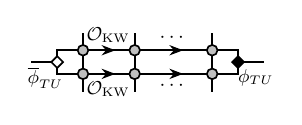} = \adjincludegraphics[valign=c, scale=1, trim={10, 10, 10, 10}]{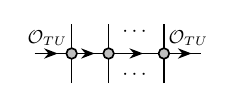},
\label{eq: KW zipper 1b}
\\
\adjincludegraphics[valign=c, scale=1, trim={10, 10, 10, 10}]{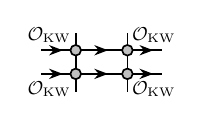} &= \adjincludegraphics[valign=c, scale=1, trim={10, 10, 10, 10}]{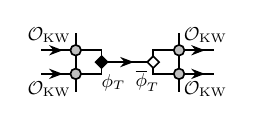} + \adjincludegraphics[valign=c, scale=1, trim={10, 10, 10, 10}]{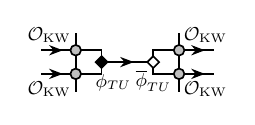} \;,
\label{eq: KW zipper 1c}
\end{align}
where $\mathcal{O}_T$ and $\mathcal{O}_{TU}$ are the local MPO tensors of $T$ and $TU$ defined by
\begin{equation}
\mathcal{O}_T = \; \adjincludegraphics[valign=c, scale=1, trim={10, 10, 10, 10}]{tikz/out/OT.pdf}\;, \qquad
\mathcal{O}_{TU} = \; \adjincludegraphics[valign=c, scale=1, trim={10, 10, 10, 10}]{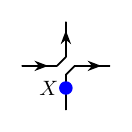}\;.
\end{equation}
The above equations~\eqref{eq: KW zipper 1b} and \eqref{eq: KW zipper 1c} readily follow from the following set of equalities:
\begin{align}
\adjincludegraphics[valign=c, scale=1, trim={10, 10, 10, 10}]{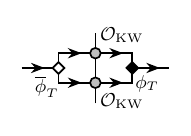} = \mathcal{O}_T, \qquad
&\adjincludegraphics[valign=c, scale=1, trim={10, 10, 10, 10}]{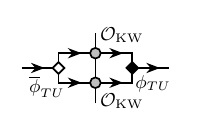} = \mathcal{O}_{TU}, \qquad
\adjincludegraphics[valign=c, scale=1, trim={10, 10, 10, 10}]{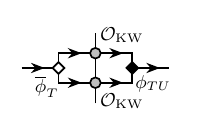} = \adjincludegraphics[valign=c, scale=1, trim={10, 10, 10, 10}]{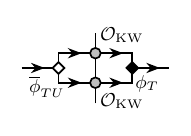} = 0,
\label{eq: KW zipper orthogonality}
\\
&\adjincludegraphics[valign=c, scale=1, trim={10, 10, 10, 10}]{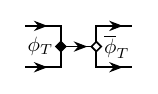}
+
\adjincludegraphics[valign=c, scale=1, trim={10, 10, 10, 10}]{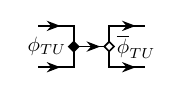}
=
\adjincludegraphics[valign=c, scale=1, trim={10, 10, 10, 10}]{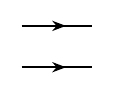} \;.
\label{eq: KW zipper completeness}
\end{align}
We note that the second line is an immediate consequence of the completeness relation
\begin{equation}
\adjincludegraphics[valign=c, scale=1, trim={10, 10, 10, 10}]{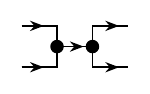}
\;+\;
\adjincludegraphics[valign=c, scale=1, trim={10, 10, 10, 10}]{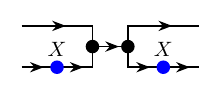}
\;=\;
\adjincludegraphics[valign=c, scale=1, trim={10, 10, 10, 10}]{tikz/out/phi_completeness3.pdf}\;.
\label{eq: identity decomposition}
\end{equation}
In what follows, we will show the remaining equation \eqref{eq: KW zipper orthogonality}.

To show \eqref{eq: KW zipper orthogonality}, we use the following expression of the composite tensor consisting of two copies of $\mathcal{O}_{\mathrm{KW}}$:
\begin{equation}
\adjincludegraphics[valign=c, scale=1, trim={10, 10, 10, 10}]{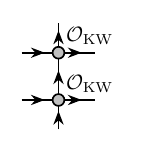}
\;=\;
\adjincludegraphics[valign=c, scale=1, trim={10, 10, 10, 10}]{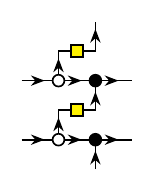}
\;=\;
\adjincludegraphics[valign=c, scale=1, trim={10, 10, 10, 10}]{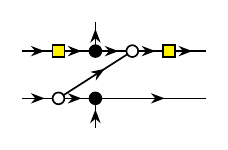}
\;.
\label{eq: OKW sq}
\end{equation}
By plugging~\eqref{eq: OKW sq} and \eqref{eq: phi T} into the left-hand side of the first equality of~\eqref{eq: KW zipper orthogonality}, we find
\begin{equation}
\text{LHS}
=\;
\adjincludegraphics[valign=c, scale=1, trim={10, 10, 10, 10}]{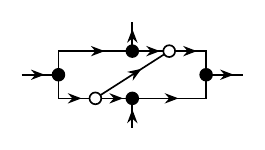}
\;=\;
\adjincludegraphics[valign=c, scale=1, trim={10, 10, 10, 10}]{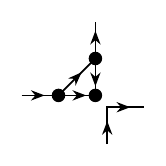}
\;=
\text{RHS}.
\end{equation}
Here, we used
\begin{equation}
\adjincludegraphics[valign=c, scale=1, trim={10, 10, 10, 10}]{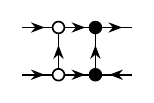} \;=\; \adjincludegraphics[valign=c, scale=1, trim={10, 10, 10, 10}]{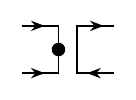}\;, \qquad
\text{where } \adjincludegraphics[valign=c, scale=1, trim={10, 10, 10, 10}]{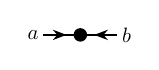} \coloneq \delta_{a, b} \quad \text{for $a, b = 0, 1$.}
\label{eq: ladder}
\end{equation}
Similarly, by plugging~\eqref{eq: OKW sq} and \eqref{eq: phi TU} into the left-hand side of the second equality of~\eqref{eq: KW zipper orthogonality}, we find
\begin{equation}
\text{LHS}
=\;
\adjincludegraphics[valign=c, scale=1, trim={10, 10, 10, 10}]{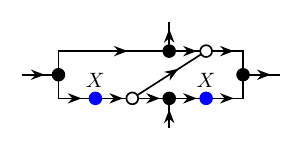}
\;=\;
\adjincludegraphics[valign=c, scale=1, trim={10, 10, 10, 10}]{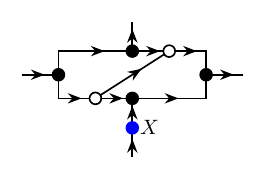}
\;=
\text{RHS}.
\end{equation}
Here, we used the following commutation relation of the Pauli-$X$ operator with the copy and multiplication tensors:
\begin{equation}
\adjincludegraphics[valign=c, scale=1, trim={10, 10, 10, 10}]{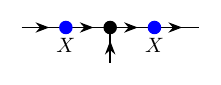} = \adjincludegraphics[valign=c, scale=1, trim={10, 10, 10, 10}]{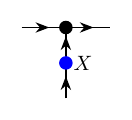}\;, \qquad
\adjincludegraphics[valign=c, scale=1, trim={10, 10, 10, 10}]{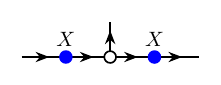} = \adjincludegraphics[valign=c, scale=1, trim={10, 10, 10, 10}]{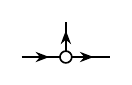}\;.
\end{equation}
A similar computation also shows the last two equalities in \eqref{eq: KW zipper orthogonality} as follows:
\begin{align}
\adjincludegraphics[valign=c, scale=1, trim={10, 10, 10, 10}]{tikz/out/KW_zipper_orthogonality1.pdf}
&\;=\; \adjincludegraphics[valign=c, scale=1, trim={10, 10, 10, 10}]{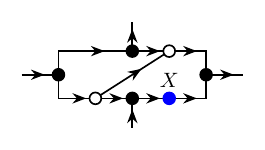}
\;=\; \adjincludegraphics[valign=c, scale=1, trim={10, 10, 10, 10}]{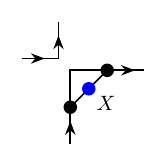}
\;= 0, \\
\adjincludegraphics[valign=c, scale=1, trim={10, 10, 10, 10}]{tikz/out/KW_zipper_orthogonality2.pdf}
&\;=\; \adjincludegraphics[valign=c, scale=1, trim={10, 10, 10, 10}]{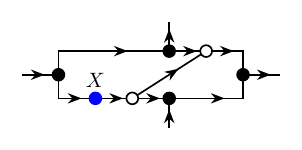}
\;=\; \adjincludegraphics[valign=c, scale=1, trim={10, 10, 10, 10}]{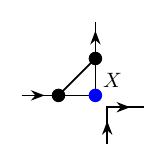}
\;= 0.
\end{align}

\vspace*{\baselineskip}
\noindent{\bf Broken zipper condition.}
The fusion and splitting tensors defined by \eqref{eq: phi T} and \eqref{eq: phi TU} satisfy the broken zipper condition.
More specifically, these tensors satisfy the zipper condition
\begin{alignat}{2}
&\adjincludegraphics[valign=c, scale=1, trim={10, 10, 10, 10}]{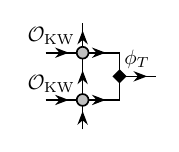}  = \adjincludegraphics[valign=c, scale=1, trim={10, 10, 10, 10}]{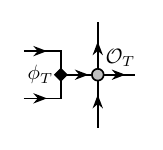}, \qquad
&&\adjincludegraphics[valign=c, scale=1, trim={10, 10, 10, 10}]{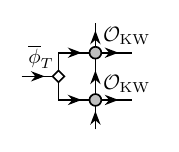} = \adjincludegraphics[valign=c, scale=1, trim={10, 10, 10, 10}]{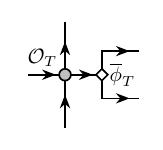},
\label{eq: KW unbroken zipper T}
\\
&\adjincludegraphics[valign=c, scale=1, trim={10, 10, 10, 10}]{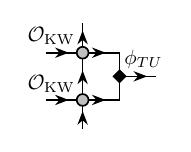}  = \adjincludegraphics[valign=c, scale=1, trim={10, 10, 10, 10}]{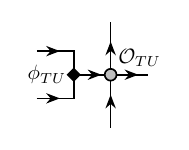}, \qquad
&&\adjincludegraphics[valign=c, scale=1, trim={10, 10, 10, 10}]{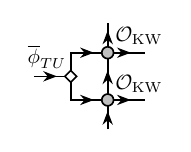} = \adjincludegraphics[valign=c, scale=1, trim={10, 10, 10, 10}]{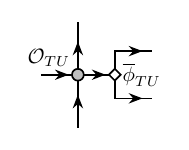},
\label{eq: KW unbroken zipper TU}
\end{alignat}
which is stronger than the broken zipper condition.
The above equalities immediately follow from \eqref{eq: KW zipper orthogonality} and \eqref{eq: KW zipper completeness}.

\vspace*{\baselineskip}
\noindent{\bf Two-sided zipper condition.}
The fusion and splitting tensors defined by \eqref{eq: phi T} and \eqref{eq: phi TU} also satisfy the following two-sided zipper condition:
\begin{alignat}{2}
&\adjincludegraphics[valign=c, scale=1, trim={10, 10, 10, 10}]{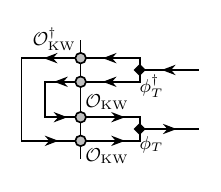} \;=\; \adjincludegraphics[valign=c, scale=1, trim={10, 10, 10, 10}]{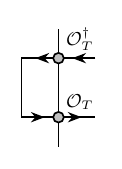}\;, \qquad
&&\adjincludegraphics[valign=c, scale=1, trim={10, 10, 10, 10}]{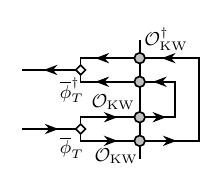} \;=\; \adjincludegraphics[valign=c, scale=1, trim={10, 10, 10, 10}]{tikz/out/KW_two_sided_zipper4.pdf}\;,
\label{eq: KW two-sided zipper T}
\\
&\adjincludegraphics[valign=c, scale=1, trim={10, 10, 10, 10}]{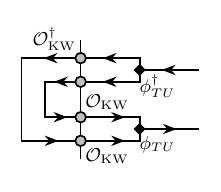} \;=\; \adjincludegraphics[valign=c, scale=1, trim={10, 10, 10, 10}]{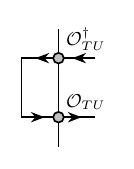}\;, \qquad
&&\adjincludegraphics[valign=c, scale=1, trim={10, 10, 10, 10}]{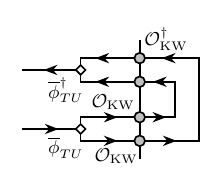} \;=\; \adjincludegraphics[valign=c, scale=1, trim={10, 10, 10, 10}]{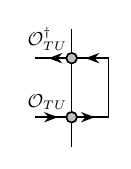}\;.
\label{eq: KW two-sided zipper TU}
\end{alignat}
The above equalities immediately follow from the zipper conditions \eqref{eq: KW unbroken zipper T} and \eqref{eq: KW unbroken zipper TU} together with the following equalities:
\begin{equation}
\adjincludegraphics[valign=c, scale=1, trim={10, 10, 10, 10}]{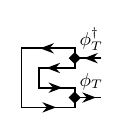} =\; \adjincludegraphics[valign=c, scale=1, trim={10, 10, 10, 10}]{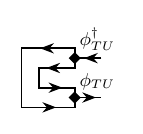} =\; \adjincludegraphics[valign=c, scale=1, trim={10, 10, 10, 10}]{tikz/out/phi_id_bar.pdf}\;, \qquad
\adjincludegraphics[valign=c, scale=1, trim={10, 10, 10, 10}]{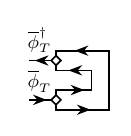} \;= \adjincludegraphics[valign=c, scale=1, trim={10, 10, 10, 10}]{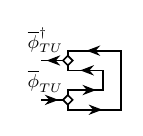} \;= \adjincludegraphics[valign=c, scale=1, trim={10, 10, 10, 10}]{tikz/out/phi_id.pdf}\;.
\end{equation}
Similarly, one can also show that
\begin{alignat}{2}
&\adjincludegraphics[valign=c, scale=1, trim={10, 10, 10, 10}]{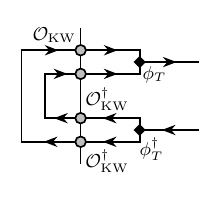} \;=\; \adjincludegraphics[valign=c, scale=1, trim={10, 10, 10, 10}]{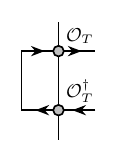}\;, \qquad
&&\adjincludegraphics[valign=c, scale=1, trim={10, 10, 10, 10}]{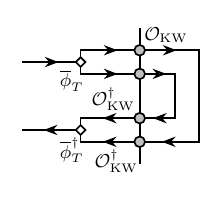} \;=\; \adjincludegraphics[valign=c, scale=1, trim={10, 10, 10, 10}]{tikz/out/KW_dagger_two_sided_zipper4.pdf}\;,
\label{eq: KW dagger two-sided zipper T}
\\
&\adjincludegraphics[valign=c, scale=1, trim={10, 10, 10, 10}]{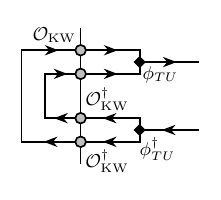} \;=\; \adjincludegraphics[valign=c, scale=1, trim={10, 10, 10, 10}]{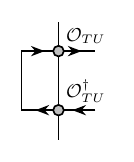}\;, \qquad
&&\adjincludegraphics[valign=c, scale=1, trim={10, 10, 10, 10}]{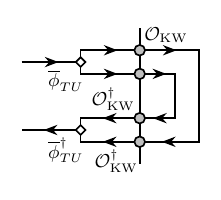} \;=\; \adjincludegraphics[valign=c, scale=1, trim={10, 10, 10, 10}]{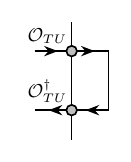}\;.
\label{eq: KW dagger two-sided zipper TU}
\end{alignat}
These equalities immediately follow from the zipper conditions \eqref{eq: KW unbroken zipper T} and \eqref{eq: KW unbroken zipper TU} together with the following equalities:
\begin{equation}
\adjincludegraphics[valign=c, scale=1, trim={10, 10, 10, 10}]{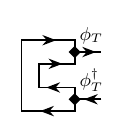} =\; \adjincludegraphics[valign=c, scale=1, trim={10, 10, 10, 10}]{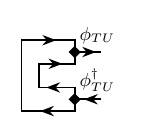} =\; \adjincludegraphics[valign=c, scale=1, trim={10, 10, 10, 10}]{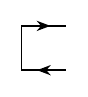}\;, \qquad
\adjincludegraphics[valign=c, scale=1, trim={10, 10, 10, 10}]{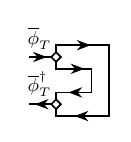} \;= \adjincludegraphics[valign=c, scale=1, trim={10, 10, 10, 10}]{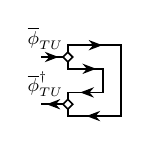} \;= \adjincludegraphics[valign=c, scale=1, trim={10, 10, 10, 10}]{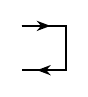}\;.
\end{equation}

\vspace*{\baselineskip}
\noindent{\bf Lattice quantum dimension and index.}
Let us compute the lattice quantum dimension and the index of the Kramers-Wannier MPO tensor $\mathcal{O}_{\mathrm{KW}}$.
To this end, we first compute the left and right dimensions of $\mathcal{O}_{\mathrm{KW}}$.
The left dimension of $\mathcal{O}_{\mathrm{KW}}$ can be computed as
\begin{equation}
\ldim(\mathcal{O}_{\mathrm{KW}}) = \frac{1}{\dim(\mathcal{H}_{\mathrm{o}})} \rank \left( \adjincludegraphics[valign=c, scale=1, trim={10, 10, 10, 10}]{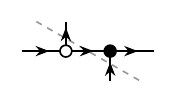} \right) = 2,
\label{eq: ldim OKW}
\end{equation}
where $\mathcal{H}_{\mathrm{o}} = \mathbb{C}^2$ is the on-site Hilbert space.
Here, the first equality follows from the fact that the Hadamard gate $H$ does not affect the rank because it is invertible.
The second equality follows from the fact that the diagram on the left-hand side is invertible as a linear map from the bottom left to the top right,\footnote{More specifically, the diagram on the left-hand side is a controlled-$X$ operator acting on the left virtual bond and the bottom physical leg.} and hence it has full rank.
Similarly, the right dimension of $\mathcal{O}_{\mathrm{KW}}$ can be computed as
\begin{equation}
\rdim(\mathcal{O}_{\mathrm{KW}}) = \frac{1}{\dim(\mathcal{H}_{\mathrm{o}})} \rank \left( \adjincludegraphics[valign=c, scale=1, trim={10, 10, 10, 10}]{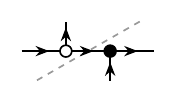} \right) = 1.
\label{eq: rdim OKW}
\end{equation}
Equations~\eqref{eq: ldim OKW} and \eqref{eq: rdim OKW} imply that the lattice quantum dimension and the index of $\mathcal{O}_{\mathrm{KW}}$ are given by
\begin{equation}
\qdim_{\mathrm{lat}} (\mathcal{O}_{\mathrm{KW}}) = \sqrt{2}, \qquad
\ind(\mathcal{O}_{\mathrm{KW}}) = \sqrt{2}.
\label{eq: ind OKW}
\end{equation}
We note that the lattice quantum dimension of $\mathcal{O}_{\mathrm{KW}}$ agrees with the quantum dimension of the unique non-invertible simple object of a $\mathbb{Z}_2$ Tambara-Yamagami category \cite{TY1998}.
The left and right dimensions computed above agree with the dimensions of the defect Hilbert spaces associated with the Kramers-Wannier duality defect \cite{Seiberg:2024gek}.\footnote{See also \cite{Schutz:1992wy, Grimm:1992ni, Grimm:2001dr, Oshikawa:1996dj, Ho:2014vla, Hauru:2015abi} for earlier works on duality-twisted boundary conditions.}
Furthermore, the index computed above agrees with the index of the Kramers-Wannier duality operator viewed as a QCA on the algebra of $\mathbb{Z}_2$-symmetric local operators \cite{Ma:2024ypm} or a QCA on a fusion spin chain \cite{Jones:2023imy}.

\vspace*{\baselineskip}
\noindent{\bf Another fusion rule.}
In the above discussion, we considered the fusion of two copies of the Kramer-Wannier MPO.
Similarly, one can also consider the fusion of the Kramers-Wannier MPO and its Hermitian conjugate.
Their fusion rule on a periodic chain is given by \cite{Li:2023ani}
\begin{equation}
\mathsf{D}_{\mathrm{KW}}^{\dagger} \mathsf{D}_{\mathrm{KW}} = 1+U.
\end{equation}
We note that the right-hand side does not involve the lattice translation.
The fusion and splitting tensors for each fusion channel are given by \cite{Lu:2025rwd}
\begin{alignat}{2}
&\phi_1 = \frac{1}{\sqrt{2}} ~ \adjincludegraphics[valign=c, scale=1, trim={10, 10, 10, 10}]{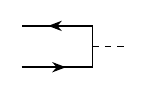}\;, \qquad
&&\overline{\phi}_1  = \frac{1}{\sqrt{2}} ~ \adjincludegraphics[valign=c, scale=1, trim={10, 10, 10, 10}]{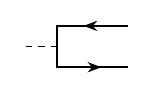}\;,
\label{eq: phi 1}
\\
&\phi_{U} = \frac{1}{\sqrt{2}} ~ \adjincludegraphics[valign=c, scale=1, trim={10, 10, 10, 10}]{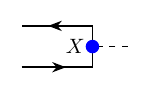}\;, \qquad
&&\overline{\phi}_{U} = \frac{1}{\sqrt{2}} ~ \adjincludegraphics[valign=c, scale=1, trim={10, 10, 10, 10}]{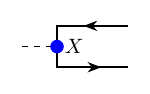}\;.
\label{eq: phi U}
\end{alignat}
Here, the dashed lines represent the one-dimensional bond Hilbert space for the on-site symmetry operators $1$ and $U$.
As opposed to the previous case, the above fusion and splitting tensors do not satisfy the strict zipper condition.
Nevertheless, a direct computation shows that they do satisfy the broken zipper condition and the two-sided zipper condition.
We will not provide a derivation of these conditions because the computation is straightforward.

\subsection{Kramers-Wannier symmetry for $\mathbb{Z}_N$ gauging}
\label{sec: ZN Kramers-Wannier symmetry}
Let us generalize the discussion in the previous subsection to the Kramers-Wannier self-duality for gauging a $\mathbb{Z}_N$ symmetry, where $N$ is an arbitrary positive integer.
We will see that the Kramers-Wannier duality operator for the $\mathbb{Z}_N$ gauging is a topological injective MPO, which we refer to as the $\mathbb{Z}_N$ Kramers-Wannier duality operator.
We will then write down the fusion and splitting tensors for the fusion channels of two copies of the $\mathbb{Z}_N$ Kramers-Wannier duality operator and show that they satisfy the broken zipper condition and the two-sided zipper condition.
These results also hold for the Kramers-Wannier duality operator for a general finite abelian group because any finite abelian group is isomorphic to the direct product of finite cyclic groups.\footnote{Recently, the Kramers-Wannier duality operator was generalized to finite semisimple self-dual Hopf $*$-algebras~\cite{Lu:2026rhb}. Although the Hopf Kramers-Wannier operator in~\cite{Lu:2026rhb} is written as an MPO, the fusion and splitting tensors for the fusion channels of two copies of it have not been constructed yet. We leave the study of this example for the future.}

\vspace*{\baselineskip}
\noindent{\bf The $\mathbb{Z}_N$ Kramers-Wannier duality operator.}
We first recall the definition of the $\mathbb{Z}_N$ Kramers-Wannier duality operator.
To this end, we consider a one-dimensional chain of $N$-dimensional qudits on a periodic lattice of $L$ sites.
The clock and shift operators on each site are denoted by $Z_i$ and $X_i$, respectively.
We take a basis of the Hilbert space on the entire lattice as $\{\ket{a_1, a_2, \cdots, a_L} \coloneq \bigotimes_{i = 1, 2, \cdots, L} \ket{a_i}_i \mid a_i = 0, 1, \cdots, N-1\}$, where $\ket{a_i}_i$ is the eigenstate of $Z_i$ with eigenvalue $e^{2\pi i a_i/N}$.
The action of the $\mathbb{Z}_N$ Kramers-Wannier duality operator $\mathsf{D}_{\mathrm{KW}_N}$ is then given by \cite{Tantivasadakarn:2022hgp, Cao:2024qjj, ParayilMana:2024txy}
\begin{equation}
\mathsf{D}_{\mathrm{KW}_N} \ket{a_1, a_2, \cdots, a_L} = \left( \bigotimes_{i = 1, 2, \cdots, L} H_i \right) \ket{a_1-a_L, a_2-a_1, \cdots, a_L-a_{L-1}},
\label{eq: ZN KW}
\end{equation}
where $a_i - a_{i-1}$ is defined modulo $N$, and $H_i$ is the generalized Hadamard gate on site $i$ defined by~\cite{Wang:2020ife}
\begin{equation}
H \ket{a} = \frac{1}{\sqrt{N}} \sum_{b = 0}^{N-1} e^{2\pi i a b/N} \ket{b}.
\end{equation}
We note that $H^2\ket{a} = \ket{N-a}$ and in particular $H^4 = 1$.
A direct computation shows that the $\mathbb{Z}_N$ Kramers-Wannier duality operator $\mathsf{D}_{\mathrm{KW}_N}$ obeys the fusion rule \cite{Cao:2024qjj, ParayilMana:2024txy}
\begin{equation}
\mathsf{D}_{\mathrm{KW}_N} \mathsf{D}_{\mathrm{KW}_N} = T \sum_{n = 0}^{N-1} U_n,
\label{eq: ZN KW fusion rule}
\end{equation}
where $T$ is the lattice translation operator defined by $T \ket{a_1, a_2, \cdots, a_L} = \ket{a_L, a_1, \cdots, a_{L-1}}$, and $U_n \coloneq \bigotimes_{i = 1, 2, \cdots, L} X_i^n$ is the $\mathbb{Z}_N$ symmetry operator.

\vspace*{\baselineskip}
\noindent{\bf MPO representation.}
The $\mathbb{Z}_N$ Kramers-Wannier duality operator~\eqref{eq: ZN KW} can be written as an MPO as follows \cite{Tantivasadakarn:2022hgp, Lu:2026rhb}:
\begin{equation}
\mathsf{D}_{\mathrm{KW}_N} = \; \adjincludegraphics[valign=c, scale=1, trim={10, 10, 10, 10}]{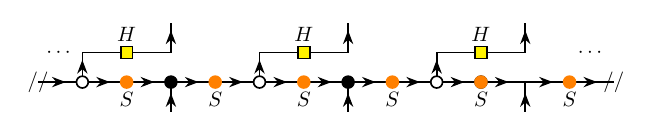} \;.
\label{eq: ZN KW MPO}
\end{equation}
Here, the bond Hilbert space is $\mathbb{C}^N$ whose basis is again labeled by $0, 1, \cdots, N-1$, and we defined
\begin{equation}
S: \ket{a} \to \ket{N-a}.
\end{equation}
We note that $S = H^2$ and in particular $SH = HS = H^{-1}$.
The black and white dots in \eqref{eq: ZN KW MPO} represent the copy and multiplication tensors~\eqref{eq: copy and multiplication} for the $\mathbb{Z}_N$ variables.
We note that the copy tensor and the multiplication tensor are intertwined by the generalized Hadamard gate up to a scalar as follows:
\begin{equation}
\adjincludegraphics[valign=c, scale=1, trim={10, 10, 10, 10}]{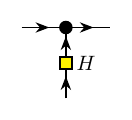}  = \frac{1}{\sqrt{N}} \; \adjincludegraphics[valign=c, scale=1, trim={10, 10, 10, 10}]{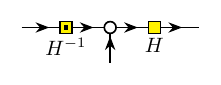}\;, \qquad
\adjincludegraphics[valign=c, scale=1, trim={10, 10, 10, 10}]{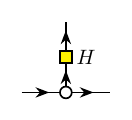} = \sqrt{N} \; \adjincludegraphics[valign=c, scale=1, trim={10, 10, 10, 10}]{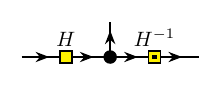}\;.
\label{eq: ZN H intertwiner}
\end{equation}
The MPO representation~\eqref{eq: ZN KW MPO} shows that the local MPO tensor of the $\mathbb{Z}_N$ Kramers-Wannier duality operator is given by
\begin{equation}
\mathcal{O}_{\mathrm{KW}_N} = \adjincludegraphics[valign=c, scale=1, trim={10, 10, 10, 10}]{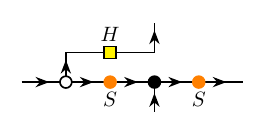}\;.
\label{eq: OKWN}
\end{equation}
This MPO tensor is injective as in the case of $N=2$.

For later use, let us write down the left and right fixed points of the transfer matrix $\mathsf{T}[\mathcal{O}_{\mathrm{KW}_N}]$.
To this end, we first write the Hermitian conjugate of the MPO tensor~\eqref{eq: ZN KW MPO} as follows:
\begin{equation}
\mathcal{O}_{\mathrm{KW}_N}^{\dagger} = \adjincludegraphics[valign=c, scale=1, trim={10, 10, 10, 10}]{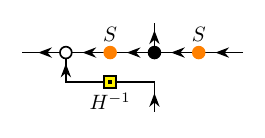}\;.
\label{eq: ZN KW MPO dagger}
\end{equation}
Based on \eqref{eq: OKWN} and \eqref{eq: ZN KW MPO dagger}, the transfer matrix $\mathsf{T}[\mathcal{O}_{\mathrm{KW}_N}]$ can be written as
\begin{equation}
\mathsf{T}[\mathcal{O}_{\mathrm{KW}_N}]  = \; \adjincludegraphics[valign=c, scale=1, trim={10, 10, 10, 10}]{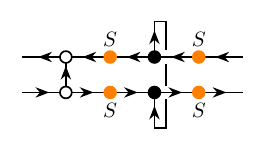} \;.
\end{equation}
The left and right fixed points of this transfer matrix are both given by the identity, that is,
\begin{equation}
\Lambda_{\mathrm{KW}_N}^l = \adjincludegraphics[valign=c, scale=1, trim={10, 10, 10, 10}]{tikz/out/Lambda_KW2.pdf}\;, \qquad
\Lambda_{\mathrm{KW}_N}^r = \adjincludegraphics[valign=c, scale=1, trim={10, 10, 10, 10}]{tikz/out/Lambda_KW1.pdf}\;.
\end{equation}
It is straightforward to show that the above $\Lambda_{\mathrm{KW}_N}^l$ and $\Lambda_{\mathrm{KW}_N}^r$ are positive definite left and right eigenvectors of $\mathsf{T}[\mathcal{O}_{\mathrm{KW}_N}]$ and hence the fixed points.

\vspace*{\baselineskip}
\noindent{\bf The $\mathbb{Z}_N$ Kramers-Wannier MPO is topological.}
One can show that the injective MPO tensor $\mathcal{O}_{\mathrm{KW}_N}$ defined by \eqref{eq: OKWN} is topological, that is, it satisfies
\begin{equation}
\adjincludegraphics[valign=c, scale=1, trim={10, 10, 10, 10}]{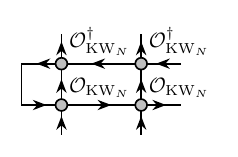}
\; = \;
\adjincludegraphics[valign=c, scale=1, trim={10, 10, 10, 10}]{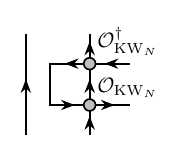},
\qquad
\adjincludegraphics[valign=c, scale=1, trim={10, 10, 10, 10}]{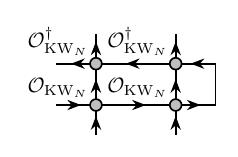}
\; = \;
\adjincludegraphics[valign=c, scale=1, trim={10, 10, 10, 10}]{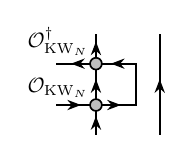}
\;,
\end{equation}
\begin{equation}
\adjincludegraphics[valign=c, scale=1, trim={10, 10, 10, 10}]{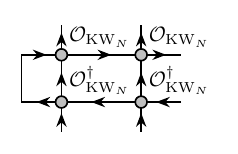}
\; = \;
\adjincludegraphics[valign=c, scale=1, trim={10, 10, 10, 10}]{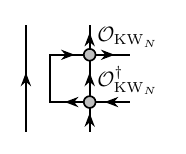},
\qquad
\adjincludegraphics[valign=c, scale=1, trim={10, 10, 10, 10}]{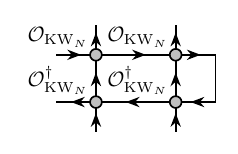}
\; = \;
\adjincludegraphics[valign=c, scale=1, trim={10, 10, 10, 10}]{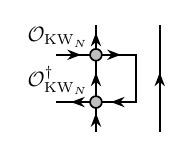}
\;.
\end{equation}
We will omit the derivation of the above equalities because it is parallel to the case of $N=2$.

\vspace*{\baselineskip}
\noindent{\bf Fusion and splitting tensors.}
The fusion rule~\eqref{eq: ZN KW fusion rule} implies that the square of the $\mathbb{Z}_N$ Kramers-Wannier duality operator has $N$ fusion channnels $\{TU_n \mid n = 0, 1, \cdots, N-1\}$.
The fusion and splitting tensors for each fusion channel $TU_n$ can be written explicitly as
\begin{equation}
\phi_{TU_n} = \adjincludegraphics[valign=c, scale=1, trim={10, 10, 10, 10}]{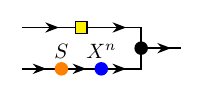}\;, \qquad
\overline{\phi}_{TU_n} = \adjincludegraphics[valign=c, scale=1, trim={10, 10, 10, 10}]{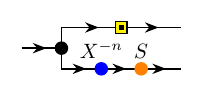}\;,
\label{eq: phi TUn}
\end{equation}
where the black dot represents the copy tensor and the yellow squares with and without a centered dot represent $H^{-1}$ and $H$, respectively.
We note that the splitting tensor is the Hermitian conjugate of the fusion tensor, i.e., $\overline{\phi}_{TU_n} = \phi_{TU_n}^{\dagger}$.
As we will see below, the above tensors satisfy the defining equations of the fusion and splitting tensors, that is,
\begin{align}
\adjincludegraphics[valign=c, scale=1, trim={10, 10, 10, 10}]{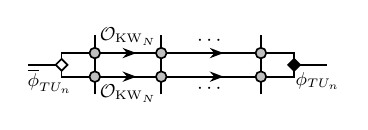}
&\;=\;
\adjincludegraphics[valign=c, scale=1, trim={10, 10, 10, 10}]{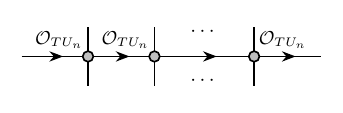}\;,
\label{eq: ZN KW zipper 1}
\\
\adjincludegraphics[valign=c, scale=1, trim={10, 10, 10, 10}]{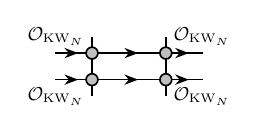}
&\;=
\sum_{n=0}^{N-1} \; \adjincludegraphics[valign=c, scale=1, trim={10, 10, 10, 10}]{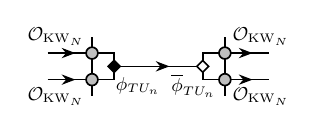} \;,
\label{eq: ZN KW zipper 2}
\end{align}
where $\mathcal{O}_{TU_n}$ is the MPO tensor of $TU_n$ defined by
\begin{equation}
\mathcal{O}_{TU_n} = \; \adjincludegraphics[valign=c, scale=1, trim={10, 10, 10, 10}]{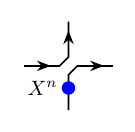}\;.
\end{equation}
As in the case of $N=2$, one can easily see that \eqref{eq: ZN KW zipper 1} and \eqref{eq: ZN KW zipper 2} follow from the following set of equalities:
\begin{equation}
\adjincludegraphics[valign=c, scale=1, trim={10, 10, 10, 10}]{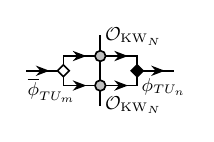} = \delta_{n, m} \; \adjincludegraphics[valign=c, scale=1, trim={10, 10, 10, 10}]{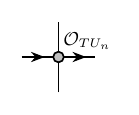}, \qquad
\sum_{n=0}^{N-1} \adjincludegraphics[valign=c, scale=1, trim={10, 10, 10, 10}]{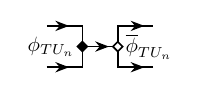} =\; \adjincludegraphics[valign=c, scale=1, trim={10, 10, 10, 10}]{tikz/out/phi_completeness3.pdf} \;.
\label{eq: ZN KW zipper derivation}
\end{equation}
Here, the second equality of \eqref{eq: ZN KW zipper derivation} is an immediate consequence of the completeness relation
\begin{equation}
\sum_{n=0}^{N-1} \; \adjincludegraphics[valign=c, scale=1, trim={10, 10, 10, 10}]{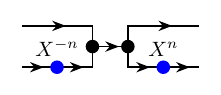} \;=\; \adjincludegraphics[valign=c, scale=1, trim={10, 10, 10, 10}]{tikz/out/phi_completeness3.pdf} \;.
\end{equation}
In what follows, we will show the first equality of~\eqref{eq: ZN KW zipper derivation}.

To show the first equality of \eqref{eq: ZN KW zipper derivation}, we use the following expression of the composite tensor consisting of two copies of $\mathcal{O}_{\mathrm{KW}_N}$:
\begin{equation}
\adjincludegraphics[valign=c, scale=1, trim={10, 10, 10, 10}]{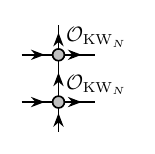}
\;=\;
\adjincludegraphics[valign=c, scale=1, trim={10, 10, 10, 10}]{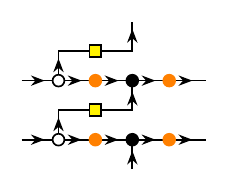}
\;=\;
\adjincludegraphics[valign=c, scale=1, trim={10, 10, 10, 10}]{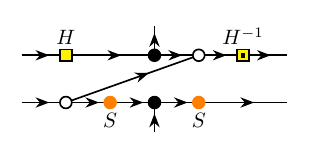}
\;.
\label{eq: ZN OKW sq}
\end{equation}
Here, the last equality follows from \eqref{eq: ZN H intertwiner} and $SH = H^{-1}$.
By plugging \eqref{eq: ZN OKW sq} and \eqref{eq: phi TUn} into the left-hand side of the first equality of \eqref{eq: ZN KW zipper derivation}, we find
\begin{equation}
\text{LHS}
= \adjincludegraphics[valign=c, scale=1, trim={10, 10, 10, 10}]{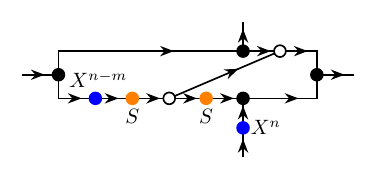}
= \adjincludegraphics[valign=c, scale=1, trim={10, 10, 10, 10}]{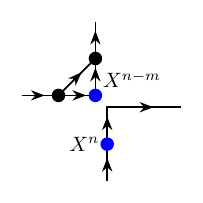}
= \text{RHS}.
\end{equation}
Here, the first equality follows from
\begin{equation}
\adjincludegraphics[valign=c, scale=1, trim={10, 10, 10, 10}]{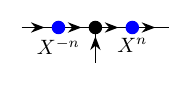} = \adjincludegraphics[valign=c, scale=1, trim={10, 10, 10, 10}]{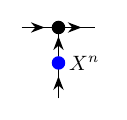}\;, \quad
\adjincludegraphics[valign=c, scale=1, trim={10, 10, 10, 10}]{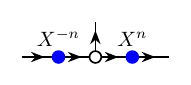} = \adjincludegraphics[valign=c, scale=1, trim={10, 10, 10, 10}]{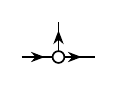}\;, \quad
\adjincludegraphics[valign=c, scale=1, trim={10, 10, 10, 10}]{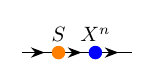} = \adjincludegraphics[valign=c, scale=1, trim={10, 10, 10, 10}]{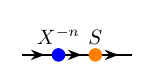}\;,
\label{eq: shift commutation relation}
\end{equation}
while the second equality follows from
\begin{equation}
\adjincludegraphics[valign=c, scale=1, trim={10, 10, 10, 10}]{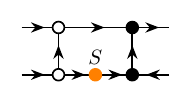} \;=\; \adjincludegraphics[valign=c, scale=1, trim={10, 10, 10, 10}]{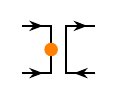}\;, \qquad
\text{where } \adjincludegraphics[valign=c, scale=1, trim={10, 10, 10, 10}]{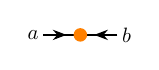} \coloneq \delta_{a, N-b} \quad \text{for $a ,b = 0, 1, \cdots, N-1$}.
\label{eq: ZN ladder}
\end{equation}

\vspace*{\baselineskip}
\noindent{\bf Broken zipper condition.}
The fusion and splitting tensors defined by \eqref{eq: phi TUn} satisfy the broken zipper condition.
More specifically, these tensors satisfy the zipper condition
\begin{equation}
\adjincludegraphics[valign=c, scale=1, trim={10, 10, 10, 10}]{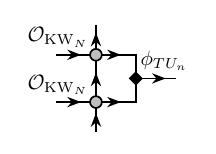} = \adjincludegraphics[valign=c, scale=1, trim={10, 10, 10, 10}]{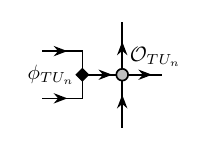}\;, \qquad
\adjincludegraphics[valign=c, scale=1, trim={10, 10, 10, 10}]{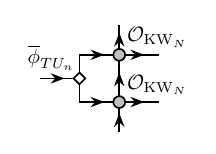} = \adjincludegraphics[valign=c, scale=1, trim={10, 10, 10, 10}]{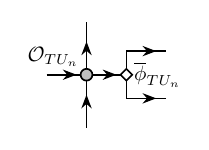}\;,
\label{eq: KWN unbroken zipper}
\end{equation}
which is stronger than the broken zipper condition.
The above equation immediately follows from \eqref{eq: ZN KW zipper derivation}.

\vspace*{\baselineskip}
\noindent{\bf Two-sided zipper condition.}
The fusion and splitting tensors defined by \eqref{eq: phi TUn} also satisfy the following two-sided zipper condition:
\begin{alignat}{2}
&\adjincludegraphics[valign=c, scale=1, trim={10, 10, 10, 10}]{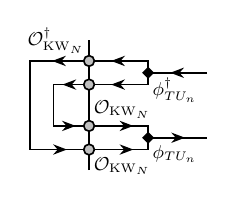} \;=\; \adjincludegraphics[valign=c, scale=1, trim={10, 10, 10, 10}]{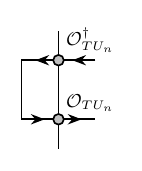}\;, \qquad
&&\adjincludegraphics[valign=c, scale=1, trim={10, 10, 10, 10}]{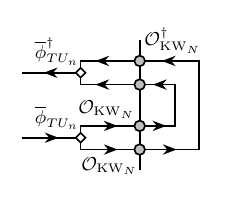} \;=\; \adjincludegraphics[valign=c, scale=1, trim={10, 10, 10, 10}]{tikz/out/KWN_two_sided_zipper4.pdf} \;,
\\
&\adjincludegraphics[valign=c, scale=1, trim={10, 10, 10, 10}]{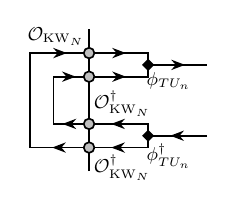} \;=\; \adjincludegraphics[valign=c, scale=1, trim={10, 10, 10, 10}]{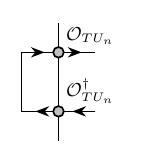}\;, \qquad
&&\adjincludegraphics[valign=c, scale=1, trim={10, 10, 10, 10}]{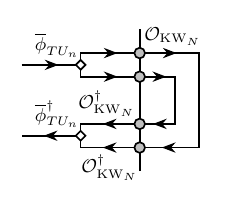} \;=\; \adjincludegraphics[valign=c, scale=1, trim={10, 10, 10, 10}]{tikz/out/KWN_dagger_two_sided_zipper4.pdf} \;.
\end{alignat}
The above equations immediately follow from the zipper condition \eqref{eq: KWN unbroken zipper} together with the following equalities:
\begin{equation}
\adjincludegraphics[valign=c, scale=1, trim={10, 10, 10, 10}]{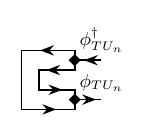} =\; \adjincludegraphics[valign=c, scale=1, trim={10, 10, 10, 10}]{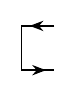}\;,
\quad
\adjincludegraphics[valign=c, scale=1, trim={10, 10, 10, 10}]{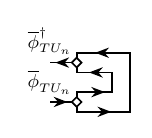} \;= \adjincludegraphics[valign=c, scale=1, trim={10, 10, 10, 10}]{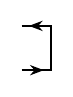}\;,
\qquad
\adjincludegraphics[valign=c, scale=1, trim={10, 10, 10, 10}]{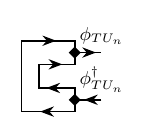} =\; \adjincludegraphics[valign=c, scale=1, trim={10, 10, 10, 10}]{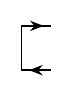}\;,
\quad
\adjincludegraphics[valign=c, scale=1, trim={10, 10, 10, 10}]{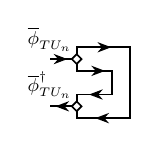} \;= \adjincludegraphics[valign=c, scale=1, trim={10, 10, 10, 10}]{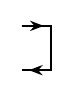}\;.
\end{equation}

\vspace*{\baselineskip}
\noindent{\bf Lattice quantum dimension and index.}
The lattice quantum dimension and the index of the $\mathbb{Z}_N$ Kramers-Wannier MPO $\mathcal{O}_{\mathrm{KW}_N}$ can be computed in the same way as in the case of $N=2$.
Concretely, one can compute the left and right dimensions of $\mathcal{O}_{\mathrm{KW}_N}$ as
\begin{equation}
\ldim(\mathcal{O}_{\mathrm{KW}_N}) = N, \qquad
\rdim(\mathcal{O}_{\mathrm{KW}_N}) = 1.
\end{equation}
Thus, the lattice quantum dimension and the index of $\mathcal{O}_{\text{KW}_N}$ are given by
\begin{equation}
\qdim_{\mathrm{lat}}(\mathcal{O}_{\mathrm{KW}_N}) = \sqrt{N}, \qquad
\ind(\mathcal{O}_{\mathrm{KW}_N}) = \sqrt{N}.
\end{equation}
We note that the lattice quantum dimension of $\mathcal{O}_{\mathrm{KW}_N}$ agrees with the quantum dimension of the unique non-invertible simple object of a $\mathbb{Z}_N$ Tambara-Yamagami category \cite{TY1998}.
The index of $\mathcal{O}_{\mathrm{KW}_N}$ computed above agrees with the index of the $\mathbb{Z}_N$ Kramers-Wannier duality operator viewed as a QCA on the algebra of $\mathbb{Z}_N$-symmetric local operators \cite{Ma:2024ypm} or a QCA on a fusion spin chain \cite{Jones:2023imy}.

\section{Summary and outlook}
\label{sec: Summary and outlook}
In this paper, we studied general structures of non-invertible symmetries on a tensor product Hilbert space in 1+1 dimensions using two complementary approaches: one is an axiomatic approach based on physical assumptions, and the other is a more concrete approach based on tensor networks.
In particular, we defined an index of non-invertible symmetry operators in 1+1d and discussed its role in constraining possible fusion rules of symmetry operators on a tensor product Hilbert space.
In what follows, we provide a more detailed summary of the results.

\vspace*{\baselineskip}
\noindent{\bf Axiomatic approach.}
Assuming the existence of defect Hilbert spaces satisfying Assumptions~\ref{assump: defect Hilbert space}--\ref{assump: Compatibility with the unitary structure}, we showed that the fusion rules of a unitary fusion category $\mathcal{C}$ can be realized by symmetry operators on a tensor product Hilbert space only when $\mathcal{C}$ is integral, if we do not allow mixing with QCAs.
This result agrees with the mathematical result recently proved in \cite{Evans:2025msy}.
Furthermore, based on the same assumptions, we showed that if the fusion rules of $\mathcal{C}$ are realized up to QCAs on a tensor product Hilbert space, then $\mathcal{C}$ must be weakly integral as long as the index is homogeneous in the sense that all fusion channels of two indecomposable symmetry operators have the same index.
This is consistent with the conjecture recently proposed in \cite{Tantivasadakarn2025KITP, Tantivasadakarn2025INI}, which states that a unitary fusion category symmetry realized on a tensor product Hilbert space must be weakly integral even if we allow mixing with spacetime symmetry.
We emphasize that the homogeneity of the index has not been proven, and hence the conjecture in \cite{Tantivasadakarn2025KITP, Tantivasadakarn2025INI} is still open.

\vspace*{\baselineskip}
\noindent{\bf Tensor network approach.}
We proposed a general class of tensor network operators, called topological injective MPOs, that describe symmetry operators on a tensor product Hilbert space in 1+1d.
For these MPOs, we defined the corresponding defect Hilbert spaces and constructed local unitary operators that move the defects.
We also showed that topological injective MPOs admit sequential quantum circuit representations, which is consistent with the recent proposal in \cite{Tantivasadakarn:2025txn}.

The tensor network description allows us to study the properties of the index of non-invertible symmetry operators without relying on physical assumptions.
In particular, within the framework of tensor networks, we can formulate the homogeneity of the index as a purely algebraic problem.
Based on this formulation, we provided sufficient conditions for the homogeneity of the index of topological injective MPOs.
More specifically, we showed that the index is homogeneous if every fusion channel of two topological injective MPOs has a pair of fusion and splitting tensors (cf. Definition~\ref{def: Fusion and splitting tensors}) that satisfy the broken zipper condition (cf. Definition~\ref{def: Broken zipper condition}) and the two-sided zipper condition (cf. Definition~\ref{def: Two-sided zipper condition}).
We provided various examples of topological injective MPOs whose fusion channels have fusion and splitting tensors satisfying the broken zipper condition and the two-sided zipper condition.
However, the existence of such fusion and splitting tensors for general topological injective MPOs has not yet been proven.

\vspace*{\baselineskip}
\noindent{\bf Outlook.}
Given that all examples of topological injective MPOs discussed in this paper satisfy the sufficient conditions for the homogeneity of the index, it would be natural to ask whether the same holds for all topological injective MPOs.
Namely, we would pose the following question:

\begin{question}\label{q: fusion and splitting tensors}
For every fusion channel of topological injective MPOs, does there exist a pair of a fusion tensor and a splitting tensor that satisfy both the broken zipper condition and the two-sided zipper condition, if we block a sufficient number of physical sites?
\end{question}

Combined with the results obtained in this paper, the above question leads to the following statement:

\begin{corollary}\label{cor: weakly integral}
If the answer to Question~\ref{q: fusion and splitting tensors} is affirmative, a unitary fusion category symmetry realized by topological injective MPOs must be weakly integral even if we allow mixing with non-trivial QCAs. More specifically, if there exists a finite set of topological injective MPOs whose fusion rules agree with those of a unitary fusion category $\mathcal{C}$ up to QCAs, then $\mathcal{C}$ must be weakly integral, as long as Question~\ref{q: fusion and splitting tensors} is answered affirmatively.
\end{corollary}

To answer Question~\ref{q: fusion and splitting tensors}, it would be helpful to study the algebraic properties of topological injective MPOs in more depth.
In recent years, people have studied general algebraic structures on MPOs and their relation to pre-bialgebras \cite{Molnar:2022nmh, Liu:2025iao}.
Investigating such structures on topological injective MPOs may give us insight into the problem of the fusion and splitting tensors for these MPOs.

We note that Corollary~\ref{cor: weakly integral} does not claim that any weakly integral unitary fusion category symmetry can be realized on a tensor product Hilbert space if we allow mixing with non-trivial QCAs.\footnote{On the other hand, any unitary fusion category symmetry can be realized exactly on a Hilbert space with local constraints \cite{Feiguin:2006ydp, Buican:2017rxc, Aasen:2020jwb}. This does not contradict Corollary~\ref{cor: weakly integral} because the projection to the constrained Hilbert space cannot be written as a topological injective MPO.}
Nevertheless, it was recently shown in \cite{Lu:2026rhb} that weakly integral fusion categories describing the self-dualities under gauging non-anomalous fusion category symmetries can be realized on a tensor product Hilbert space if we allow mixing with lattice translations.
It would be interesting to construct concrete lattice models that realize more general weakly integral fusion category symmetries on a tensor product Hilbert space.\footnote{Typical examples of weakly integral fusion categories are weakly group-theoretical ones. To the best of the author's knowledge, it is an open problem whether every weakly integral fusion category is weakly group-theoretical \cite[Question 2]{Etingof:0809.3031}.}
(Note added: after the first submission of this manuscript to arXiv, concrete lattice models that realize all weakly integral fusion category symmetries mixing with non-trivial QCAs were constructed in \cite{Wen:2026ncw}.)

Another interesting direction is to consider different realizations of fusion category symmetries on a tensor product Hilbert space.
Recently, it was found in \cite{Chatterjee:2024gje, Pace:2024oys} that a symmetry with non-commutative fusion rules dictated by the Onsager algebra on the lattice can flow to a $\mathrm{U}(1) \times \mathrm{U}(1)$ symmetry with a mixed anomaly in the continuum.
It would be interesting to study whether non-invertible symmetries also admit a similar realization on the lattice.

\begin{acknowledgments}
The author thanks Christian Copetti, Kantaro Ohmori, and Shuhei Ohyama for helpful discussions.
The author is supported in part by the EPSRC Open Fellowship EP/X01276X/1 and by the Leverhulme-Peierls Fellowship funded by the Leverhulme Trust.
A part of this work was done while visiting the Okinawa Institute of Science and Technology (OIST) through the Theoretical Sciences Visiting Program (TSVP) Thematic Program on Generalized Symmetries in Quantum Matter (TP25QM).
\end{acknowledgments}

\appendix

\section{Choice of $\Lambda^l$ and $\Lambda^r$ in Definition \ref{def: Topological injective MPO}}
\label{sec: Choice of Lambda}
In this appendix, we show that if there exist positive definite linear maps $\Lambda^l$ and $\Lambda^r$ that satisfy \eqref{eq: topological injective MPO}, the fixed points of the transfer matrix $\mathsf{T}[\mathcal{O}]$ must also satisfy \eqref{eq: topological injective MPO}.
The same applies to \eqref{eq: O dagger topological}.
Therefore, as long as we impose the positive definiteness of $\Lambda^l$ and $\Lambda^r$ in Definition~\ref{def: Topological injective MPO}, we can choose the fixed points of $\mathsf{T}[\mathcal{O}]$ as $\Lambda^l$ and $\Lambda^r$ without loss of generality.
In what follows, we will show the above statement only for the right fixed point.
The case of the left fixed point is similar.

To show the above statement, we suppose that there exists a positive definite linear map $\widetilde{\Lambda}^r$ that satisfies the second equality of \eqref{eq: topological injective MPO}, i.e.,
\begin{equation}
\adjincludegraphics[valign=c, scale=1, trim={10, 10, 10, 10}]{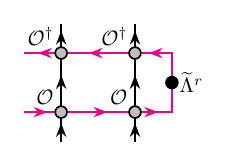}
\;\; = \;\;
\adjincludegraphics[valign=c, scale=1, trim={10, 10, 10, 10}]{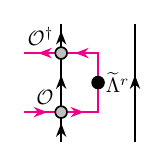} \;.
\label{eq: Lambda tilde topological}
\end{equation}
If we take the trace over both physical legs, the above equation reduces to
\begin{equation}
\mathsf{T}[\mathcal{O}] \mathsf{T}[\mathcal{O}] \widetilde{\Lambda}^r = \dim(\mathcal{H}_{\mathrm{o}}) \mathsf{T}[\mathcal{O}] \widetilde{\Lambda}^r,
\label{eq: TOTO Lambda}
\end{equation}
where $\mathsf{T}[\mathcal{O}]$ is the transfer matrix of $\mathcal{O}$.
Equation~\eqref{eq: TOTO Lambda} shows that $\mathsf{T}[\mathcal{O}]\widetilde{\Lambda}^r$ is an eigenvector of $\mathsf{T}[\mathcal{O}]$ with eigenvalue $\dim(\mathcal{H}_{\mathrm{o}})$.
On the other hand, $\mathsf{T}[\mathcal{O}]\widetilde{\Lambda}^r$ is positive definite because the transfer matrix of an injective MPO tensor preserves the positive definiteness; see, e.g., \cite[Section V]{Molnar:2018hls}.
Since a positive definite eigenvector of $\mathsf{T}[\mathcal{O}]$ is unique and is given by the fixed point due to \cite[Theorem 2.4]{Evans1978}, we find that $\mathsf{T}[\mathcal{O}]\widetilde{\Lambda}^r$ is the right fixed point of $\mathsf{T}[\mathcal{O}]$.
Therefore, $\widetilde{\Lambda}^r$ can be written as
\begin{equation}
\widetilde{\Lambda}^r = \Lambda^r +\Delta,
\label{eq: Lambda tilde def}
\end{equation}
where $\Lambda^r$ is the right fixed point of $\mathsf{T}[\mathcal{O}]$ and $\Delta$ is an eigenvector of $\mathsf{T}[\mathcal{O}]$ with eigenvalue zero.
Now, we plug \eqref{eq: Lambda tilde def} into \eqref{eq: Lambda tilde topological} and write down equations for $\Lambda^r$ and $\Delta$.
To this end, we use the decomposition
\begin{equation}
\widetilde{\mathsf{T}}[\mathcal{O}] \coloneq \adjincludegraphics[valign=c, scale=1, trim={10, 10, 10, 10}]{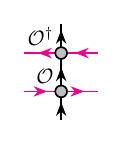}
= \frac{1}{\dim(\mathcal{H}_{\mathrm{o}})} \mathsf{T}[\mathcal{O}] \otimes \id_{\mathcal{H}_{\mathrm{o}}} + \sum_{\alpha} S_{\alpha} \otimes \sigma_{\alpha},
\label{eq: TO tilde}
\end{equation}
where $\sigma_{\alpha}$'s are linearly independent traceless operators acting on the physical leg and $S_{\alpha}$'s are operators acting on the pair of the top and bottom virtual bonds.
Using the above decomposition, we can compute the left-hand side and the right-hand side of \eqref{eq: Lambda tilde topological} as follows:
\begin{align}
\text{LHS} &= \widetilde{\mathsf{T}}[\mathcal{O}]^{(1)} \Lambda^r \otimes \id_{\mathcal{H}_{\mathrm{o}}}^{(2)} + \widetilde{\mathsf{T}}[\mathcal{O}]^{(1)} \sum_{\alpha} S_{\alpha}(\Lambda^r + \Delta) \otimes \sigma_{\alpha}^{(2)}, \\
\text{RHS} &= \widetilde{\mathsf{T}}[\mathcal{O}]^{(1)} \Lambda^r \otimes \id_{\mathcal{H}_{\mathrm{o}}}^{(2)} + \sum_{\alpha} S_{\alpha} \Delta \otimes \sigma_{\alpha}^{(1)} \otimes \id_{\mathcal{H}_{\mathrm{o}}}^{(2)}.
\end{align}
Here, the operators with superscripts 1 and 2 act on the left and right physical legs, respectively.
Due to the above equations, the condition~\eqref{eq: Lambda tilde topological} reduces to
\begin{equation}
\widetilde{\mathsf{T}}[\mathcal{O}]^{(1)} \sum_{\alpha} S_{\alpha}(\Lambda^r + \Delta) \otimes \sigma_{\alpha}^{(2)} = \sum_{\alpha} S_{\alpha} \Delta \otimes \sigma_{\alpha}^{(1)} \otimes \id_{\mathcal{H}_{\mathrm{o}}}^{(2)}.
\end{equation}
Since $\sigma_{\alpha}$'s are linearly independent traceless operators, the above equation is equivalent to 
\begin{equation}
S_{\alpha} \Delta = 0, \qquad \widetilde{\mathsf{T}}[\mathcal{O}] S_{\alpha} \Lambda^r = 0, \qquad \forall \alpha.
\end{equation}
In particular, if there exists a pair $(\Lambda^r, \Delta)$ that satisfies the above equation, the pair $(\Lambda^r, 0)$ also satisfies the same equation.
Recalling that $\widetilde{\Lambda}^r$ is defined by \eqref{eq: Lambda tilde def}, we conclude that if there exists a positive definite $\widetilde{\Lambda}^r$ that satisfies \eqref{eq: Lambda tilde topological}, the fixed point $\Lambda^r$ also satisfies the same equation.

\section{Equivalence of Definitions \ref{def: Topological injective MPO} and \ref{def: Topological injective MPO 2}}
\label{sec: Equivalence of definitions}
In Section~\ref{sec: Definitions}, we provided two definitions of a topological injective MPO, namely, Definitions~\ref{def: Topological injective MPO} and \ref{def: Topological injective MPO 2}.
In this appendix, we show that these definitions are equivalent.
More specifically, we show that \eqref{eq: O dagger topological} and \eqref{eq: zigzag def} are equivalent given that the injective MPO tensor $\mathcal{O}$ satisfies \eqref{eq: topological injective MPO}.
As shown in Section~\ref{sec: Properties of topological injective MPOs}, equation~\eqref{eq: O dagger topological} together with \eqref{eq: topological injective MPO} implies \eqref{eq: zigzag def}.
Thus, in what follows, it suffices to show that equation~\eqref{eq: zigzag def} together with \eqref{eq: topological injective MPO} implies \eqref{eq: O dagger topological}.

To show \eqref{eq: O dagger topological}, we recall that $\mathcal{O}$ can be decomposed into two three-leg tensors as in~\eqref{eq: AB decomposition}.
Using this decomposition, we can write the zigzag relations in \eqref{eq: zigzag def} as
\begin{equation}
\adjincludegraphics[valign=c, scale=1, trim={10, 10, 10, 10}]{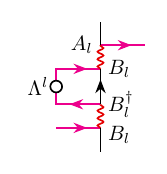}
\;=\; \delta_l(\mathcal{O}) \;\; \adjincludegraphics[valign=c, scale=1, trim={10, 10, 10, 10}]{tikz/out/AB_l.pdf} \;, \qquad
\adjincludegraphics[valign=c, scale=1, trim={10, 10, 10, 10}]{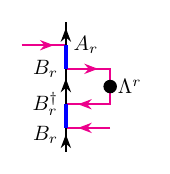}
\;=\; \delta_r(\mathcal{O}) \;\; \adjincludegraphics[valign=c, scale=1, trim={10, 10, 10, 10}]{tikz/out/AB_r.pdf} \;,
\end{equation}
where we used \eqref{eq: A Lambda A}.
Since $A_l$ and $A_r$ have left inverses and $B_l$ and $B_r$ have right inverses by construction, the above equation implies
\begin{equation}
\adjincludegraphics[valign=c, scale=1, trim={10, 10, 10, 10}]{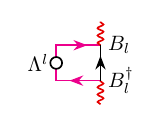}
\;=\; \delta_l(\mathcal{O}) \;\; \adjincludegraphics[valign=c, scale=1, trim={10, 10, 10, 10}]{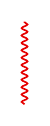} \;, \qquad
\adjincludegraphics[valign=c, scale=1, trim={10, 10, 10, 10}]{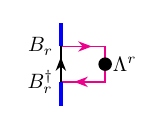}
\;=\; \delta_r(\mathcal{O}) \;\; \adjincludegraphics[valign=c, scale=1, trim={10, 10, 10, 10}]{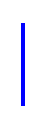} \;.
\label{eq: B Lambda B}
\end{equation}
Using this equation, we can show the first equality of \eqref{eq: O dagger topological} by a direct computation as follows:
\begin{equation}
\adjincludegraphics[valign=c, scale=1, trim={10, 10, 10, 10}]{tikz/out/dagger_topological_injective_MPO_l1.pdf}
\;=\; \delta_l(\mathcal{O}) \; \adjincludegraphics[valign=c, scale=1, trim={10, 10, 10, 10}]{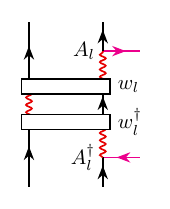}
\;=\; \delta_l(\mathcal{O}) \; \adjincludegraphics[valign=c, scale=1, trim={10, 10, 10, 10}]{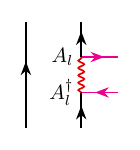}
\;=\; \adjincludegraphics[valign=c, scale=1, trim={10, 10, 10, 10}]{tikz/out/dagger_topological_injective_MPO_l2.pdf} \;.
\end{equation}
Here, $w_l$ is defined by \eqref{eq: wl wr} and we used the fact that it is unitary \eqref{eq: wl unitary}.\footnote{We note that the derivation of the unitarity of $w_l$ does not rely on \eqref{eq: O dagger topological}: it follows only from \eqref{eq: topological injective MPO}.}
A similar computation also shows the second equality of \eqref{eq: O dagger topological}.
Thus, we find that Definitions~\ref{def: Topological injective MPO} and \ref{def: Topological injective MPO 2} are equivalent.

\section{Determination of $\delta_l(\mathcal{O})$ and $\delta_r(\mathcal{O})$ in Definition \ref{def: Topological injective MPO 2}}
\label{sec: Derivation of delta lr}
In this appendix, following the proof of \cite[Proposition 1]{Franco-Rubio:2025qss}, we show that $\delta_l(\mathcal{O})$ and $\delta_r(\mathcal{O})$ in the zigzag equation~\eqref{eq: zigzag def} are uniquely given by the positive real numbers defined by~\eqref{eq: delta lr}.
To this end, we recall that the three-leg tensors $B_l$ and $B_r$ in the decomposition~\eqref{eq: AB decomposition} satisfy \eqref{eq: B Lambda B}.
By taking the trace of both sides of this equation, we obtain
\begin{align}
\delta_l(\mathcal{O}) &= \frac{1}{\lrank(\mathcal{O})} \tr{\adjincludegraphics[valign=c, scale=1, trim={10, 10, 10, 10}]{tikz/out/delta_l2.pdf}}
\;=\; \frac{1}{\lrank(\mathcal{O})} \tr{\adjincludegraphics[valign=c, scale=1, trim={10, 10, 10, 10}]{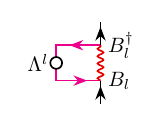}}, \label{eq: delta l trace} \\
\delta_r(\mathcal{O}) &= \frac{1}{\rrank(\mathcal{O})} \tr{\adjincludegraphics[valign=c, scale=1, trim={10, 10, 10, 10}]{tikz/out/delta_r2.pdf}}
\;=\; \frac{1}{\rrank(\mathcal{O})} \tr{\adjincludegraphics[valign=c, scale=1, trim={10, 10, 10, 10}]{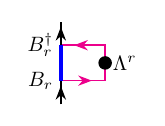}}. \label{eq: delta r trace}
\end{align}
On the other hand, equation~\eqref{eq: bubble removal} together with \eqref{eq: A Lambda A} implies that $B_l$ and $B_r$ satisfy
\begin{equation}
\adjincludegraphics[valign=c, scale=1, trim={10, 10, 10, 10}]{tikz/out/delta_l4.pdf}
\:=\; \gamma \;\; \adjincludegraphics[valign=c, scale=1, trim={10, 10, 10, 10}]{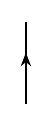}
\;=\; \adjincludegraphics[valign=c, scale=1, trim={10, 10, 10, 10}]{tikz/out/delta_r4.pdf}.
\end{equation}
By plugging this into \eqref{eq: delta l trace} and \eqref{eq: delta r trace}, we obtain
\begin{equation}
\delta_l(\mathcal{O}) = \frac{\gamma}{\ldim(\mathcal{O})}, \qquad
\delta_r(\mathcal{O}) = \frac{\gamma}{\rdim(\mathcal{O})},
\end{equation}
which shows \eqref{eq: delta lr}.

\section{Derivation of \eqref{eq: RepA MPO dagger}}
\label{sec: RepA MPO dagger}
In this appendix, we show that the MPO tensor~\eqref{eq: RepA MPO} labeled by a finite dimensional $*$-representation $(V, \rho)$ of a finite dimensional semisimple Hopf $*$-algebra $A$ satisfies \eqref{eq: RepA MPO dagger}.
Equivalently, we show
\begin{equation}
\begin{aligned}
\adjincludegraphics[valign=c, scale=1, trim={10, 10, 10, 10}]{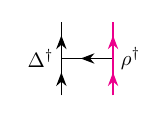}
\;&=\; \adjincludegraphics[valign=c, scale=1, trim={10, 10, 10, 10}]{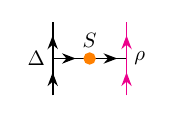}\;, \quad \text{i.e.,} \\
(\Delta^{\dagger} \otimes \id_V) \circ (\id_A \otimes \rho^{\dagger}) &= (\id_A \otimes \rho) \circ (\id_A \otimes S \otimes \id_V) \circ (\Delta \otimes \id_V),
\end{aligned}
\label{eq: RepA MPO dagger derivation}
\end{equation}
where the Hermitian conjugate is defined with respect to the inner product~\eqref{eq: inner product on A}.
In the above equation, the representation map $\rho$ is regarded as a linear map from $A \otimes V$ to $V$.
The same linear map regarded as a map from $A$ to $\End(V)$ via the canonical isomorphism $\Hom(A \otimes V, V) \cong \Hom(A, \End(V))$ will be denoted by $\varrho$.\footnote{In Section~\ref{sec: Non-anomalous fusion category symmetries}, both $\rho$ and $\varrho$ are denoted by $\rho$ by abuse of notation.}
We note that $\rho$ and $\varrho$ are related by $\rho(a \otimes v) = \varrho(a)(v)$ for all $a \in A$ and $v \in V$.
In what follows, the inner product on a Hilbert space $X$ is denoted by $\braket{x | y}_X$ for $x, y \in X$, where $X$ is either $A$, $V$, $A \otimes A$, or $A \otimes V$.
Furthermore, orthonormal bases of $A$ and $V$ are denoted by $\{a_i \mid i = 1, 2, \cdots, \dim(A)\}$ and $\{v_{\mu} \mid \mu = 1, 2 \cdots, \dim(V)\}$, respectively.

To show \eqref{eq: RepA MPO dagger derivation}, we compute the components of the linear maps on both sides.
On the one hand, the components of the linear map on the left-hand side can be computed as
\begin{equation}
\begin{aligned}
& \quad \braket{a_i \otimes v_{\mu} | (\Delta^{\dagger} \otimes \id_V) \circ (\id_A \otimes \rho^{\dagger}) (a_j \otimes v_{\nu})}_{A \otimes V} \\
&= \sum_{k = 1}^{\dim(A)} \braket{a_i | \Delta^{\dagger}(a_j \otimes a_k)}_A \braket{a_k \otimes v_{\mu} | \rho^{\dagger}(v_{\nu})}_{A \otimes V}
= \sum_{k = 1}^{\dim(A)} \braket{\Delta(a_i) | a_j \otimes a_k}_{A \otimes A} \braket{\rho(a_k \otimes v_{\mu}) | v_{\nu}}_V \\
&= \sum_{k = 1}^{\dim(A)} \braket{\Delta(a_i) | a_j \otimes a_k}_{A \otimes A} \braket{v_{\mu} | \varrho(a_k)^{\dagger}(v_{\nu})}_V
= \sum_{k = 1}^{\dim(A)} \braket{\Delta(a_i) | a_j \otimes a_k}_{A \otimes A} \braket{v_{\mu} | \varrho(a_k^*)(v_{\nu})}_V, 
\end{aligned}
\end{equation}
where the last equality follows from the fact that $(V, \rho)$ is a $*$-representation.
On the other hand, the component of the linear map on the right-hand side can be computed as
\begin{equation}
\begin{aligned}
&\quad \braket{a_i \otimes v_{\mu} | (\id_A \otimes \rho) \circ (\id_A \otimes S \otimes \id_V) \circ (\Delta \otimes \id_V) (a_j \otimes v_{\nu})}_{A \otimes V} \\
&= \sum_{k=1}^{\dim(A)} \braket{a_i \otimes a_k | (\id_A \otimes S) \circ \Delta(a_j)}_{A \otimes A} \braket{v_{\mu} | \varrho(a_k)(v_{\nu})}_V.
\end{aligned}
\end{equation}
The above equations imply that \eqref{eq: RepA MPO dagger derivation} holds if we have
\begin{equation}
\braket{\Delta(a_i) | a_j \otimes a_k}_{A \otimes A} = \braket{a_i \otimes a_k | (\id_A \otimes S) \circ \Delta(a_j)}_{A \otimes A}, \qquad \forall i, j, k,
\label{eq: RepA MPO dagger derivation 2}
\end{equation}
in a basis where $a_i^* = a_i$ for all $i = 1, 2, \cdots, \dim(A)$.\footnote{Such a basis always exists. Indeed, since $A$ is semisimple, it is isomorphic to a direct sum of endomorphism algebras $\bigoplus \End(V_X)$, where the direct sum is taken over all (isomorphism classes of) irreducible $*$-representations $(V_X, \rho_X) \in \mathrm{Rep}(A)$. The $*$-operation and the inner product on $\bigoplus \End(V_X)$ are given by the Hermitian conjugation and the Hilbert-Schmidt inner product, respectively. We can then take $\{a_i \mid i = 1, 2, \cdots, \dim(A)\}$ to be a Hermitian orthonormal basis of $\bigoplus \End(V_X)$.}
In what follows, we will show that \eqref{eq: RepA MPO dagger derivation 2} holds in any orthonormal basis of $A$.

To show \eqref{eq: RepA MPO dagger derivation 2}, we note that the left-hand side of \eqref{eq: RepA MPO dagger derivation 2} can be written as
\begin{equation}
\braket{\Delta(a_i) | a_j \otimes a_k}_{A \otimes A} = \braket{(\id_A \otimes a^k) \circ \Delta(a_i) | a_j}_A = \braket{\widehat{\Delta}(a^k)(a_i) | a_j}_A,
\label{eq: RepA MPO dagger derivation 2a}
\end{equation}
where $\{a^k \mid k = 1, 2, \cdots, \dim(A)\}$ is the dual basis of $\{a_k \mid k = 1, 2, \cdots, \dim(A)\}$, and $\widehat{\Delta}: A^* \to \End(A)$ is defined by
\begin{equation}
\widehat{\Delta} (\phi) \coloneq (\id_A \otimes \phi) \circ \Delta, \qquad \forall \phi \in A^*.
\end{equation}
One can show that $(A, \widehat{\Delta})$ is a $*$-representation of the dual Hopf $*$-algebra $A^*$,\footnote{This immediately follows from Example 3.1.6 and Proposition 3.1.7 of \cite{Timmermann2008}.} where the $*$-operation on $A^*$ is defined by $\phi^* \coloneq \phi \circ S$ for all $\phi \in A^*$.
Therefore, it follows that
\begin{equation}
\braket{\widehat{\Delta}(a^k)(a_i) | a_j}_A
= \braket{a_i | \widehat{\Delta}((a^k)^*)(a_j)}_A
= \braket{a_i \otimes a_k | (\id_A \otimes S) \circ \Delta (a_j)}_A.
\label{eq: RepA MPO dagger derivation 2b}
\end{equation}
Equations~\eqref{eq: RepA MPO dagger derivation 2a} and \eqref{eq: RepA MPO dagger derivation 2b} show \eqref{eq: RepA MPO dagger derivation 2}.
Thus, we find that \eqref{eq: RepA MPO dagger derivation} holds.

\bibliographystyle{ytphys}
\bibliography{bibliography}

\end{document}